# Raster Domain Text Steganography: A Unified Framework for Multimodal Secure Embedding

*A V Uday Kiran Kandala,
Queen Mary University of London, London, United Kingdom.
Email: a.kandala@se25.qmul.ac.uk

**ABSTRACT:** This work introduces a unified raster domain steganographic framework, termed as the Glyph Perturbation Cardinality (GPC) framework, capable of embedding heterogeneous data such as text, images, audio, and video directly into the pixel space of rendered textual glyphs. Unlike linguistic or structural text based steganography, the proposed method operates exclusively after font rasterization, modifying only the bitmap produced by a deterministic text rendering pipeline. Each glyph functions as a covert encoding unit, where a payload value is expressed through the cardinality of minimally perturbed interior ink pixels. These minimal intensity increments remain visually imperceptible while forming a stable and decodable signal. The framework is demonstrated for text to text embedding and generalized to multimodal inputs by normalizing image intensities, audio derived scalar features, and video frame values into bounded integer sequences distributed across glyphs. Decoding is achieved by re-rasterizing the cover text, subtracting canonical glyph rasters, and recovering payload values via pixel count analysis. The approach is computationally lightweight, and grounded in deterministic raster behavior, enabling ordinary text to serve as a visually covert medium for multimodal data embedding.

## 1. INTRODUCTION:

Text based steganography has historically focused on manipulating linguistic or structural elements of natural language. Early systems embedded information through syntactic transformations, synonym substitution, lexical choice constraints, whitespace modulation, or invisible control characters, enabling low capacity communication channels tightly bound to grammatical and semantic consistency. Later work expanded to semantic rewriting, paraphrasing models, and limited font level variations, yet these approaches remain constrained by the inherent rigidity of natural language and the narrow representational bandwidth of symbolic text. A parallel line of research sought to encode non-textual data such as binary payloads, feature vectors, or compressed multimedia by treating text as a sequence of abstract symbols, using techniques including whitespace patterns, invisible Unicode ranges, homoglyph substitution, and reversible font shaping tricks.

Early whitespace based approaches manipulated visible document structure, for example by encoding information through variations in spacing or line layout [1]. While it is simple and language agnostic, these methods offered limited embedding capacity and were highly vulnerable to normalization and reformatting. Subsequent linguistic approaches, such as synonym based substitution [2], attempted to exploit lexical redundancy, but often suffered from semantic drift and degradation in plausibility when contextual constraints were violated. Researchers later explored syntactic transformation based methods [3], which embed data by selecting among predefined grammatical structures; however, their reliance on rule based NLP limited linguistic coverage and increased sensitivity to text modification.

Beyond linguistic manipulation, structural text steganography has exploited document layout rather than textual content, for example by encoding information through relative paragraph sizes using splitting, merging, and resizing operations [4]. Zheng et al. [6] introduced glyph based text steganography, encoding information via controlled perturbations along a learned font manifold, enabling robust extraction from vector documents, raster images, and printed text while preserving semantic content. Hybrid approaches have combined linguistic structuring with format based encoding, such as POS-based word layering coupled with subtle RGB modulation of characters to increase embedding capacity while maintaining visual imperceptibility [7]. More recent hybrid schemes integrate preprocessing of the secret message with the embedding process, for example by combining multilayer encoding, format preserving encryption, and Huffman compression prior to embedding data using invisible Unicode characters [8]. Earlier work also explored zero distortion text steganography, where the cover text serves solely as a reference and secret bits are represented via location matrices, often combined with abbreviation based compression and chaotic index encryption to increase capacity without modifying visible text [9].

With the rise of machine learning approaches, learning based steganographic carriers have emerged. Adversarial OCR based methods embed information by inducing controlled recognition errors in optical character recognition systems, enabling covert transmission that remains difficult to detect by alternative recognizers [5]. Neural linguistic steganography has framed embedding as a controlled text generation problem, for example using character level LSTM language models to encode secret bits through probabilistic symbol selection, achieving higher capacity and lower perplexity than word level models [12]. Early coverless text steganography relied on probabilistic language models such as Markov chains to guide text generation and bit embedding [16], laying the groundwork for subsequent neural and large language model based approaches.

Recent work has further advanced LLM based linguistic steganography by incorporating controllability and semantic constraints. Shi et al. [17] proposed an emotionally controllable framework integrating named entity recognition and sentiment analysis to improve imperceptibility and contextual alignment. To mitigate semantic drift in long text generation, Li et al. [18] introduced a semantic controllable framework guided by knowledge graph triplets and prompt engineering. Huang et al. [19] formulated LLM based text steganography as an entropy maximization problem under a perceptual KL-divergence constraint, significantly improving embedding efficiency without sacrificing fluency. Addressing the lack of token level probability access in commercial LLMs, Wu et al. [20] proposed a black box generative steganography method based on keyword construction, encrypted mappings, and rejection sampling.

Beyond text modification or generation, coverless approaches have explored cross modal semantic relationships, such as MM-Stega, which embeds information in the relevance between paired texts and images without altering either modality [13]. Comprehensive surveys and evaluations [10,11,14,15] have systematically compared format based, linguistic, hybrid, and generative text steganography techniques, consistently identifying trade offs among embedding capacity, robustness, and detectability, and highlighting persistent challenges such as semantic incoherence and vulnerability to steganalysis in many classical and hybrid schemes.

Across prior work, text steganography has remained constrained by the symbolic rigidity of natural language. Linguistic and semantic methods struggle to preserve contextual coherence, format based schemes are fragile under normalization and re-rendering, and statistical or neural generators often introduce detectable distributional artifacts. Even recent learning based approaches largely operate at the logical text level, manipulating tokens or probabilities rather than the visual form ultimately perceived by readers. As a result, the discrete nature of text symbols and the limited redundancy of Unicode sequences impose a persistent bottleneck on achieving high capacity, low detectability embedding.

In contrast, the present work shifts the embedding process into the raster domain of rendered text. Rather than modifying characters, Unicode points, or linguistic structures, the proposed method operates strictly after font rasterization, perturbing carefully selected pixels within the bitmap representation of each glyph under a fixed and deterministic rendering pipeline. Operating at this stage removes linguistic constraints and yields an embedding mechanism that is independent of text encoding, tokenization, or layout normalization.

By treating each rendered glyph as a stable pixel container, the framework encodes information through a perturbation cardinality, the number of minimally modified interior ink pixels within a glyph. These minimal perturbations remain visually imperceptible under normal viewing conditions, yet are recoverable by re-rasterizing the cover text and analyzing localized pixel differences. Because the same mechanism applies uniformly to normalized numerical representations, the proposed framework supports multimodal embedding of text, images, audio, and video within ordinary rendered text. This work establishes raster domain glyph perturbation as a distinct and underexplored direction in text steganography, demonstrating its feasibility and generality, while detailed optimization, security analysis, and modality specific refinements are deferred to future work.

## 2. DETERMINISTIC RASTERIZATION AND PERTURBATION PIPELINE:

All embedding and decoding operations in the proposed framework rely on a strictly deterministic text rasterization pipeline. Prior to perturbation, the cover text is rendered using a fixed font file, font size, resolution, and software rasterization backend, producing a stable grayscale bitmap for each glyph. To ensure reproducibility, the rendering process is constrained to a controlled configuration in which sources of platform dependent variability such as subpixel positioning, font hinting, and adaptive anti aliasing are either fixed by the rendering backend or held constant across encoding and decoding.

Once rasterized, each glyph is treated as an isolated bitmap container. Perturbations are applied exclusively within the interior stroke region of the glyph, defined as the set of pixels whose intensity lies below a fixed foreground intensities in the canonical raster. Edge pixels and background regions are excluded to avoid visible contour distortion. A deterministic index function maps the integer payload assigned to each glyph to a specific subset of interior pixels. This mapping is fixed and reproducible, ensuring that the same payload always selects the same pixel locations for a given glyph raster. Selected pixels undergo minimal intensity adjustments, producing black to near black perturbations that remain visually imperceptible under normal viewing conditions as shown in the figure 1.

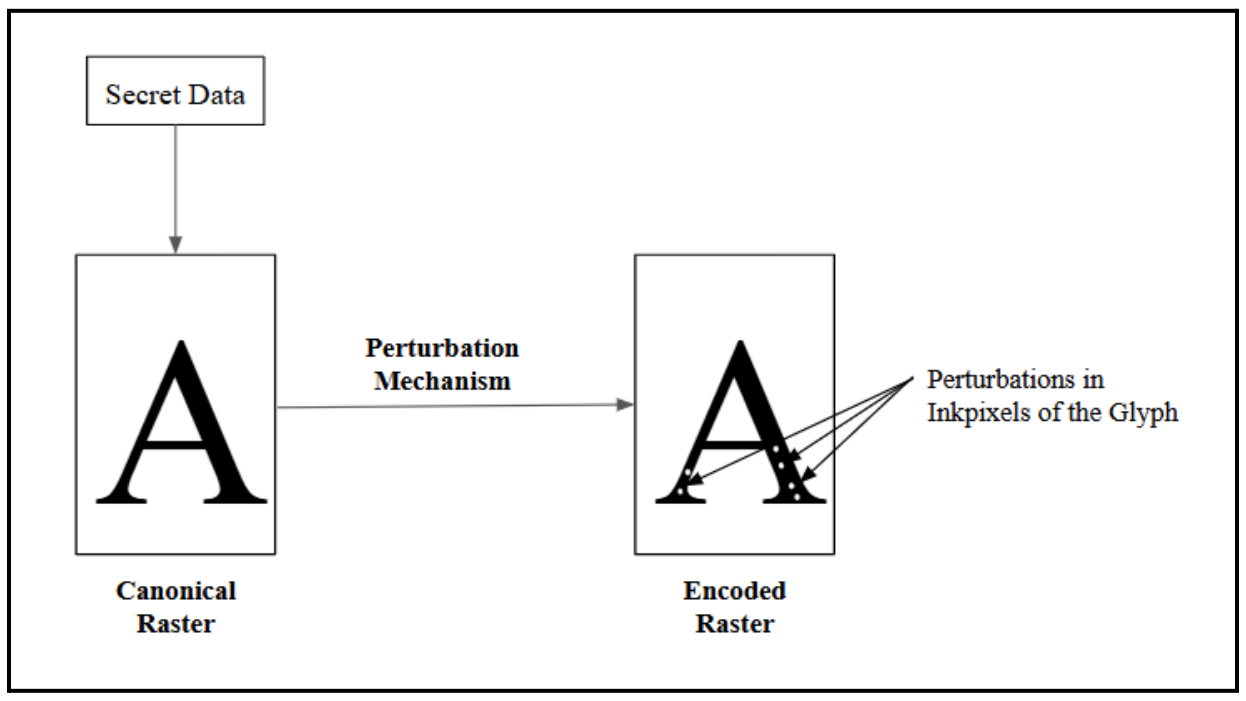

**Fig 1. Perturbating secret data in canonical raster**

Decoding proceeds by regenerating the canonical glyph rasters using the same rendering parameters and computing pixel wise differences with the encoded raster. Pixels whose intensity difference exceeds the black pixel intensities (0) are counted, and the resulting perturbation cardinality directly recovers the embedded integer. Because both the rasterization and pixel indexing processes are deterministic, the perturbation pattern is invertible as long as the raster representation is preserved bit by bit.

All embedding and recovery mechanisms described in the following sections assume this deterministic rasterization pipeline and operate entirely within the resulting pixel domain.

## 3. METHODOLOGY:

The proposed framework unifies text, image, audio, and video steganography by reducing all modalities to ordered numerical sequences that are embedded directly within the raster domain of rendered text glyphs. Rather than manipulating linguistic structure, font selection, or document layout, the method operates exclusively on the bitmap produced by a deterministic text rendering pipeline. Each glyph functions as a bounded capacity carrier, where a payload value is encoded as the cardinality of minimally perturbed interior stroke pixels. These perturbations are applied strictly within inked regions, remain visually imperceptible under normal viewing conditions, and are preserved across identical re-renderings. Once a modality is normalized into bounded integer values, the resulting sequence is distributed across glyphs in reading order, enabling multimodal embedding while preserving the visual appearance of ordinary text, as shown in Figure 2. In all modalities, rendered glyphs are treated as independent raster tiles, each serving as a fixed size container for a single payload value, and are concatenated in reading order to form the final encoded raster. Where, the glyphs are rendered as fixed size raster tiles to isolate and validate the perturbation channel and extension of the framework to naturally formatted documents with mixed case, variable kerning, and paragraph layout is left for future work.The following subsections describe the encoding and decoding pipelines for each modality, detailing how payload values are constructed, embedded, and recovered through differential raster analysis.

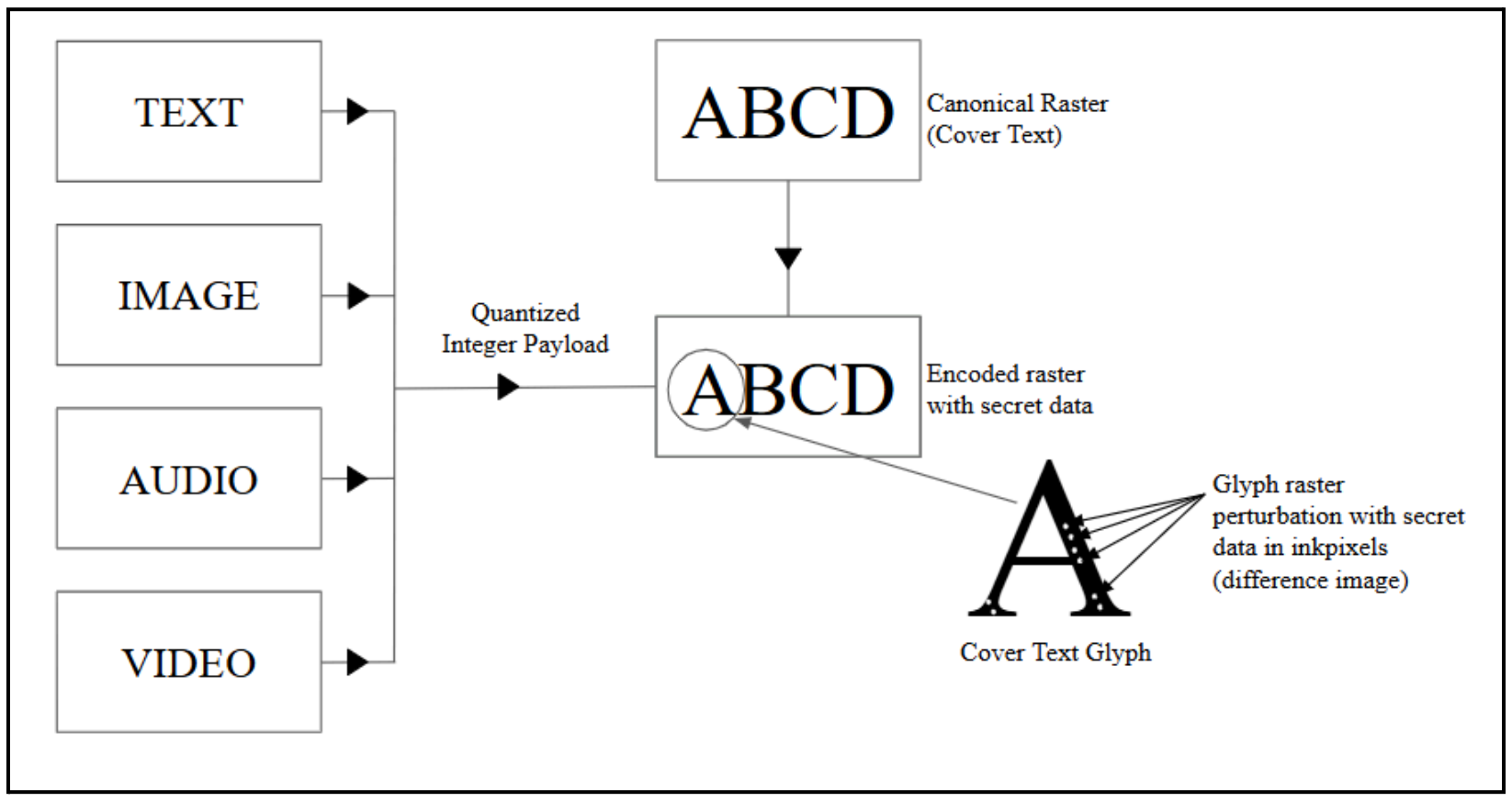

**Fig 2. Multimodal data embedding into text raster glyphs**

### 3.1. TEXT-TO-TEXT STEGANOGRAPHY USING RASTER GLYPH PERTURBATIONS:

Text-to-text embedding serves as the foundational instantiation of the proposed framework, as it directly illustrates how symbolic information is encoded within the raster domain of rendered glyphs. Each cover character is rendered into a bitmap using a deterministic text rendering pipeline with fixed font, size, rasterizer, ensuring that identical characters produce identical canonical pixel layouts across repeated renderings. Within this stable raster representation, information is embedded by applying minimal, sub perceptual perturbations to selected interior stroke pixels, leaving the visual appearance of the glyph unchanged under normal viewing conditions.

Each secret character $S \in \{A,\ldots,Z\}$ is mapped to an integer payload,

$$V = \mathrm{Ord}(S) - 64, V \in [1,26],$$

which specifies the number of ink pixels within the corresponding glyph that will be perturbed. Because all modifications occur strictly after rasterization and the perturbation magnitude is kept below human perceptual thresholds, the encoded text remains visually indistinguishable from its canonical rendering.

**3.1.1. Encoding Procedure:**

For each cover character $C_i$ , the corresponding glyph is rasterized to obtain a bitmap $G_i$ . The set of ink pixels belonging to the interior stroke region is identified and serves as the admissible perturbation space. Using a deterministic, seeded selection rule, exactly $V_i$ ink pixels are chosen and incremented by a small constant intensity Δ (typically one gray level unit in all experiments). All perturbed glyphs are then concatenated in reading order to produce the encoded text raster as shown in figure 3.

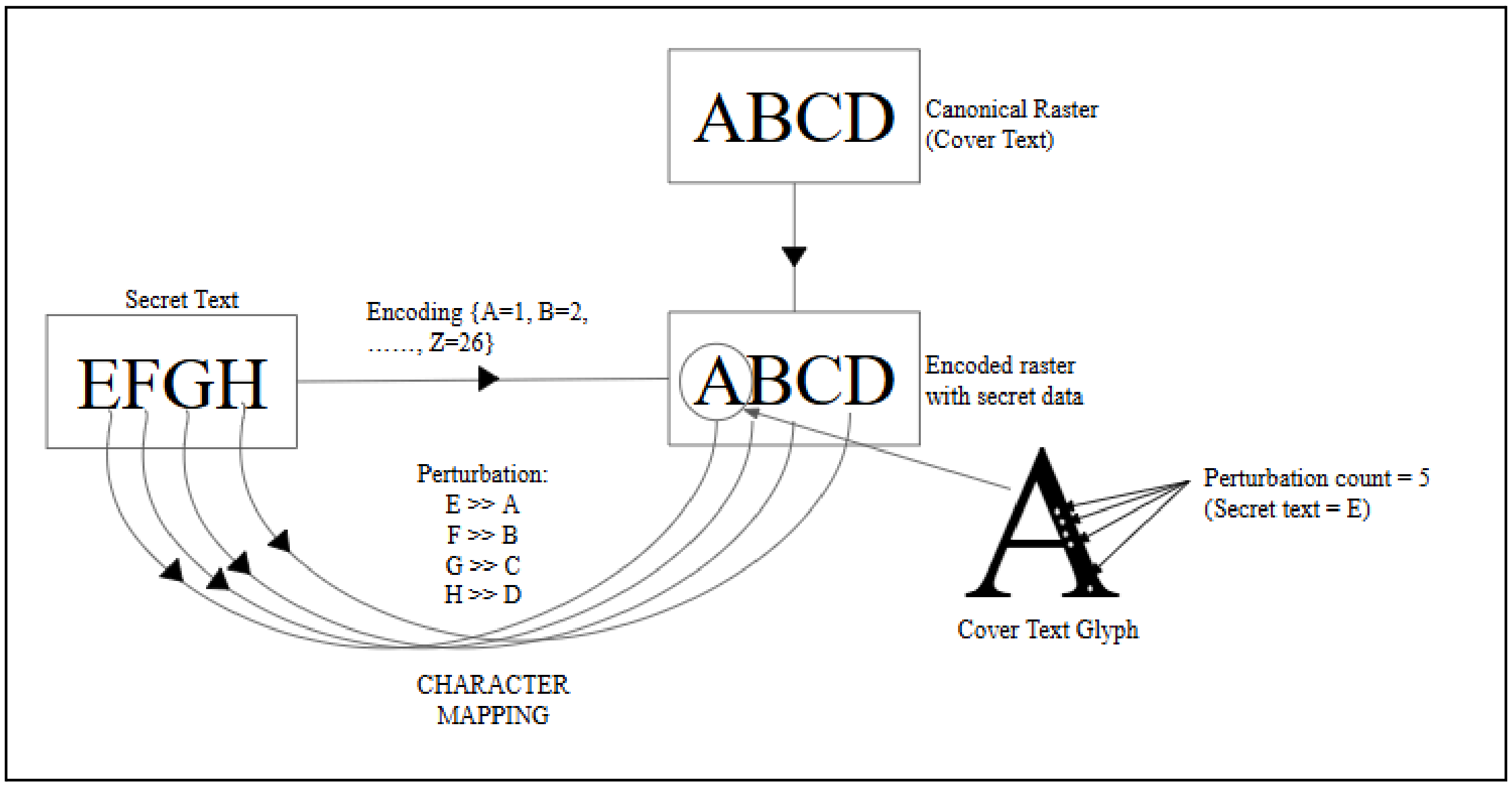

**Fig 3. Text to Text embedding through raster perturbation**

**3.1.2. Decoding Procedure:**

Decoding requires only the encoded raster and knowledge of the original cover text. The receiver re-renders the same cover text using identical rendering parameters to reconstruct the canonical glyph rasters. Because both images share the same pixel grid, decoding proceeds by computing pixel wise differences within each glyph region and counting the number of pixels whose intensity increases from 0 to 1. This count recovers the integer payload $V_i$, which is then mapped back to the corresponding character. Concatenating all recovered characters yields the decoded secret message.

**3.2. IMAGE-TO-TEXT STEGANOGRAPHY VIA RASTER GLYPH PERTURBATIONS:**

Image embedding extends the perturbation cardinality framework by first converting visual content into a bounded integer sequence compatible with the glyph level raster channel. Unlike traditional image steganography, no modifications are applied directly to image pixels. Instead, image derived values are embedded into the interior pixels of rendered text glyphs using the same raster domain perturbation mechanism described in Section 3.1.

**3.2.1. Image Preprocessing and Payload Formation:**

Let $I \in [0,255]^{H\times W\times 3}$ denote an RGB image. Each color channel is flattened independently in raster scan order to retain a consistent raster scan ordering, yielding three sequences of length H×W. Because a single glyph can encode only a limited perturbation count, each pixel intensity $x \in [0,255]$ is normalized to an integer payload value:

$$\nu_j = \left[\frac{x}{255}\right] \bullet Pmax, \; \nu_j \in [0, P_{Max}]$$

where $P_{Max}$ denotes the maximum number of admissible perturbations per glyph. In this work, Pmax=26 is used as a conservative operating point aligned with the text-to-text alphabetic mapping. However, the framework itself is not inherently restricted to this value. The resulting bounded integer sequence $V = \{\nu_j\}$ serves as the payload for embedding.

### 3.2.2. Embedding and Decoding:

Each integer payload value is embedded into a single glyph by perturbing exactly $\nu_j$ interior ink pixels, following the same deterministic selection and modification rule introduced in Section 3.1. No changes are made to glyph outlines, spacing, or textual content. Decoding proceeds identically by re-rendering the cover text using identical rendering parameters, computing pixel wise differences between canonical and encoded glyphs, and recovering perturbation counts, which are then inversely mapped to reconstruct the image intensities as shown in Figure 4.

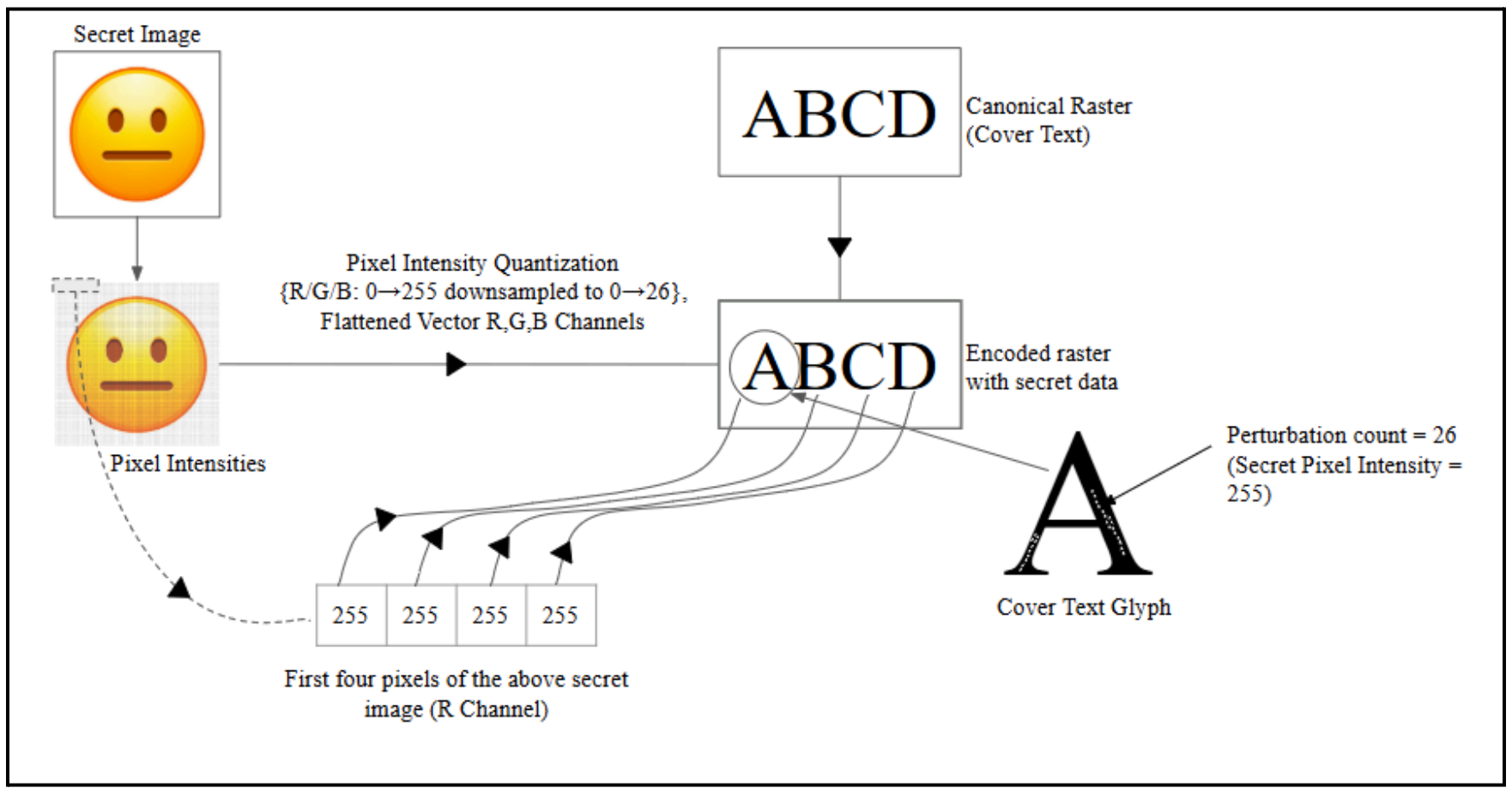

**Fig 4. Image to Text embedding through raster perturbation**

## 3.3. AUDIO-TO-TEXT STEGANOGRAPHY VIA RASTER GLYPH PERTURBATIONS:

Audio to text embedding extends the proposed raster domain framework to temporal signals by converting audio into a compact sequence of bounded integer values, which are then embedded into rendered text glyphs using the same perturbation cardinality mechanism described above. The raster embedding and decoding procedures remain unchanged, only the formation of the integer payload differs from other modalities.

### 3.3.1. Audio Payload Formation:

Let $a(t) \in [-1,1]$, $t=1,\ldots,T$, denote a single channel audio signal. Because raw audio is high dimensional and temporally dense, the signal is first segmented into short-time frames using fixed window and hop parameters. From each frame, a single scalar value $S_j$ is extracted to summarize short time signal energy, computed as the root mean square (RMS) amplitude. This yields an ordered sequence $\{S_j\}^{J}_{j=1}$.

To make each value compatible with the glyph perturbation channel, values are normalized to a bounded integer range (i.e. 0-26).

$$\nu_j = \left[\frac{Sj - Smin}{Smax - Smin} \bullet PMax\right] \; \nu_j \in [0, Pmax]$$

where Pmax (i.e.=26) denotes the maximum number of admissible perturbations per glyph, and $S_{min}$, $S_{max}$ are computed over the signal. The resulting integer sequence

$$V_i=(\nu_1,\nu_2,\ldots,\nu_j)$$

forms the payload for embedding.

**3.3.2. Embedding into Rasterized Glyphs:**

Each integer payload value $v_j$ is embedded into a single glyph using the same deterministic raster-domain procedure described in Sections 3.1 and 3.2. For each glyph, exactly $v_j$ interior ink pixels are selected deterministically and perturbed by a minimal intensity increment. The perturbed glyphs are concatenated to produce an encoded text raster that is visually indistinguishable from the original as shown in figure 5.

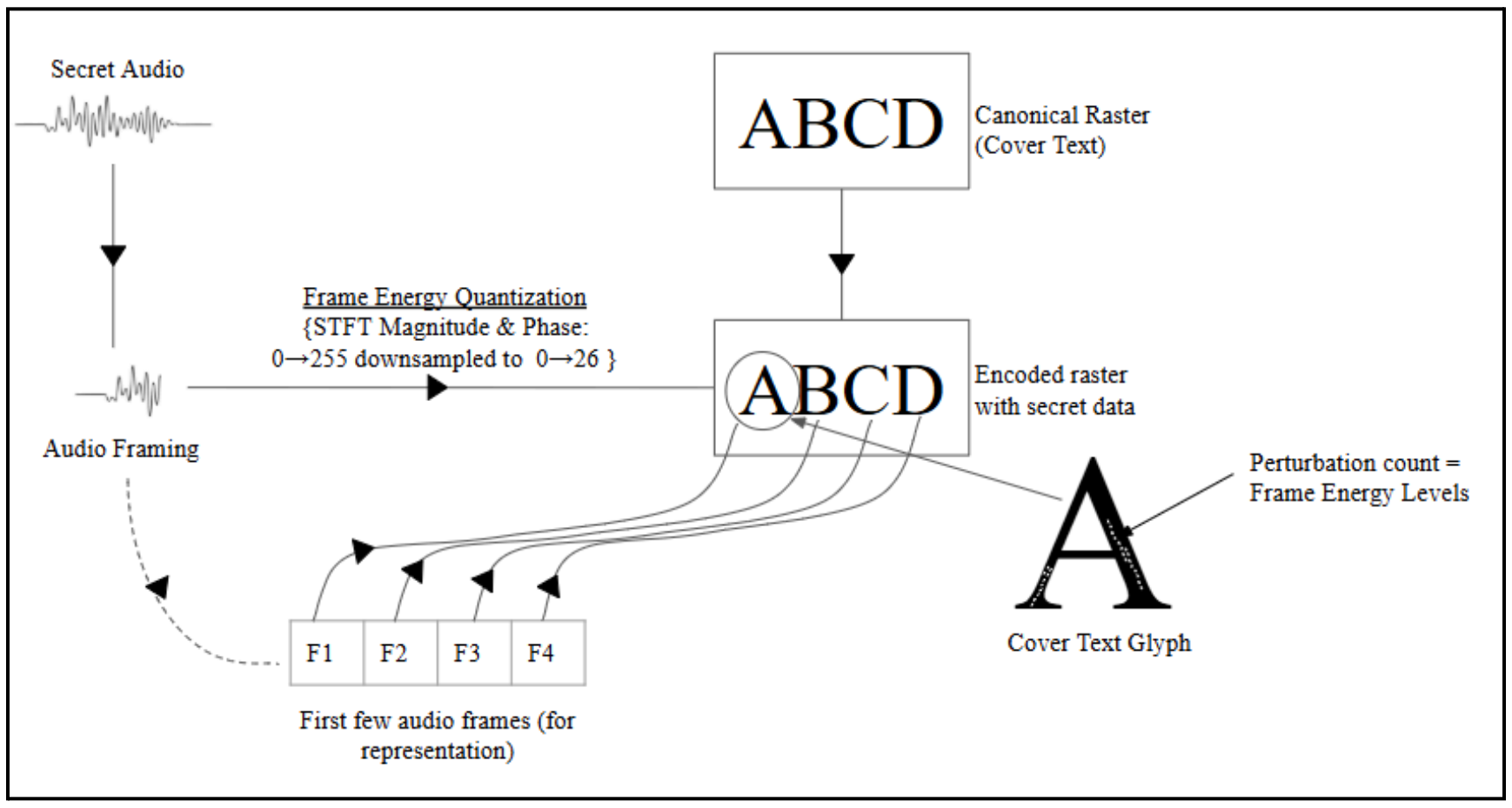

**Fig 5. Audio to Text embedding through raster perturbation**

**3.3.3. Decoding Procedure:**

Decoding follows the identical raster domain process used for text and image modalities. The receiver re-renders the cover text using identical rendering parameters to reconstruct the canonical glyph rasters. Pixel wise differences between the encoded and canonical rasters are computed within each glyph region, and the number of perturbed pixels is counted to recover the integer payload values $V_j$.

The recovered integer sequence is inverse normalized to obtain reconstructed frame level energy values $S_j$, corresponding to the original RMS amplitudes. These values are applied as frame wise amplitude scaling factors and recombined using overlap and add synthesis to produce a reconstructed time domain audio signal.

Due to aggressive low cardinality quantization (0-26 levels), the reconstructed audio is not sample perfect and does not preserve fine spectral detail. However, global temporal alignment and coarse energy structure are maintained, demonstrating reliable cross modal recovery through the raster domain perturbation channel.

## 3.4. VIDEO-TO-TEXT STEGANOGRAPHY VIA RASTER GLYPH PERTURBATIONS:

Video to text embedding extends the proposed raster domain framework to spatio temporal visual signals by converting video content into an ordered sequence of bounded integer values, which are then embedded into rendered text glyphs using the same perturbation cardinality mechanism described in the sections above. As with the audio modality, the raster embedding and decoding procedures remain unchanged, where only the formation of the integer payload differs.

**3.4.1. Video Payload Formation:**

Let a video sequence be represented as an ordered set of frames $\{F_t\}^{T}_{t=1}$, where each frame $F_t \in [0,255]^{H\times W\times 3}$ is an RGB image. To ensure a predictable and bounded payload size, each frame is temporally sampled at a fixed rate and spatially resized to a canonical resolution. No motion estimation, block aggregation, or learned feature extraction is performed. Instead, the proposed framework operates directly on pixel level intensity values. For each normalized frame, the three color channels (R, G, B) are processed independently. Pixel intensities in each channel are flattened in raster order and normalized to a bounded integer range compatible with the glyph perturbation channel:

$$\nu_j = \left( \left[ \frac{x}{255} \right] \bullet Pmax \right), \nu \in [0,\text{Pmax}],$$

Where x denotes the original pixel intensity and $P_{Max} = 26$ is the maximum number of admissible perturbations per glyph. This produces three ordered integer sequences per frame, corresponding to the R, G, and B channels, each of length H×W.

Concatenating the channel wise sequences across all frames yields a deterministic payload stream

$$V_i=(\nu_1,\nu_2,\ldots,\nu_N),$$

which preserves both spatial and temporal ordering. Each integer value in this sequence is embedded into a single rendered glyph by encoding the value as a perturbation cardinality within the glyph's interior stroke pixels. This dense, pixel level quantization strategy enables full frame video information to be transmitted through the rasterized text channel without modifying the visual appearance of the carrier text.

**3.4.2. Embedding into Rasterized Glyphs:**

Each integer payload value $V_i$ is embedded into a single rendered glyph using the same deterministic raster domain procedur. For each glyph, exactly $V_i$ interior ink pixels are selected deterministically and perturbed by a minimal intensity increment. The perturbed glyphs are concatenated in reading order to produce an encoded text raster that remains visually indistinguishable from the original paragraph.

Because video intensities are embedded sequentially and independently, no temporal coupling is introduced at the embedding stage. Each frame's contribution to the payload is localized to a corresponding segment of the text as shown in figure 6.

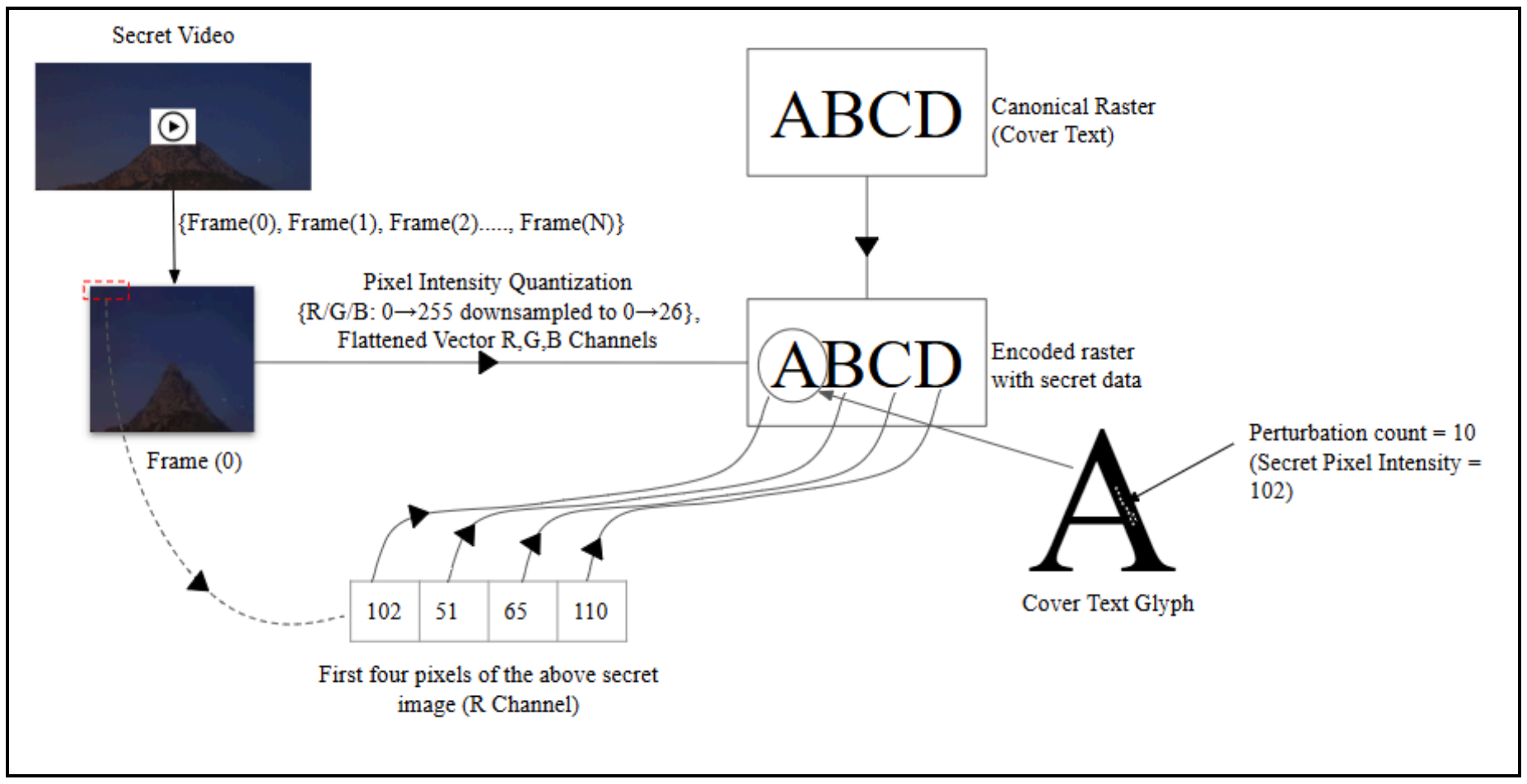

**Fig 6. Video to Text embedding through raster perturbation**

**3.4.3. Decoding Procedure:**

Decoding follows the identical raster-domain process used for text, image, and audio modalities. The receiver re-renders the cover text using identical rendering parameters to reconstruct the canonical glyph rasters. Pixel wise differences between the encoded and canonical rasters are computed within each glyph region, and the number of perturbed pixels is counted to recover the integer payload sequence.

The recovered integers are inverse normalized to reconstruct frame level visual intensities, which are then reshaped and expanded into approximate video frames using standard spatial upsampling. Reconstructed frames are temporally stitched in the original sampling order to form the recovered video sequence.

Although aggressive spatial and intensity quantization is applied, the reconstructed video preserves global scene structure, dominant color regions, and temporal coherence across frames. Error remains uniformly distributed and does not introduce text shaped artifacts or structured distortion. These results demonstrate that the perturbation cardinality raster channel supports stable and deterministic recovery of spatiotemporal information without modifying the visual appearance of the carrier text. In this work, video-to-text embedding operates on the visual stream only; any associated audio track may be processed independently using the audio-to-text pipeline described in Section 3.3.

## 4. EXPERIMENTAL SETUP AND DETERMINISTIC RENDERING CONFIGURATION:

This section describes the computational environment, rasterization parameters, and embedding protocol used across all experiments. All reported results are obtained under a fixed software pipeline to ensure repeatable glyph rasterization and reliable perturbation recovery.

### 4.1. Computational Environment:

All experiments were conducted on a single workstation using a fixed software configuration to minimize variability in text rendering and signal processing:

- Hardware: Intel(R) Core(TM) i5-9300H CPU @ 2.40 GHz
- Operating System: Windows 11
- Programming Environment: Python 3.10
- Core Libraries: NumPy, Pillow (PIL), Matplotlib
- Image Processing: OpenCV, scikit-image
- Audio / Video Processing: Librosa, FFmpeg
- Randomness Control: All pseudo-random operations seeded (seed = 42)

All experiments were executed within the same environment to ensure that repeated rasterization of identical text produced pixel identical glyph bitmaps, which is a prerequisite for perturbation count decoding. All pseudo random operations used for interior pixel selection were initialized with a fixed seed to ensure repeatable perturbation patterns across runs.

### 4.2 Glyph Rasterization Parameters:

All cover text was rendered using a fixed “TrueType” font rasterization pipeline implemented via the Pillow imaging library. Rasterization parameters were held constant across all modalities.

- Font Type: TrueType (Arial)
- Color Mode: 8 bit grayscale
- Ink Color: Black (0)
- Background Color: White (255)

Glyphs were rendered onto fixed size grayscale canvases depending on the experiment:

- Text to Text: font size 160 pt
- Image to Text: font size 36 pt
- Audio to Text: font size 36 pt

- Video to Text: font size 36 pt

Text was centered within each glyph tile using bounding box alignment. Potential antialiasing effects are mitigated during embedding by restricting all perturbations to fully inked interior stroke pixels (intensity value 0), ensuring that glyph contours and visual identity remain unchanged.

**4.3 Perturbation Model:**

The same perturbation primitive was used consistently across all modalities.

- Perturbation Operation: Increment interior ink pixel intensity by +1 gray level
- Detection Threshold: 1 gray level difference
- Eligible Pixels: Interior stroke pixels only (excluding glyph contours)
- Maximum Payload per Glyph (Pmax): 26
- Pixel Selection Rule: Deterministic pseudo-random sampling (seeded per glyph)

For a canonical glyph raster $G_i$, the set of eligible perturbation pixels is defined as:

$$I_i = \{p \mid G_i(p) = 0\}$$

Each payload value $v_i \in [0,26]$ is encoded by perturbing exactly $v_i$ pixels from $I_i$. This ensures imperceptibility, separability, and deterministic reproducibility.

**4.4. Dataset Preparation:**

Experiments were conducted across four modalities using small, controlled datasets designed to validate the proposed raster-domain perturbation framework as a proof of concept rather than as a large scale benchmark evaluation.

**Text Data:**

A set of five short words was used, serving alternately as secret text (ABCD, THERE, GOODBYE, THERE, EVENING) and cover text (ABCD, HELLO, WELCOME, WHERE, MORNING). This modality evaluates exact recoverability under deterministic rasterization and serves as a baseline for lossless embedding.

**Image Data:**

Five emoji images were selected as secret payloads. Each image was resized to a fixed spatial resolution. The three color channels (R, G, B) were processed independently, with pixel intensities normalized from the range [0,255] to bounded integer payloads in [0,26]. This produced three sequential payload streams corresponding to the color channels, preserving spatial ordering within each channel.

**Audio Data:**

Five short audio samples sourced from the Pixabay dataset were used for evaluation. All signals were resampled to 16 kHz prior to processing. Each audio signal was segmented into overlapping time domain frames using a frame length of 1024 samples and a hop size of 512 samples. For each frame, a single scalar descriptor was computed as the root-mean-square (RMS) amplitude, yielding a compact sequence of frame level energy values. These RMS values were linearly normalized and quantized to the integer range [0,26], consistent with the unified low cardinality encoding used across modalities, and the resulting integer sequence served as the payload for subsequent raster domain glyph embedding..

**Video Data:**

Five short video clips were used as secret payloads. Videos were temporally sampled and spatially normalized to a fixed resolution. Each frame was treated as an RGB image, with pixel intensities in each color channel normalized independently from [0,255] to [0,26]. Payload values were generated sequentially across frames, preserving temporal ordering.

Across all modalities, normalization and quantization mappings were invertible up to the imposed bin resolution, enabling controlled reconstruction fidelity consistent with the selected perturbation capacity.

Glyphs were rasterized at elevated resolutions during embedding to increase perturbation capacity and ensure reliable separation of perturbation counts. All encoding and decoding operations were performed in this fixed high resolution raster domain. For visualization and document presentation only, encoded glyph rasters may be rescaled to standard text sizes; decoding is never performed on rescaled representations. Glyph resolutions were chosen per modality to balance payload density and perturbation capacity, while remaining fixed within each experiment.

**4.5 Embedding Protocol:**

For each experiment:

1. A cover text of sufficient length was rasterized using the fixed glyph parameters.
2. A modality specific integer payload sequence $V_i=\{v_1, v_2, \ldots v_j\}$ was prepared.
3. Each payload value $v_i$ was embedded into the corresponding glyph by perturbing exactly $v_i$ interior pixels.
4. The resulting encoded raster was exported using a lossless image format (PNG).

No linguistic modification of the cover text was performed and all information embedding occurs exclusively in the raster domain.

**4.6 Decoding Protocol:**

Decoding was performed using only the encoded raster and the known cover text:

1. The cover text was re-rendered using identical rasterization parameters.
2. Pixel wise differences between encoded and canonical glyphs were computed.
3. Perturbed pixels were counted per glyph to recover $\hat{v}_i$.
4. Payload sequences were inverse normalized and reconstructed according to modality.

All decoding operations were fully deterministic under the fixed rendering pipeline.

**4.7 Evaluation Metrics:**

Reconstruction quality and embedding behavior were assessed using modality appropriate metrics.

- Text: Exact character recovery, CER, BER.
- Image: PSNR, SSIM, error heatmaps.
- Audio:Time domain MSE and MAE, mean payload decoding error, and qualitative spectrogram comparison.
- Video: Frame wise PSNR, SSIM, temporal consistency.

## 5. RESULTS & DISCUSSION:

### 5.1. TEXT TO TEXT:

Five different cover words were paired with five secret words and are encoded using the same glyph perturbation cardinality mechanism employed across other modalities as shown in figure 7(a) and 7(b). The encoded rasters were analysed using canonical encoded differences, perturbation counts, and glyph capacity diagnostics, all of which reveal the same stable structural behaviour. The difference maps show that perturbations appear only as scattered interior pixel activations while glyph edges and stroke boundaries remain untouched as shown in the figure 7(c), confirming that the embedding process preserves the visual identity of the rendered text. Capacity plots demonstrate that each glyph contains several hundred to thousand interior “ink” pixels as in figure 7(d), while the encoder typically uses only tens of them as in figure 7(e), where this large safety margin ensures that embedded signals remain both imperceptible and fully separable during decoding.

Across all five text pairs as shown in table 1, the recovered secret strings match the ground truth exactly, with 0% Character Error Rate (CER) and 0% Bit Error Rate (BER) in every case. Because each glyph encodes information by perturbing interior ink pixels, the maximum embeddable value depends on glyph geometry as shown in figure 7(f). Under fixed font and rasterization parameters, the number of usable interior pixels varied across glyphs, yielding observed perturbation counts from roughly 10 for sparse characters to approximately 70-76 for denser glyphs. In all experiments, Pmax was conservatively capped at 26 to maintain uniformity across modalities, despite higher per-glyph capacity being available. This variability does not affect decoding, as the receiver counts perturbations per glyph independently. Raster level deviations remain extremely small, with pixel wise MSE between encoded and canonical glyph rasters ranging from 0.00019 to 0.00112 as shown in figure 7(g), confirming that the perturbation field injects only negligible deviation relative to the canonical raster. These results demonstrate that the text text channel is perfectly reversible under deterministic rasterisation, with the perturbation cardinality signal remaining intact even at minimal embedding strength.

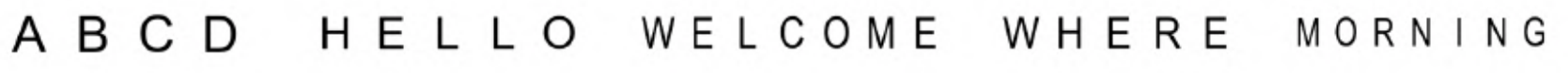

**(a)**

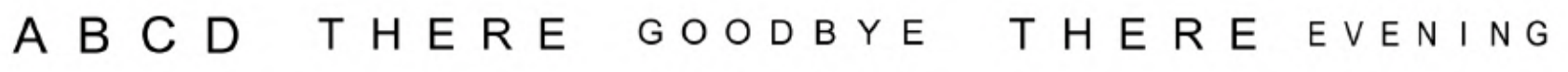

**(b)**

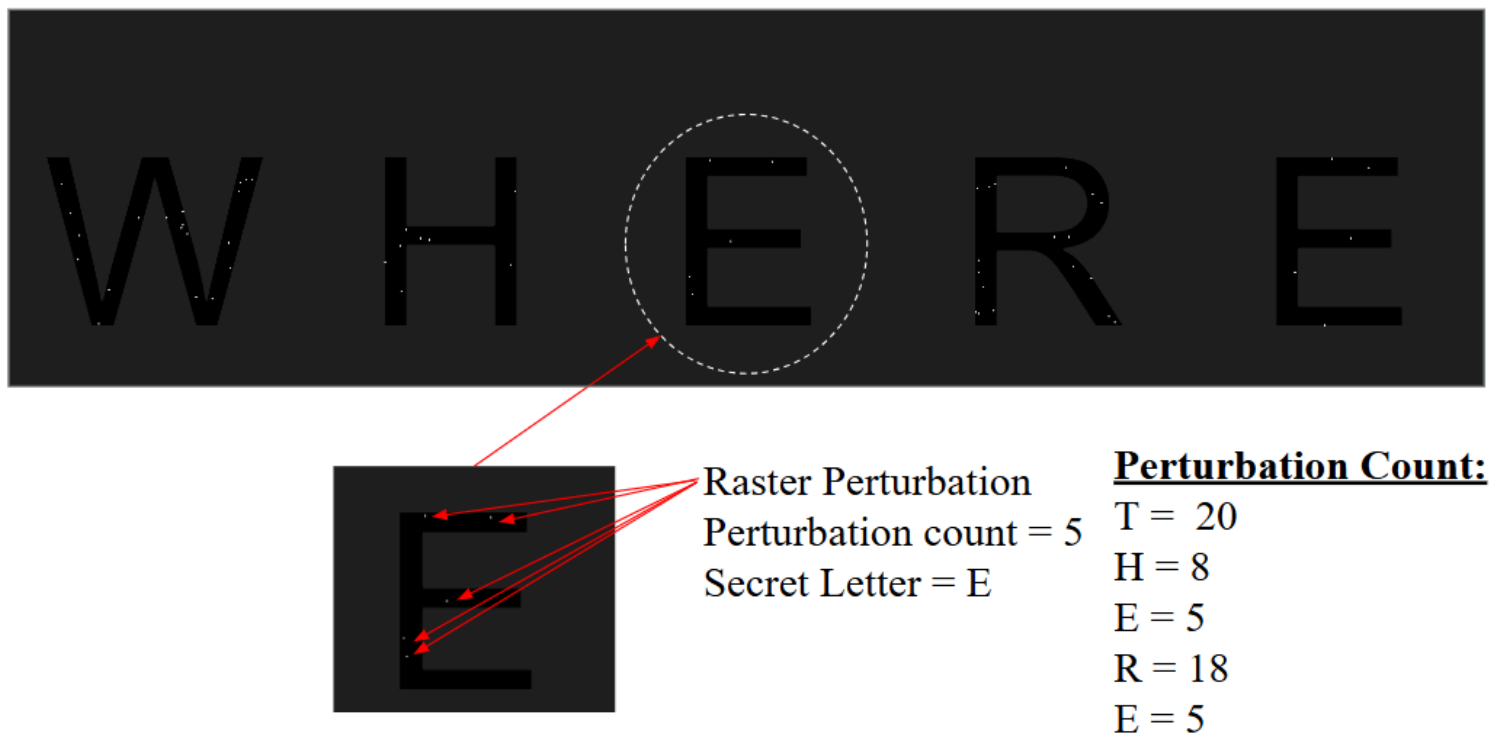

**(c)**

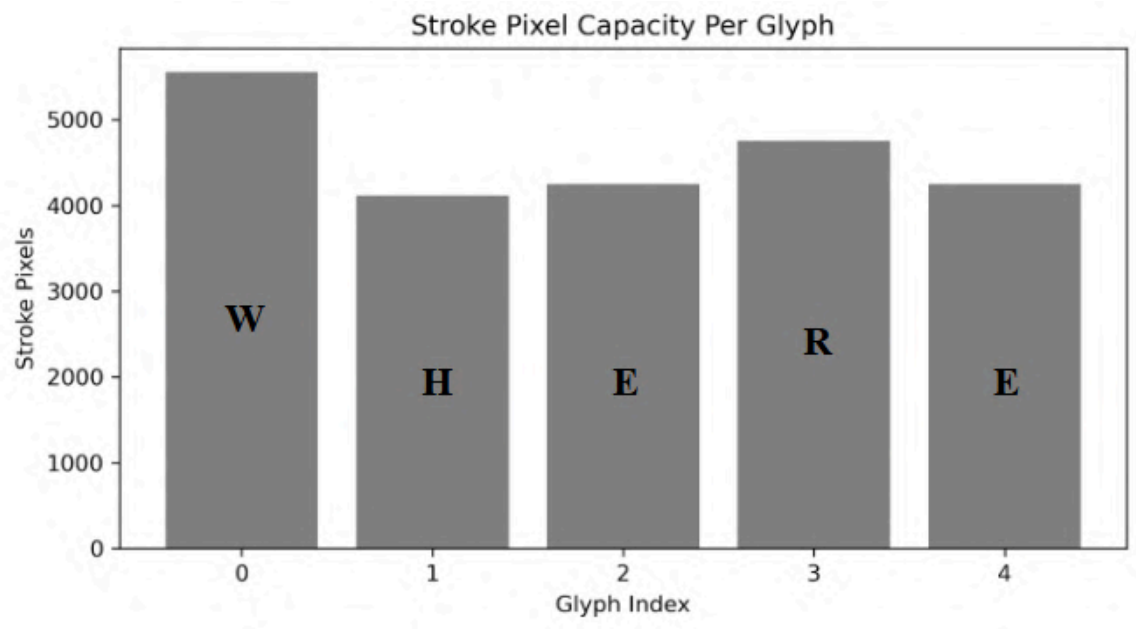

**(d)**

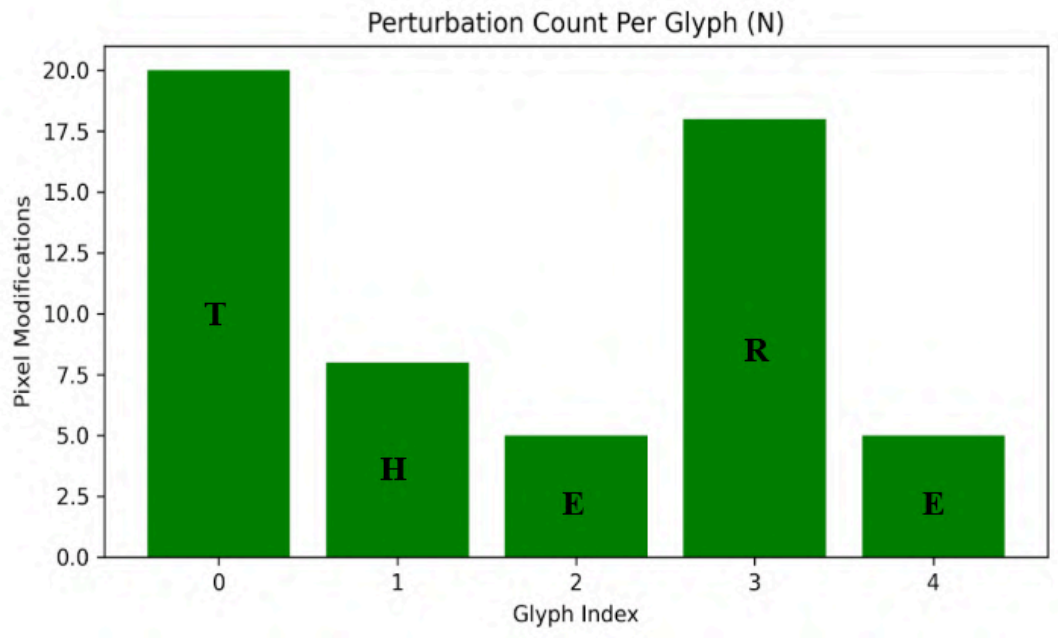

**(e)**

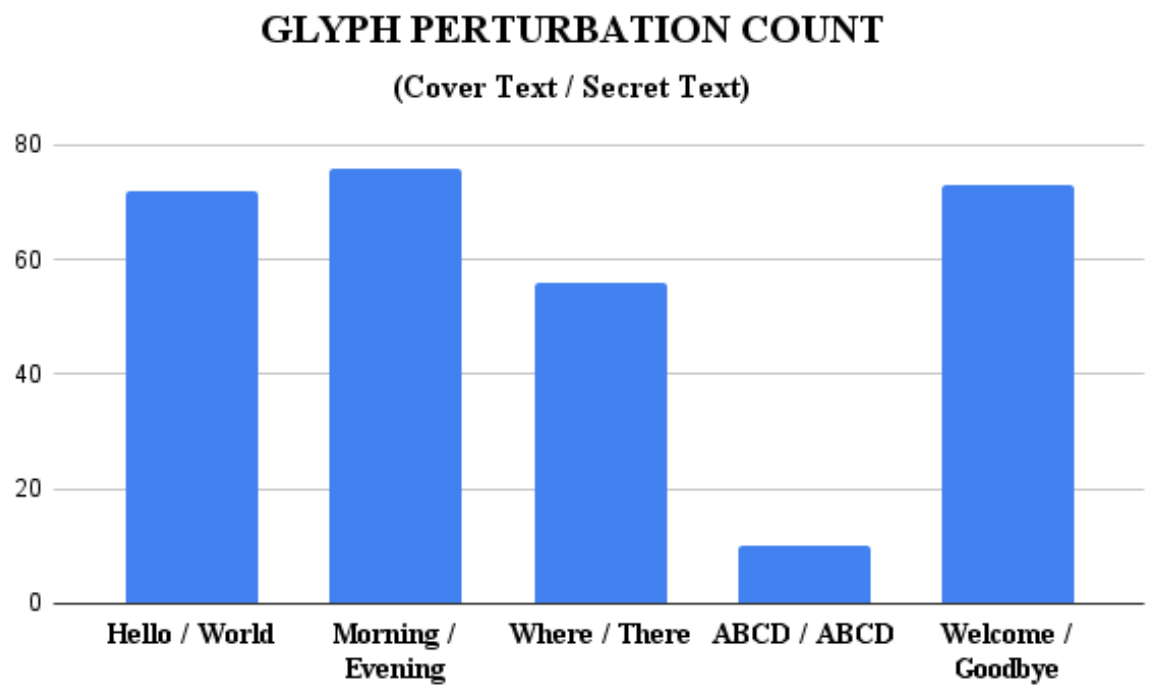

(f)

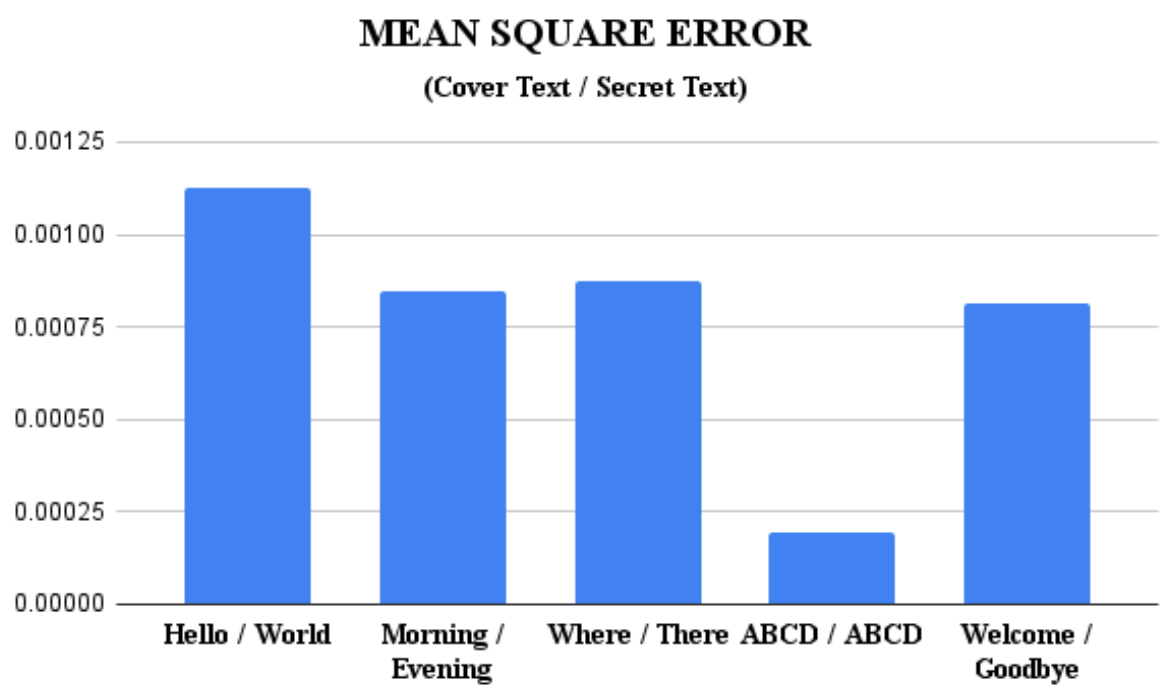

(g)

**Fig 7. (a) Cover Text, (b) Secret Text, (c) Glyph Raster Perturbations in Cover Text (d) Stroke pixel capacity per glyph (Where), (e) Perturbation count per Glyph (There), (f) Glyph Perturbation count for all words, (g) MSE for all words.**

| CANONICAL RASTER | ENCODED RASTER | SECRET TEXT | DIFFERENCE IMAGE | PERTURBATION COUNT PER GLYPH | STROKE PIXEL CAPACITY PER GLYPH |
|---|---|---|---|---|---|
| A B C D | A B C D | A B C D | | 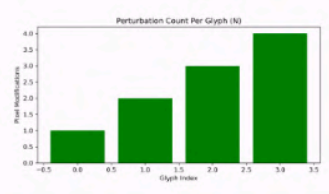 | 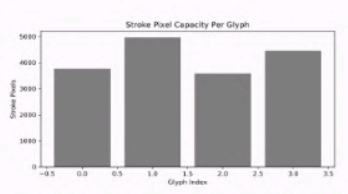 |
| H E L L O | H E L L O | T H E R E | | 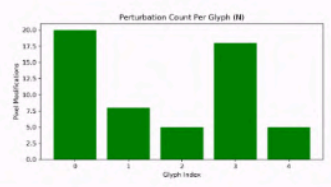 | 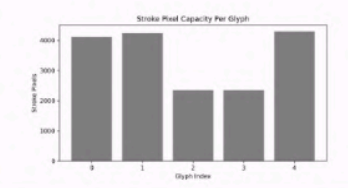 |
| W E L C O M E | W E L C O M E | G O O D B Y E | | 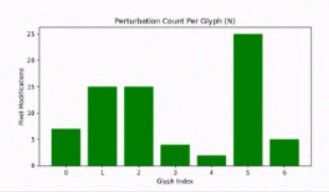 | 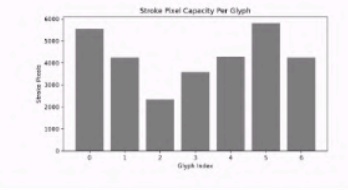 |
| W H E R E | W H E R E | T H E R E | | 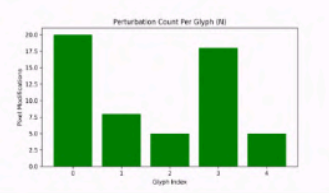 | 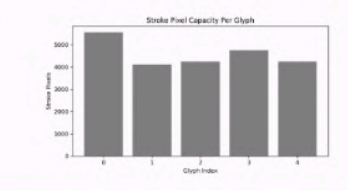 |
| M O R N I N G | M O R N I N G | E V E N I N G | | 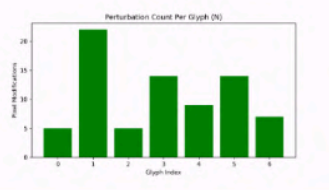 | 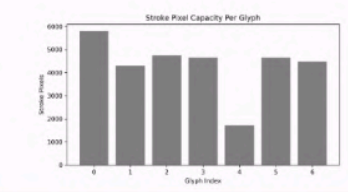 |

**Table1. Shows canonical, encoded rasters along with perturbation count per glyph and MSE.**

## 5.2. IMAGE TO TEXT:

Five distinct emoji style images (Figure 8(e)) were encoded into quantized integer sequences and embedded into rendered text using glyph level perturbation counts. All five reconstructions exhibit consistent and stable behavior of the raster domain embedding channel. The recovered integer sequences fall into sharply defined quantization bands, confirming that the perturbation mechanism acts as a deterministic 26 level encoder without introducing drift, stochastic noise, or decoding instability during rasterization or recovery (Figures 8(f-j)).

Canonical (Figure 8(c)), encoded (Figure 8(d)), and decoded raster visualizations show that perturbations remain confined to interior stroke pixels while glyph contours and outlines remain untouched, preserving the visual appearance of the encoded text and ensuring that the embedding remains hidden at the typographic level. Error heatmaps across all emoji images (Figures 8(k-o)) exhibit only low magnitude, spatially smooth deviations corresponding to expected quantization effects. Error distributions remain strongly centered within a narrow ±5 intensity range, indicating that quantization, not raster noise or perturbation instability, is the sole source of reconstruction loss.

Quantitative metrics reinforce these observations. As shown in table 2, MSE values (computed in the 0-255 pixel domain) remain tightly bounded between 4.2-5.3, MAE ranges from 1.25-1.70, and RMSE remains between 2.04-2.31, confirming that errors are small and uniformly distributed. PSNR values lie in the 40.8-41.9 dB range, indicating perceptually high quality reconstruction under standard imaging criteria. SSIM scores are similarly strong, with average SSIM spanning 0.986-0.993 and channel wise SSIM remaining near unity for all three color channels (R, G, B).

Together, these results demonstrate that the image to text raster decoding loop introduces only predictable quantization effects, with no stochastic distortion or structural degradation. In practice, this confirms that the proposed glyph perturbation mechanism provides a stable, deterministic, and visually imperceptible carrier for transmitting image content through ordinary textual renderings.

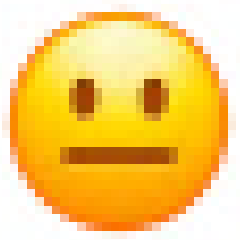

(a)

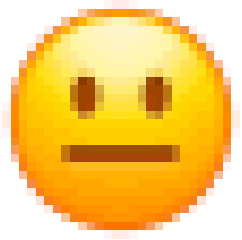

(b)

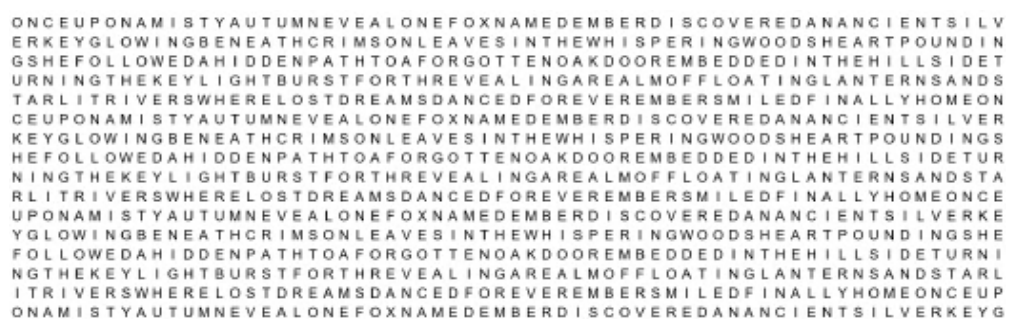

(c)

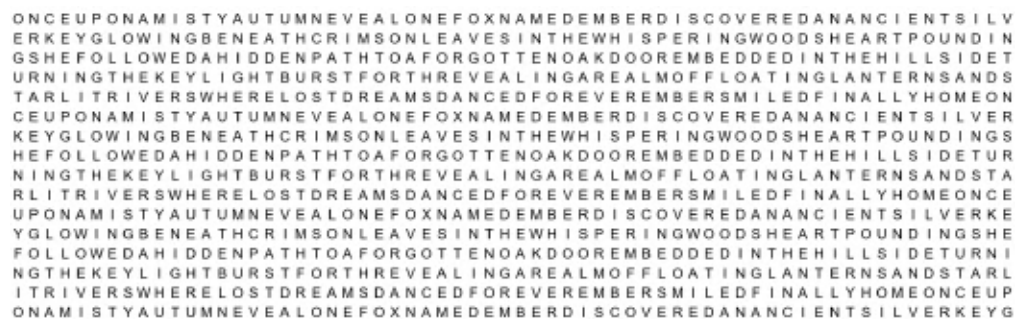

(d)

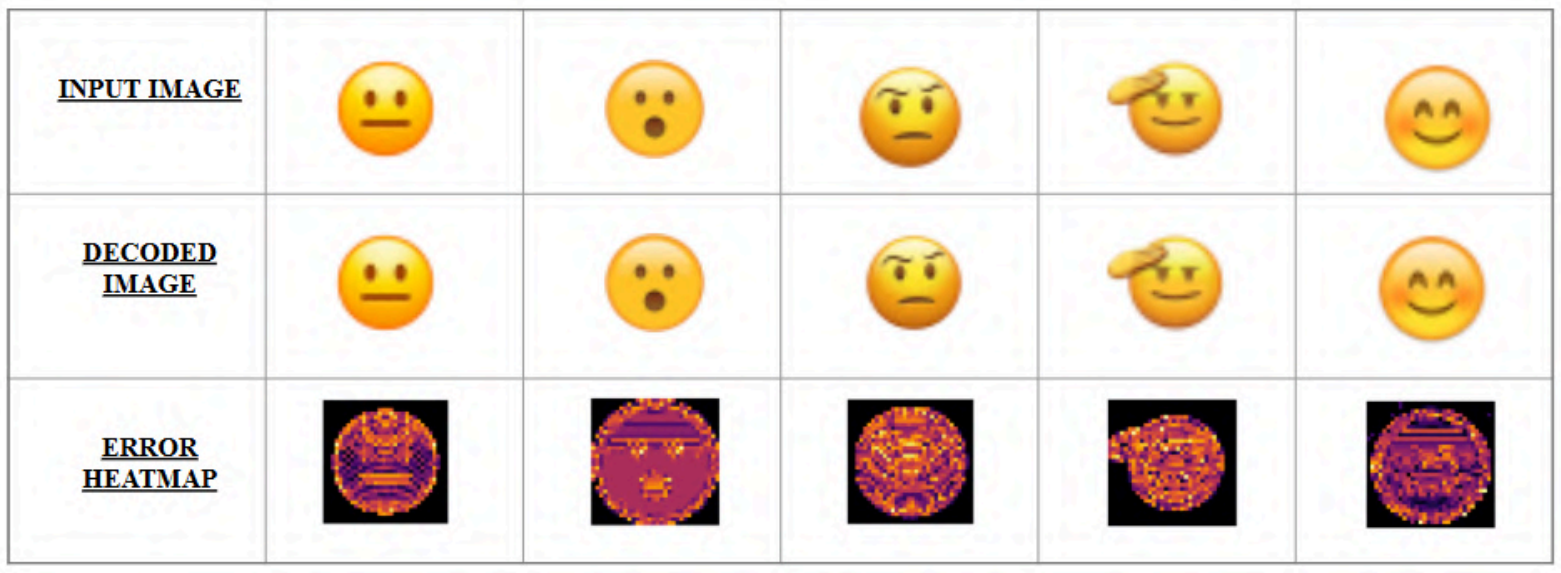

(e)

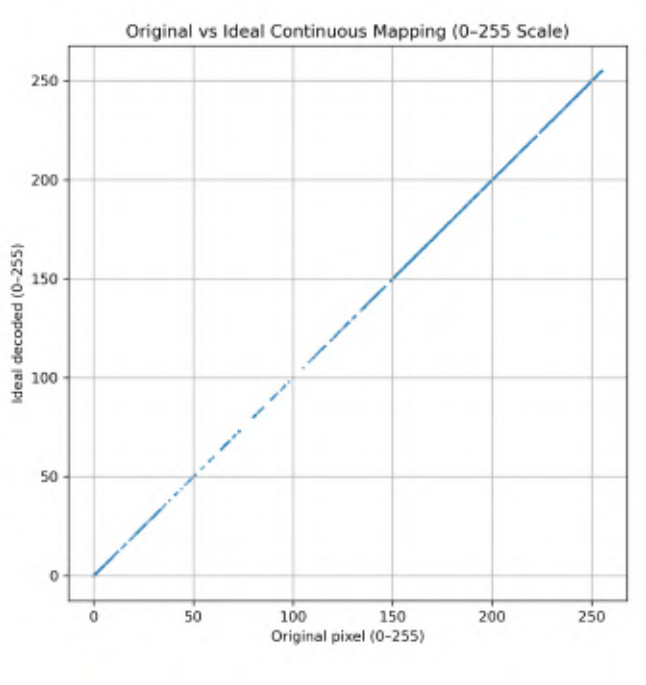

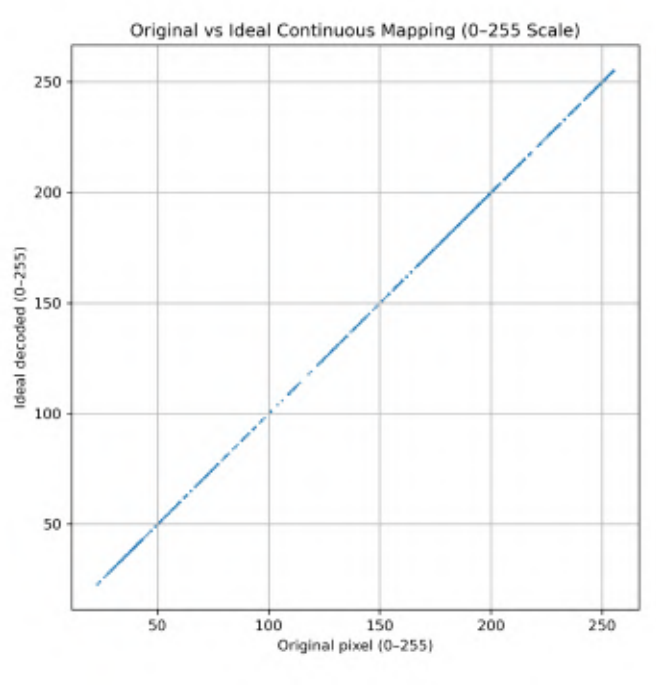

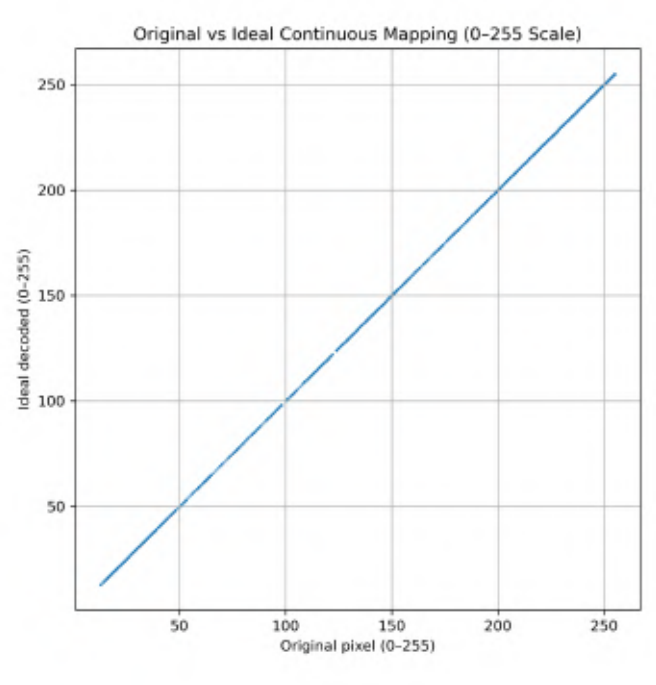

(f) (g) (h)

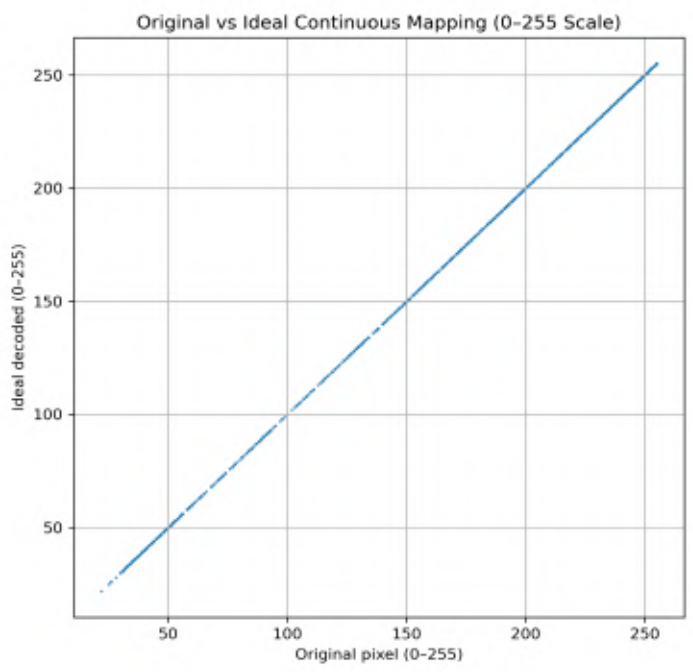

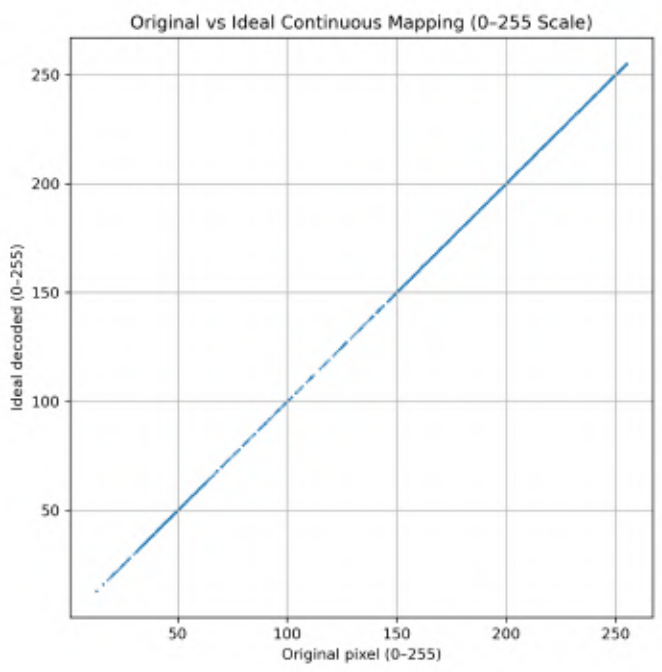

(i) (j)

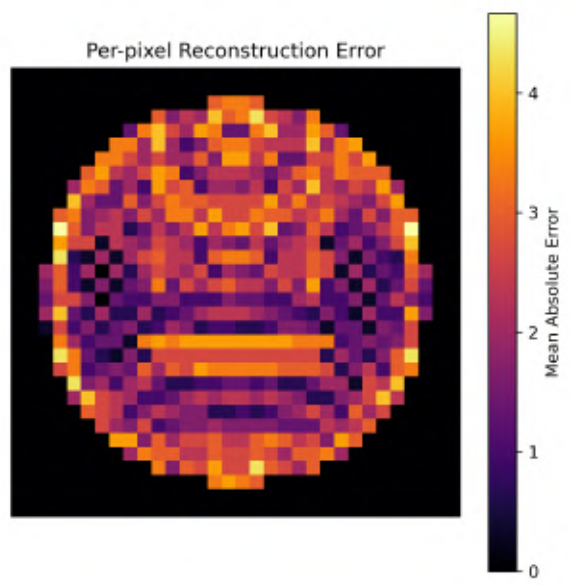

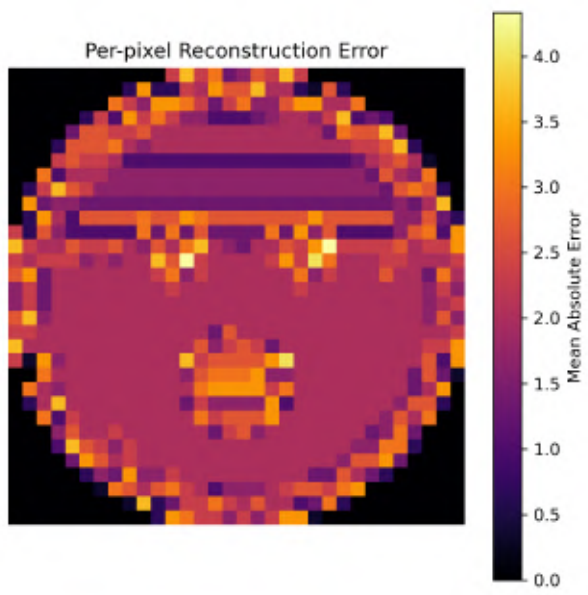

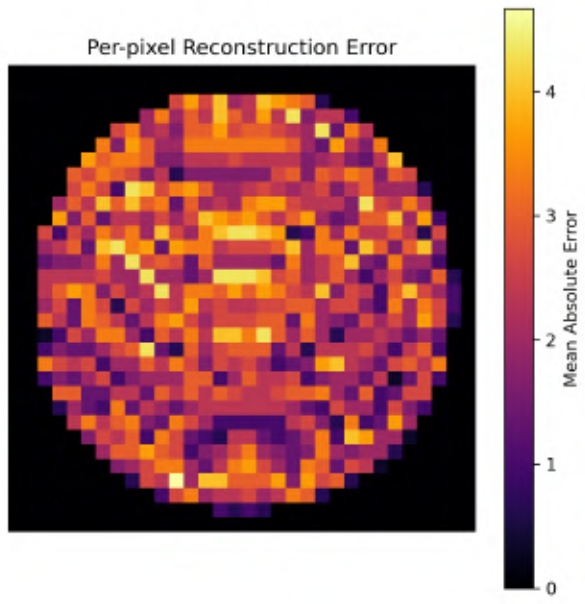

(k) (l) (m)

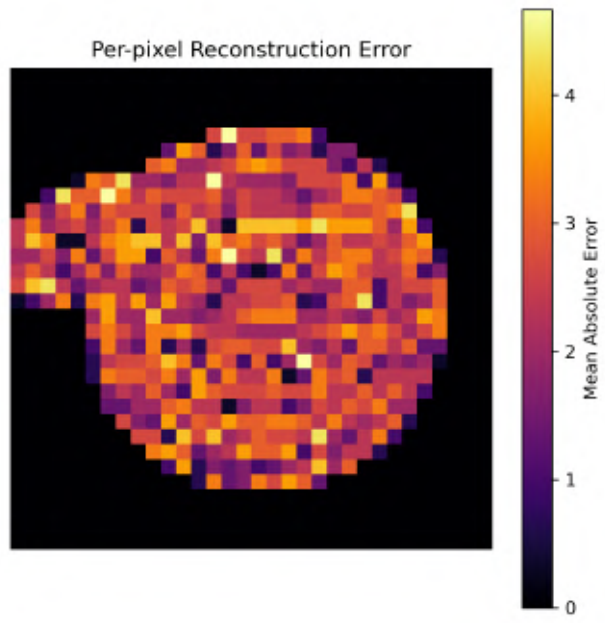

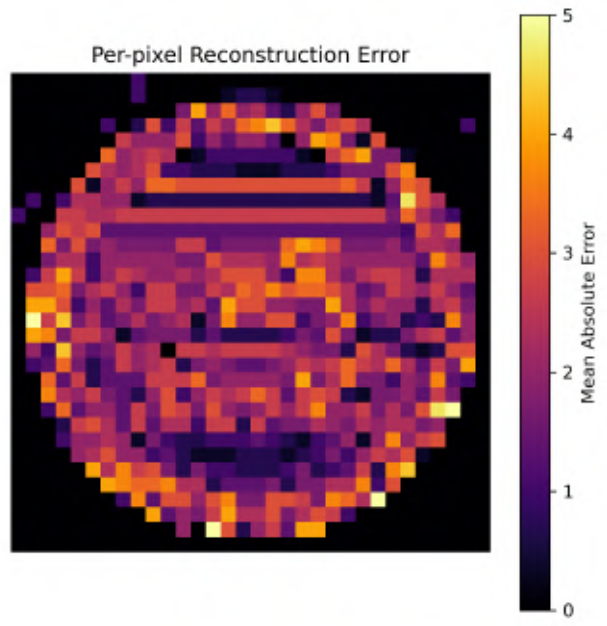

(n) (o)

**Fig 8. (a) Original Image, (b) Decoded Image, (c) Canonical Raster (Same for all Images), (d) Encoded Raster (Same for all Images), (e) Error heatmap between original and decoded images, (f)-(j) Pixel intensity plots between original and decoded images, (k)-(o) Error heatmaps between Original and Decoded Images (true scale).**

| METRICS | IMAGE 1 | IMAGE 2 | IMAGE 3 | IMAGE 4 | IMAGE 5 |
|---|---|---|---|---|---|
| **MSE** | 4.2161 | 5.3375 | 5.2438 | 4.1809 | 4.8157 |
| **MAE** | 1.2916 | 1.7014 | 1.5856 | 1.2526 | 1.4231 |
| **RMSE** | 2.0533 | 2.3103 | 2.2899 | 2.0447 | 2.1944 |
| **PSNR** | 41.8816 | 40.857 | .40.9343 | 41.918 | 41.3041 |
| **SSIM AVG** | 0.986 | 0.989 | 0.9928 | 0.9929 | 0.9914 |
| **SSIM R** | 0.9666 | 0.9869 | 0.9907 | 0.988 | 0.987 |
| **SSIM G** | 0.997 | 0.9936 | 0.9971 | 0.9977 | 0.9957 |
| **SSIM B** | 0.993 | 0.9876 | 0.9906 | 0.9931 | 0.9913 |

**Table2. Metrics for Image to Text raster perturbations.**

## 5.3. AUDIO TO TEXT:

Five short audio clips sourced from the Pixabay dataset were evaluated to characterize the behavior of the proposed audio to text raster perturbation channel under low-cardinality quantization. In this formulation, audio is not represented in the spectral domain; instead, each signal is decomposed into overlapping time domain frames, and a single scalar descriptor per frame is computed using the root mean square (RMS) energy. These frame level RMS values are normalized and quantized to the shared 0-26 integer range, producing a compact symbolic payload that is embedded into rasterized glyphs using the same perturbation mechanism employed across other modalities.

Reconstruction is performed by mapping the decoded integer values back to RMS amplitudes and rescaling the original time domain frames prior to overlap-add synthesis. Spectrograms computed from the reconstructed waveforms show strong visual similarity to the original audio across all evaluated samples (Figures 9-13). Dominant time frequency structures, temporal envelopes, and energy distributions are largely preserved, with no persistent broadband noise floor or collapse of spectral content observed under the 0-26 setting. At the waveform level, reconstructed signals remain aligned at the frame index and retain comparable dynamic range envelopes, indicating that the RMS based scalar representation provides a stable low rate description of audio content suitable for raster-domain embedding.

Quantitative results further support these observations, as summarized in table 3 and figures 9-13. Across the five samples, Mean Absolute Error (MAE) values range from 0.00432 to 0.00915, Mean Squared Error (MSE) ranges from 0.00003 to 0.00018 and the corresponding signal-to-noise ratios lie between 24 and 28.2 dB. These values indicate low average distortion despite aggressive low cardinality quantization. The proposed RMS based approach intentionally captures coarse temporal energy structure, which aligns well with the limited capacity of the 0-26 channel. As a result, the audio-to-text instantiation functions as a deterministic and perceptually stable descriptor rather than a high fidelity audio codec, while remaining fully compatible with the unified raster perturbation framework used across modalities.

| Column 1 | MAE | MSE | SNR |
|---|---|---|---|
| **Audio1** | 0.00721 | 0.00011224 | 28.288 |
| **Audio2** | 0.00915 | 0.000185 | 24.237 |
| **Audio3** | 0.00432 | 0.00003146 | 28.058 |
| **Audio4** | 0.006 | 0.00155 | 26.489 |
| **Audio5** | 0.00537 | 0.00004796 | 27.286 |

**Table 3. Metrics for Audio to Text raster perturbations.**

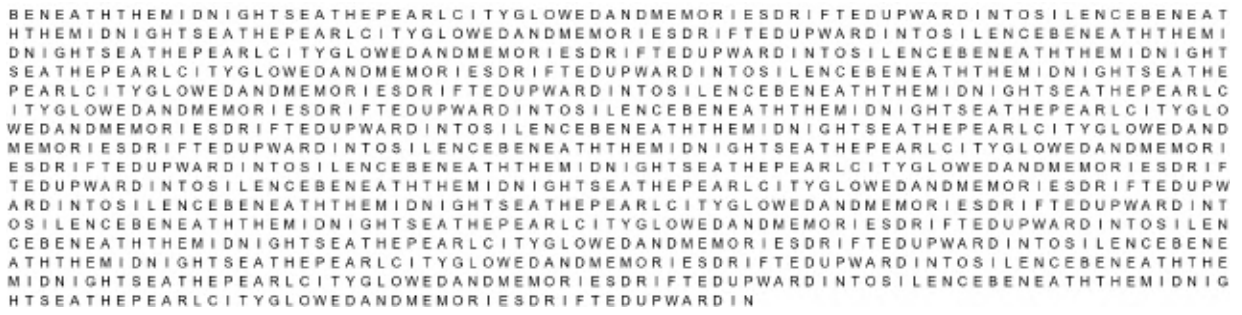

**(a)**

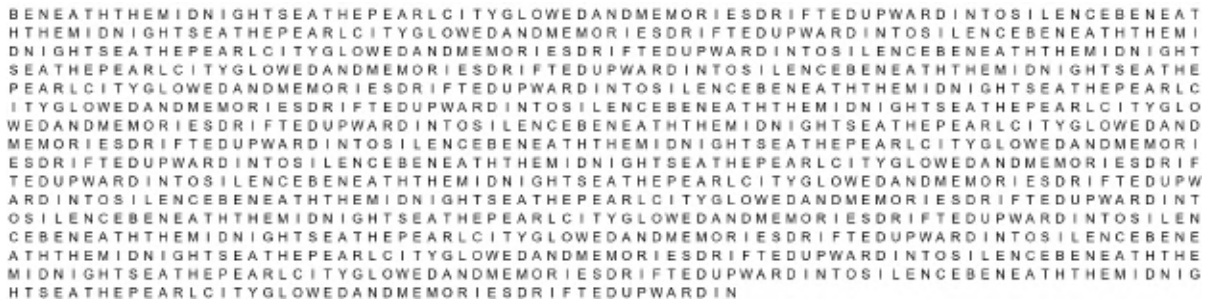

**(b)**

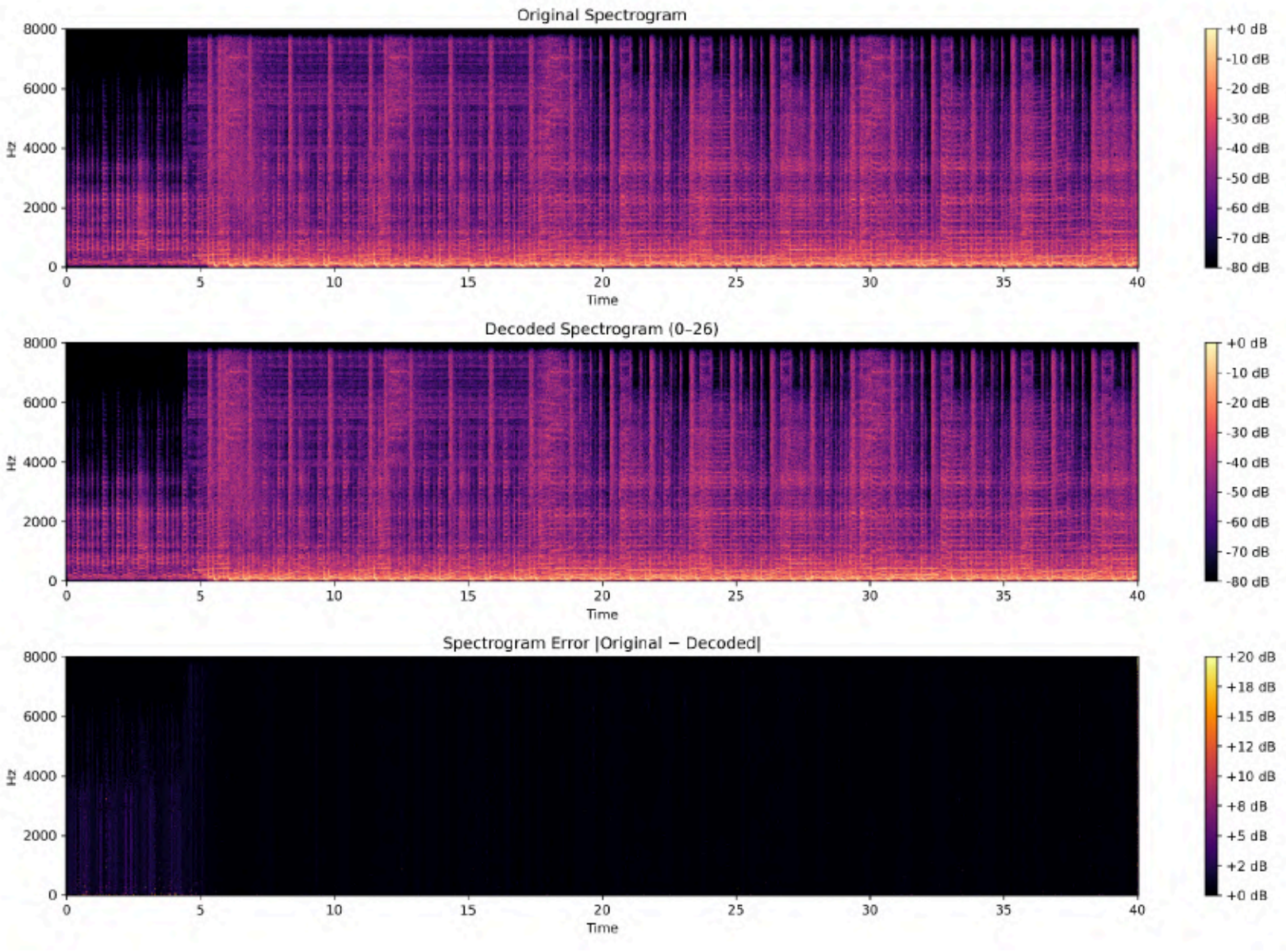

**(c)**

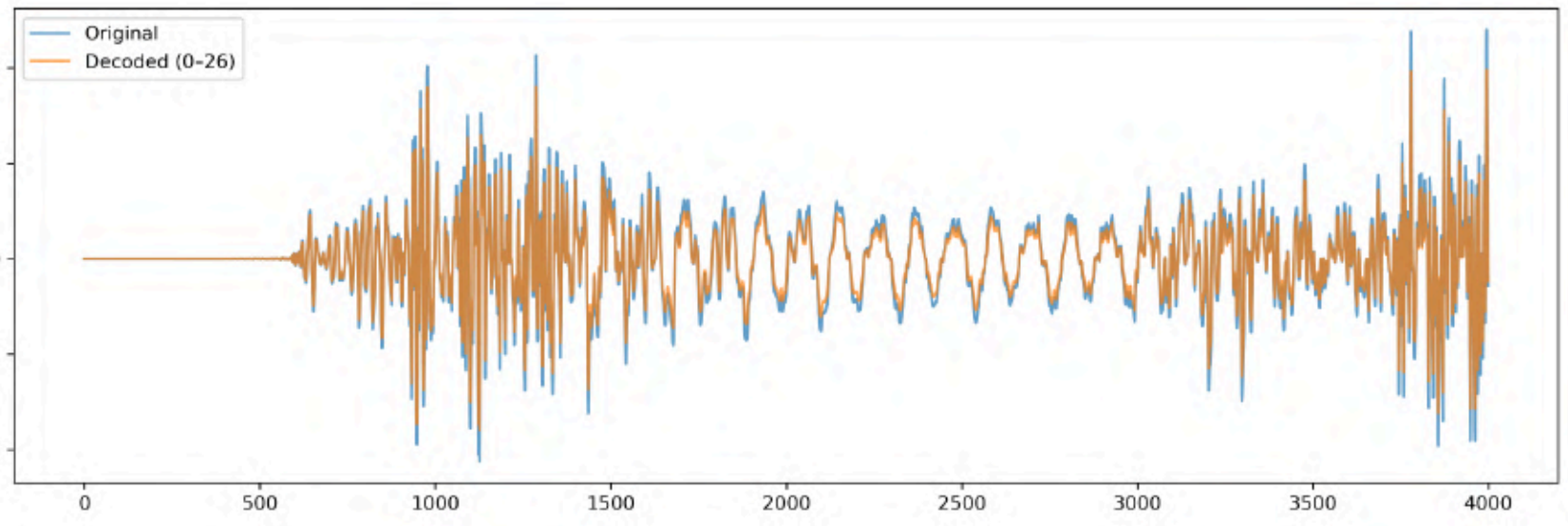

**(d)**

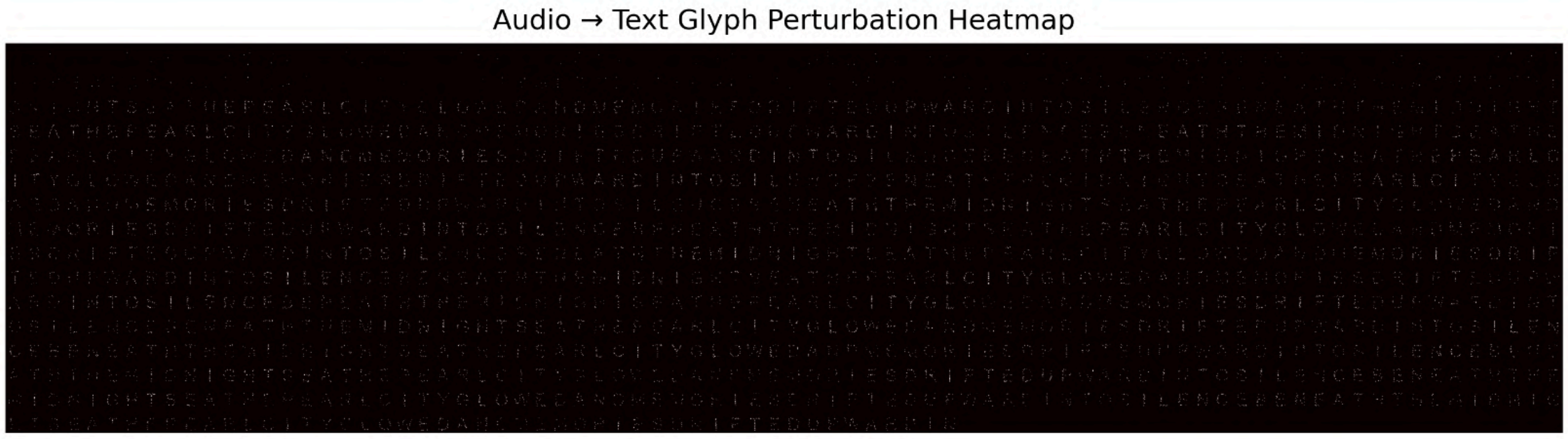

(e)

**Fig 9. Audio Sample 1: (a) Canonical Raster, (b) Encoded Raster, (c) Original and Decoded Spectrogram, (d) Original and Decoded Waveforms, (e) Glyph perturbation heatmap between canonical and encoded rasters.**

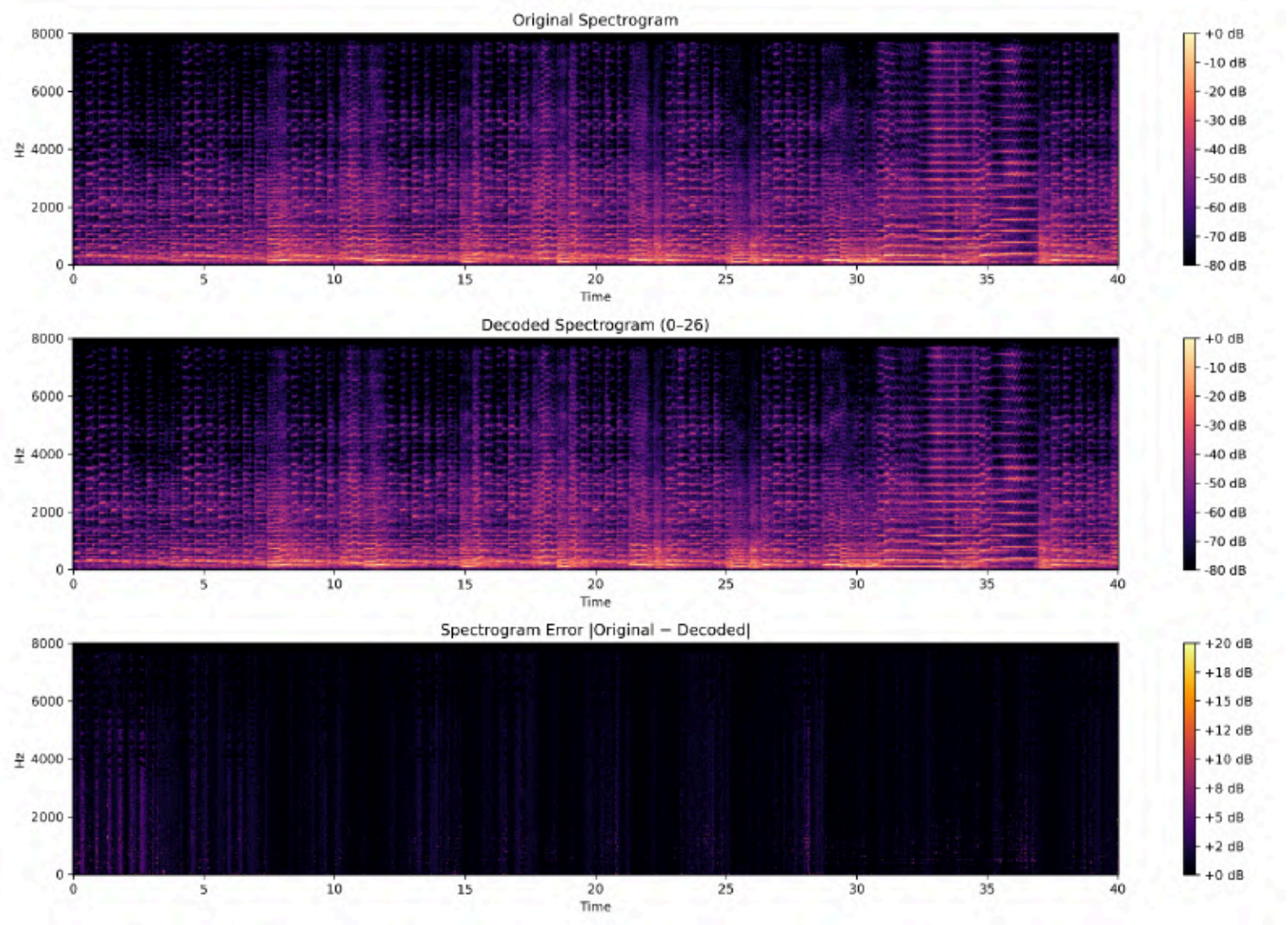

(a)

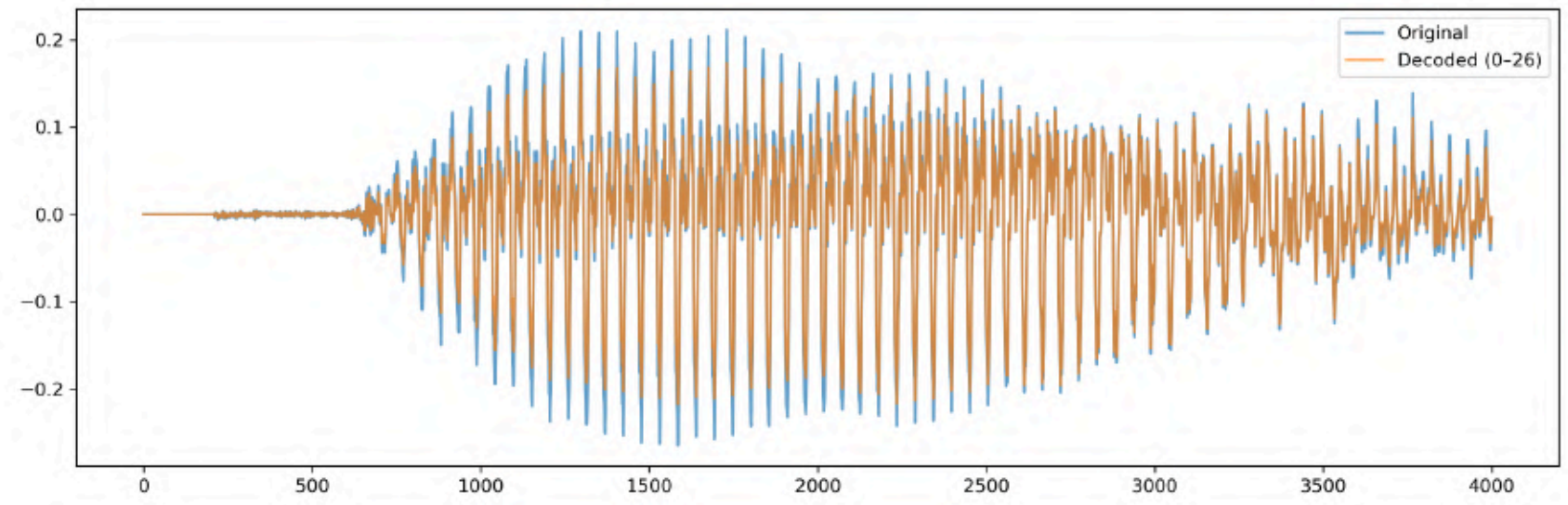

(b)

**Fig 10. Audio sample 2: (a) Original and Decoded Spectrogram, (b) Original and Decoded Waveforms**

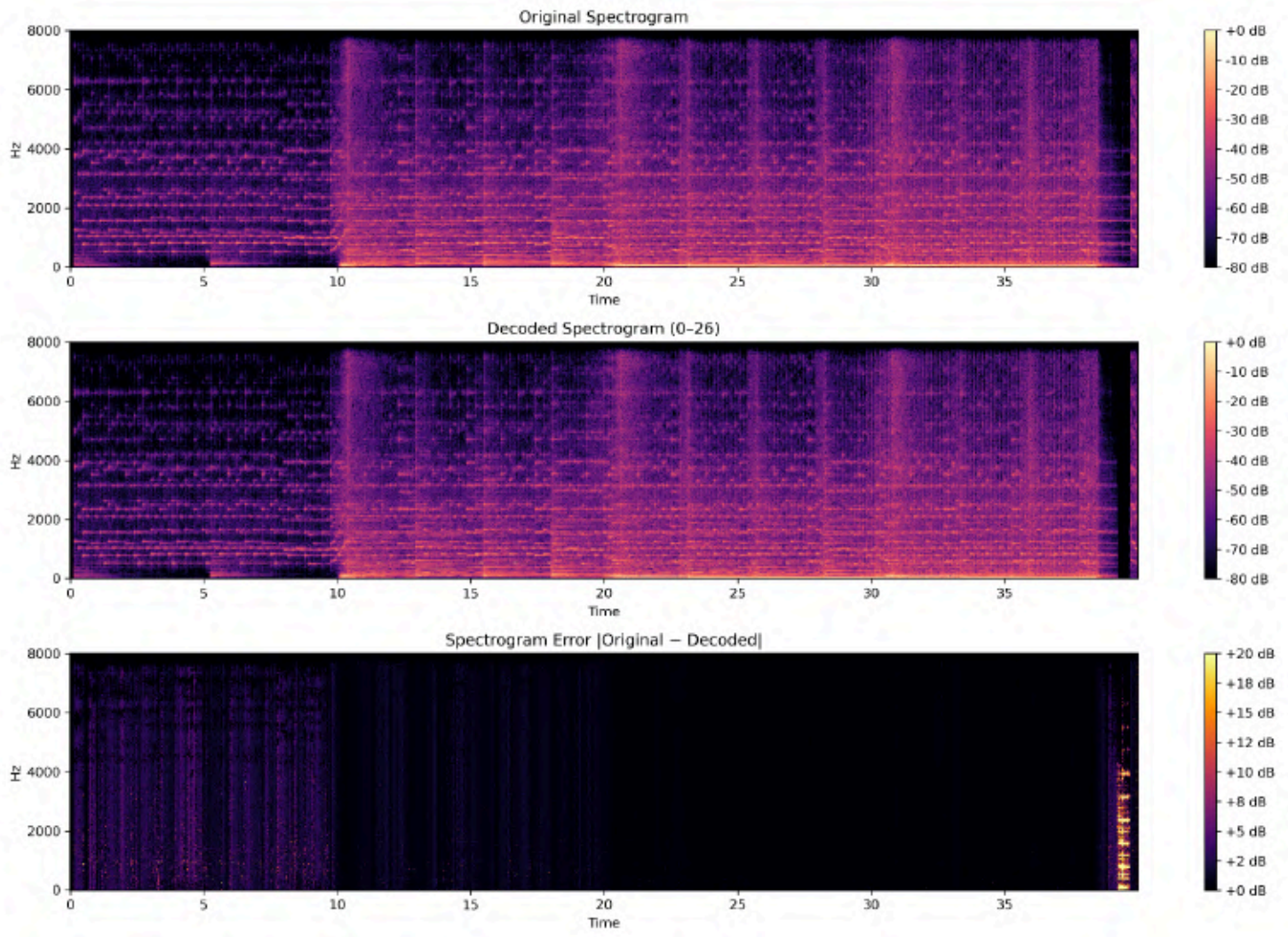

**(a)**

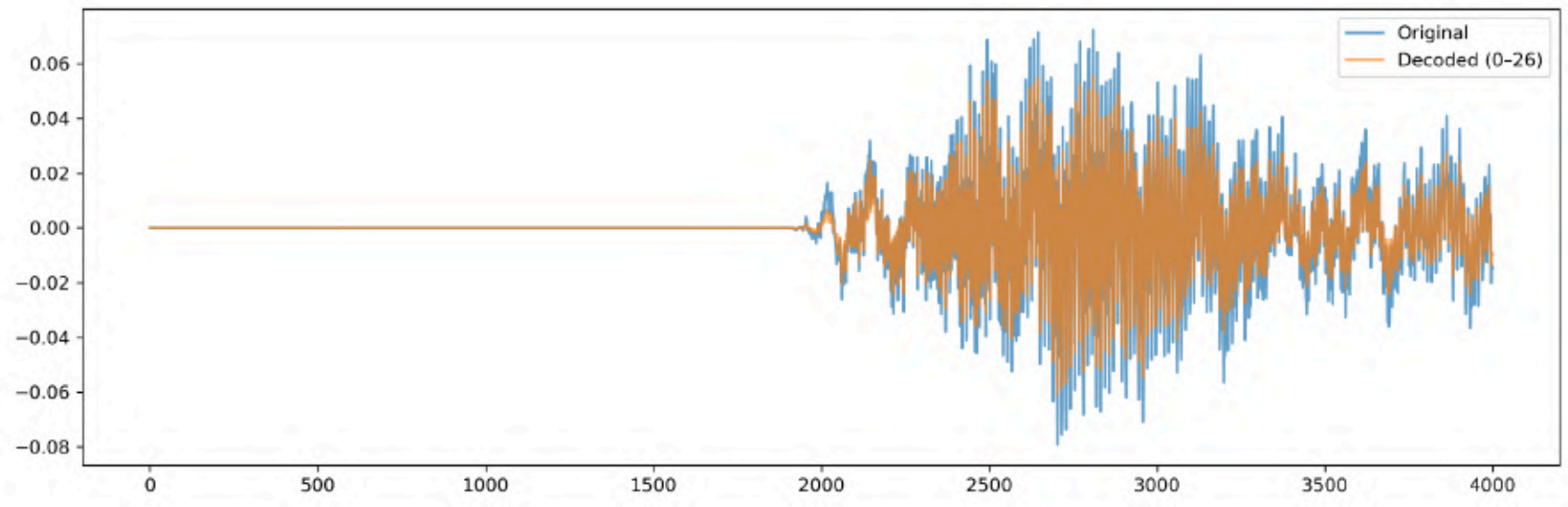

**(b)**

**Fig 11. Audio sample 3: (a) Original and Decoded Spectrogram, (b) Original and Decoded Waveforms**

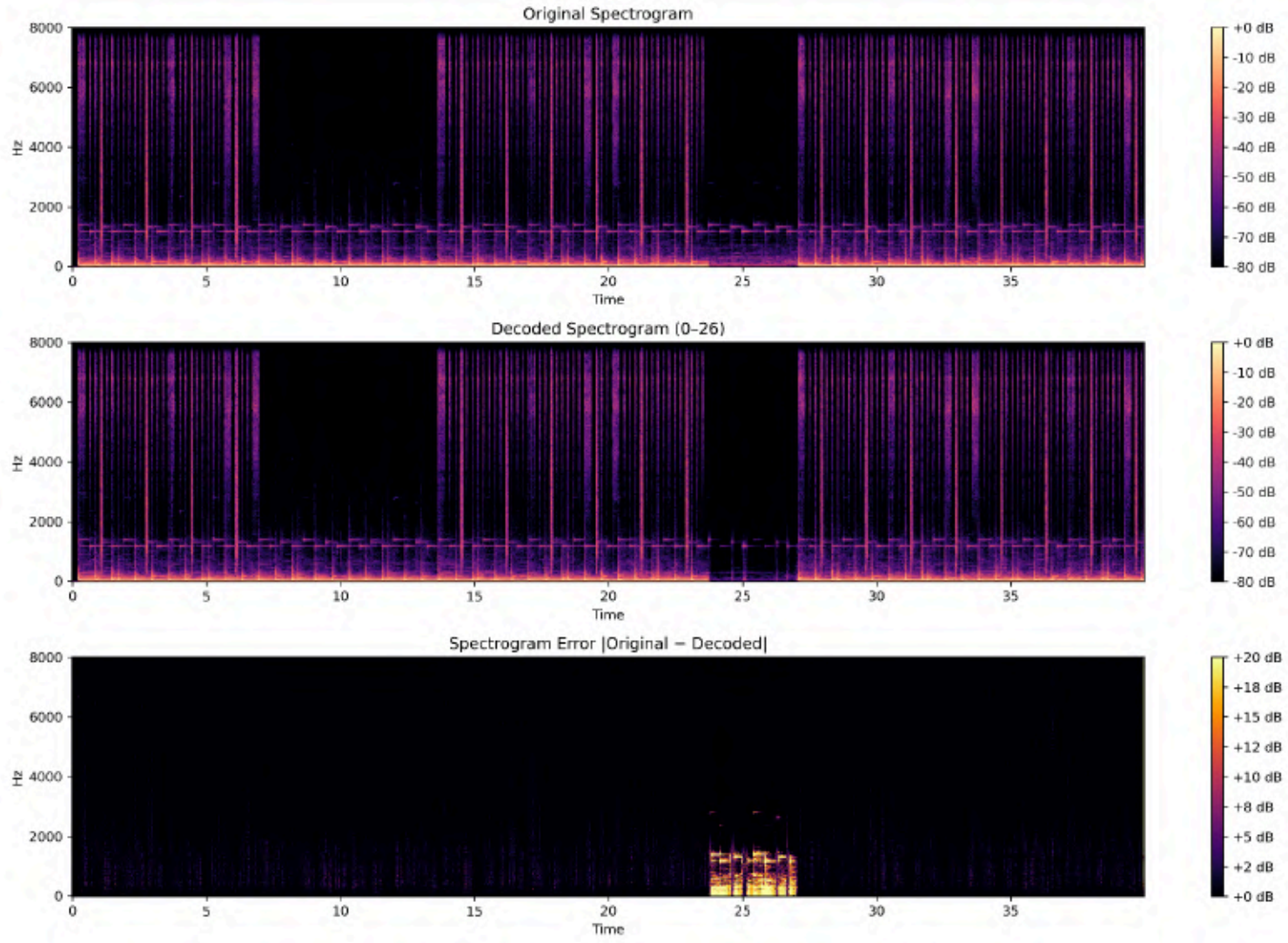

**(a)**

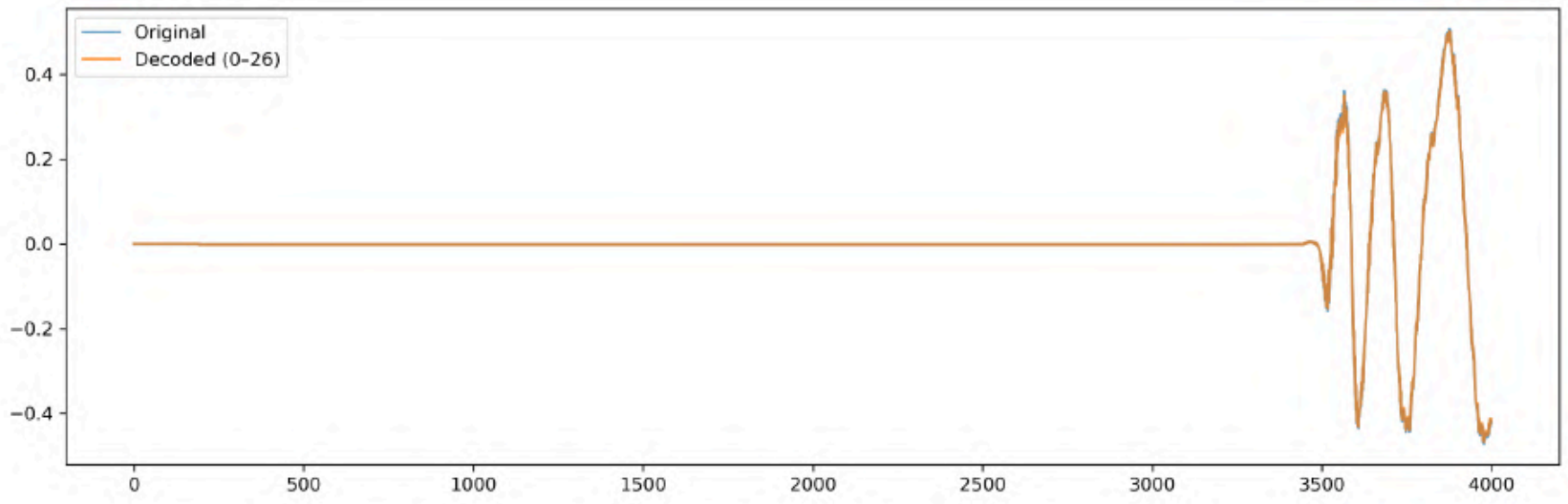

**(b)**

**Fig 12. Audio sample 4: (a) Original and Decoded Spectrogram, (b) Original and Decoded Waveforms**

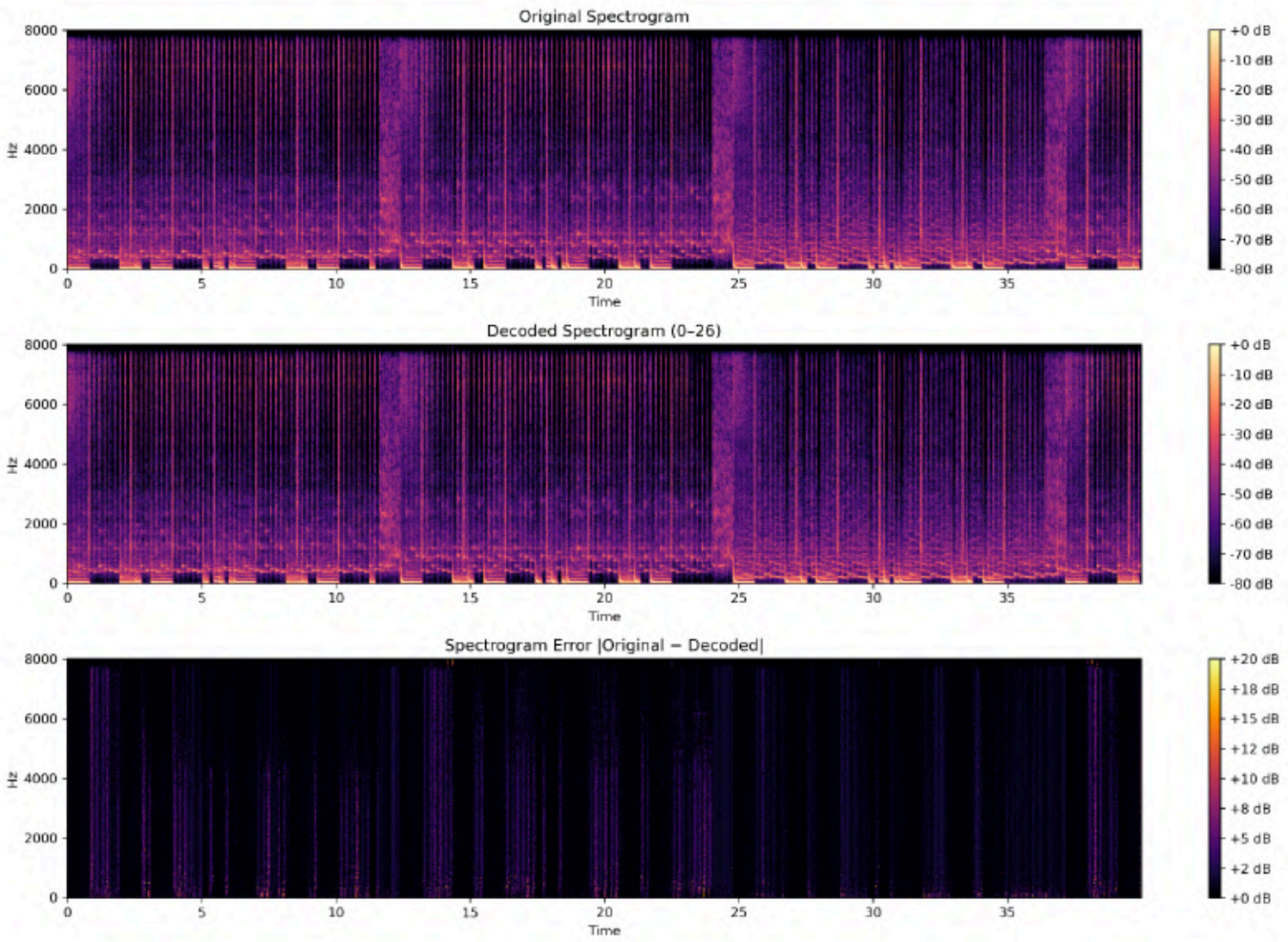

**(a)**

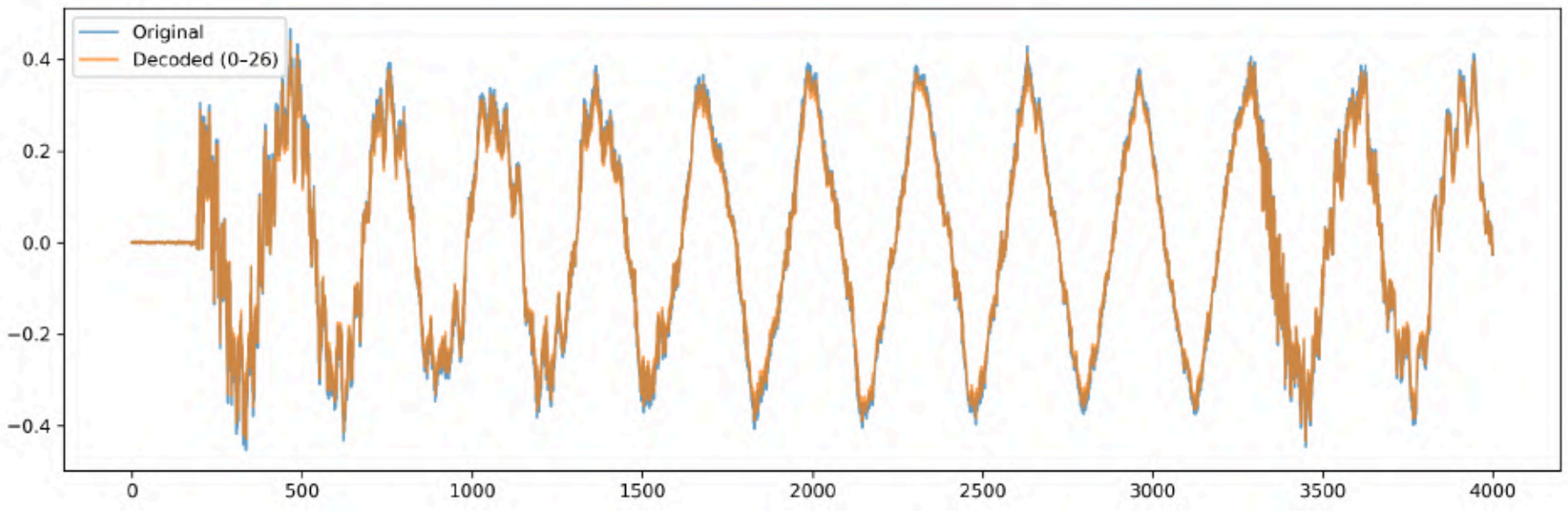

**(b)**

**Fig 13. Audio sample 5: (a) Original and Decoded Spectrogram, (b) Original and Decoded Waveforms**

### 5.4. VIDEO TO TEXT:

Five distinct videos were used to evaluate the performance of the proposed video to text encoding and reconstruction framework. Across all five test videos, the reconstruction quality remained consistently high despite the aggressive 26 level quantization imposed by the encoding scheme. PSNR values were tightly clustered between 38.7-40 dB for nearly every frame, indicating that the introduced error is small and visually negligible relative to the signal magnitude. SSIM values followed the same pattern, ranging from 0.93-0.95 in Videos 1, 2, and 5, while Videos 3 and 4 achieved slightly higher structural fidelity at 0.95-0.96, reflecting content with likely good textures and more uniform frame statistics as shown in figure 19. These metrics collectively show that the method introduces only minimal distortion and preserves overall visual structure with high reliability.

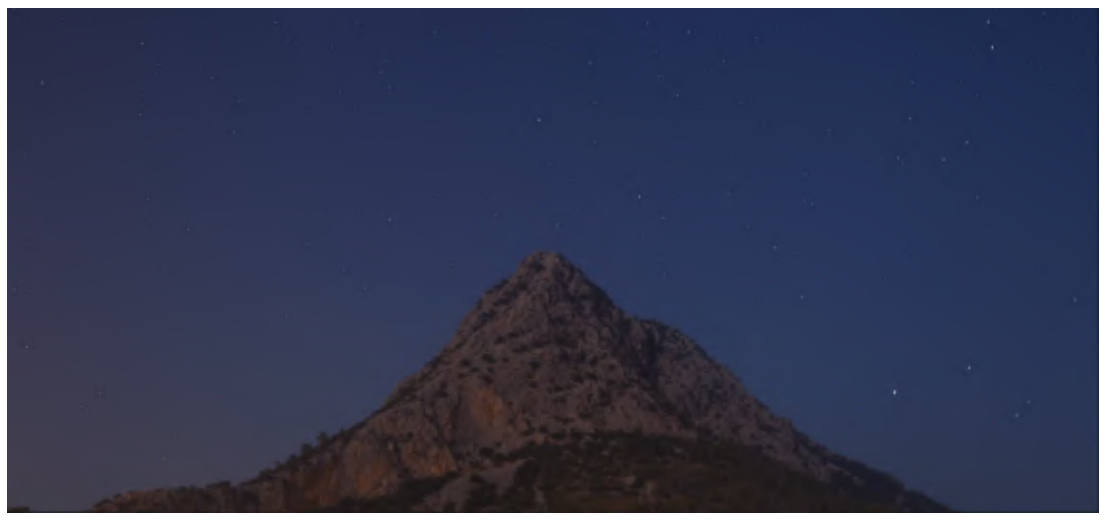

**(a)**

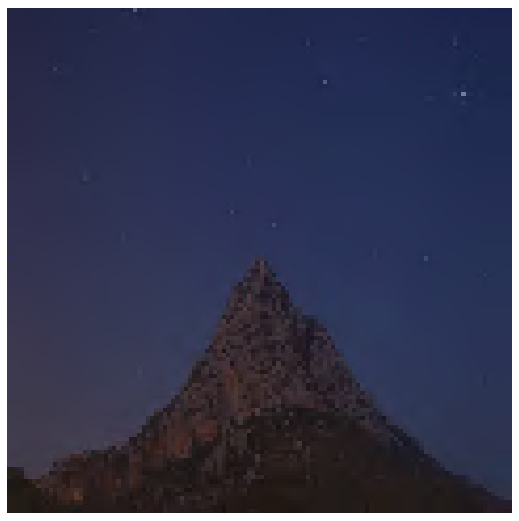

**(b)**

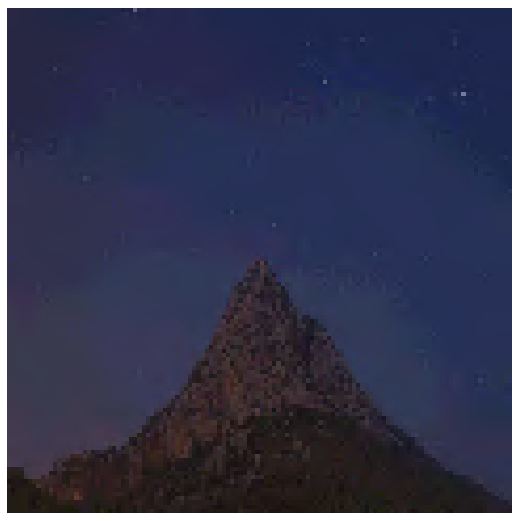

**(c)**

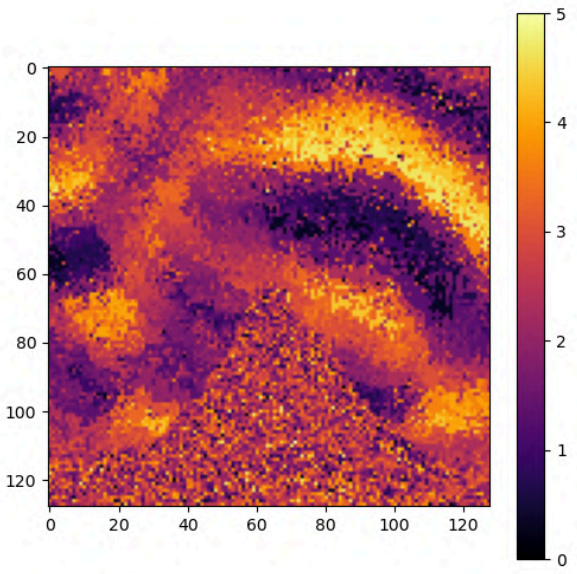

**(d)**

**Fig 14. (a) Original Frame(0) of Video-1 (Resolution 1280*720, 30FPS), (b) & (c) Canonical Frame(0), Decoded Frame (0) (Resolution 120*120, 30FPS), (d) Error heatmap for Canonical and Decoded Frame (0) of Video-1.**

Heatmaps of per pixel reconstruction error reinforce this conclusion. Across all videos, the error remained bounded within a 0-5 intensity range, confirming that nearly all deviations arise from deterministic rounding during quantization and de-quantization. Importantly, the spatial distribution of error remained diffuse and uniform rather than concentrated along edges, moving objects, or high frequency regions as shown in figure 14-18. This suggests that the encoding process does not introduce any systematic artifacts or text shaped imprints from the rasterized tiles, demonstrating that the visual embedding remains completely hidden and non disruptive to the decoded video. Overall, the quantitative metrics and visual error patterns demonstrate that the proposed raster based representation is robust across diverse video types, frame counts, and motion patterns. Even under full frame encoding, the reconstruction remains perceptually indistinguishable from the original, with no visible artifacts and only small, uniformly distributed quantization noise. These results confirm that the method can reliably encode frame information in a rasterized glyph sequence composed of independent glyph tiles without compromising structural fidelity, making it suitable for multimodal transmission, secure encoding, or cross domain representation tasks.

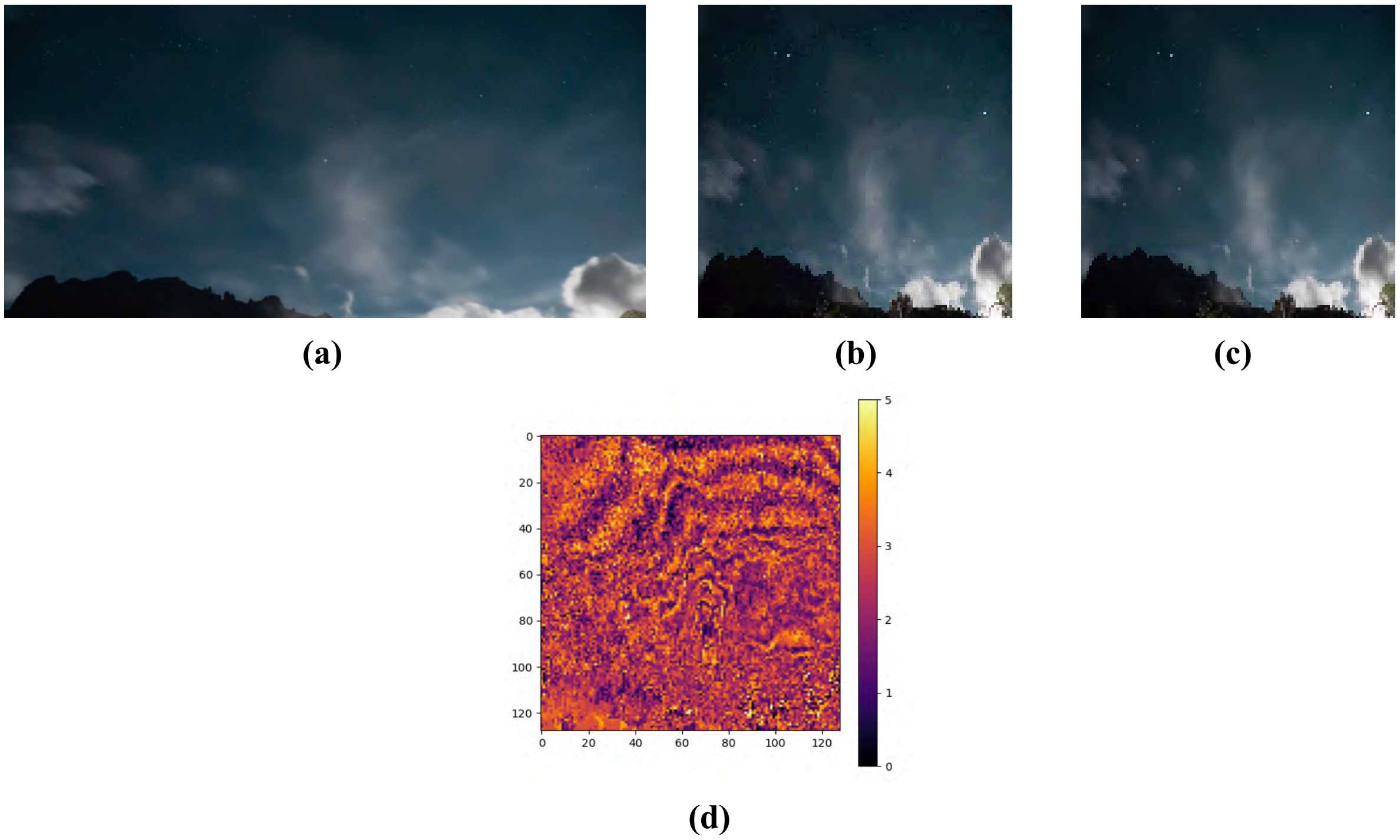

**Fig 15. (a) Original Frame(0) of Video-2 (Resolution 3840*2160, 30FPS), (b) & (c) Canonical Frame(0), Decoded Frame (0) (Resolution 120*120, 30FPS), (d) Error heatmap for Canonical and Decoded Frame (0) of Video-2.**

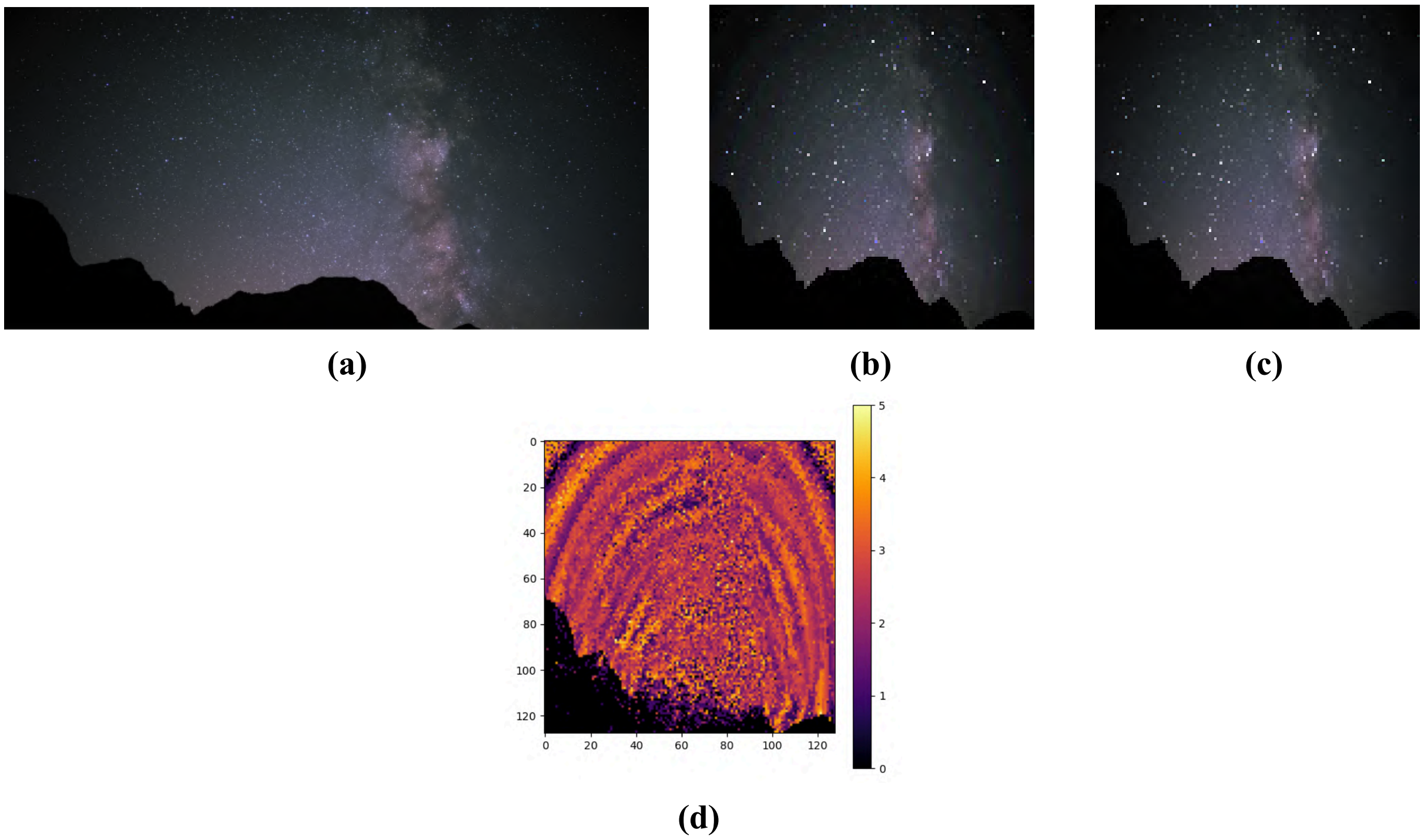

**Fig 16. (a) Original Frame(0) of Video-3 (Resolution 3840*2160, 30FPS), (b) & (c) Canonical Frame(0), Decoded Frame (0) (Resolution 120*120, 30FPS), (d) Error heatmap for Canonical and Decoded Frame (0) of Video-3.**

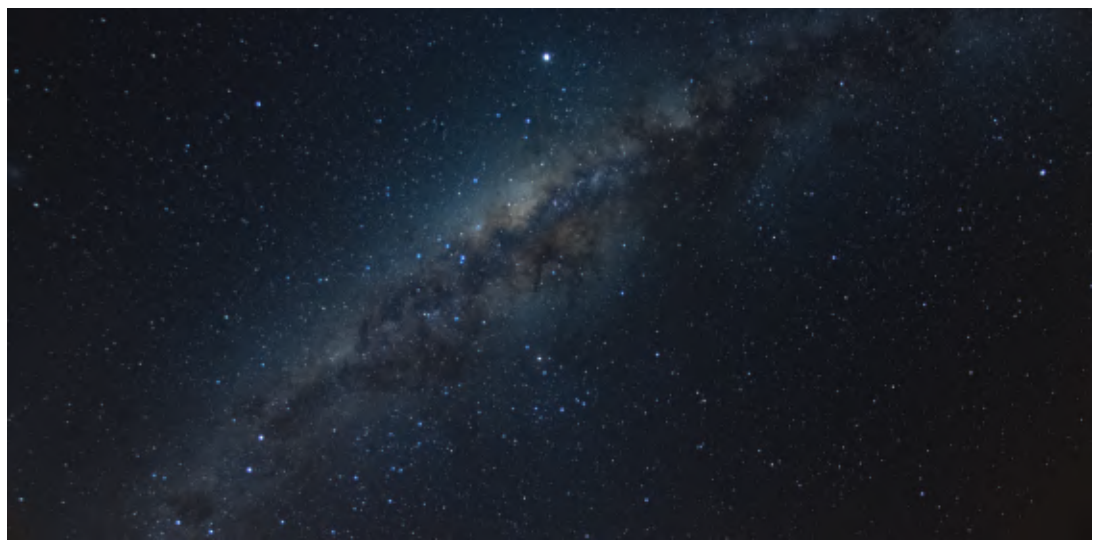

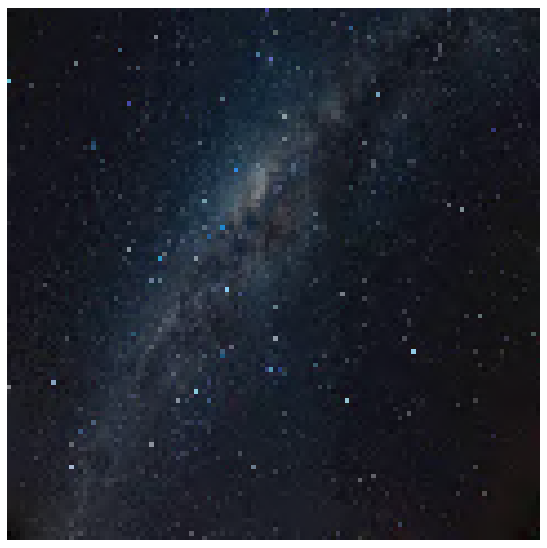

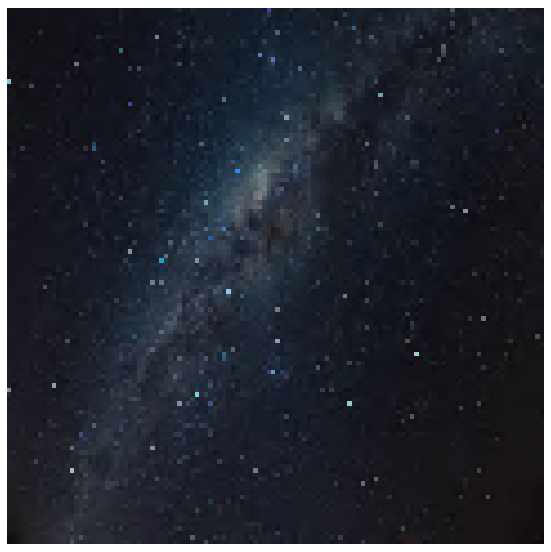

(a) (b) (c)

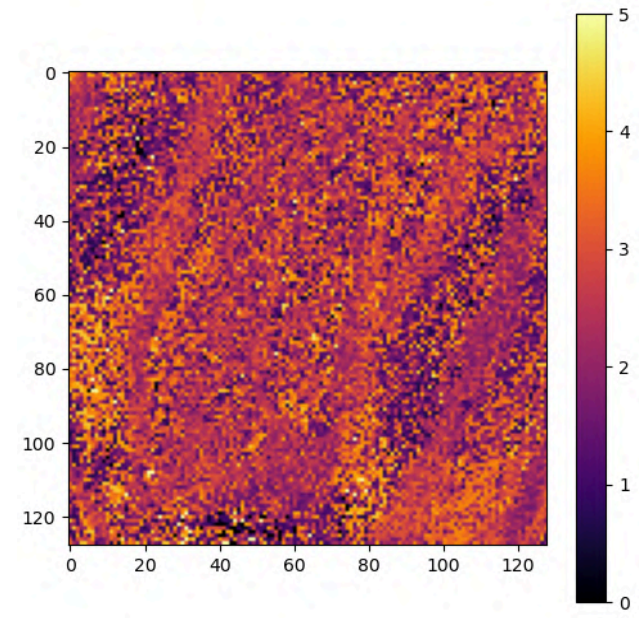

(d)

**Fig 17. (a) Original Frame(0) of Video-4 (Resolution 3840*2160, 30FPS), (b) & (c) Canonical Frame(0), Decoded Frame (0) (Resolution 120*120, 30FPS), (d) Error heatmap for Canonical and Decoded Frame (0) of Video-4.**

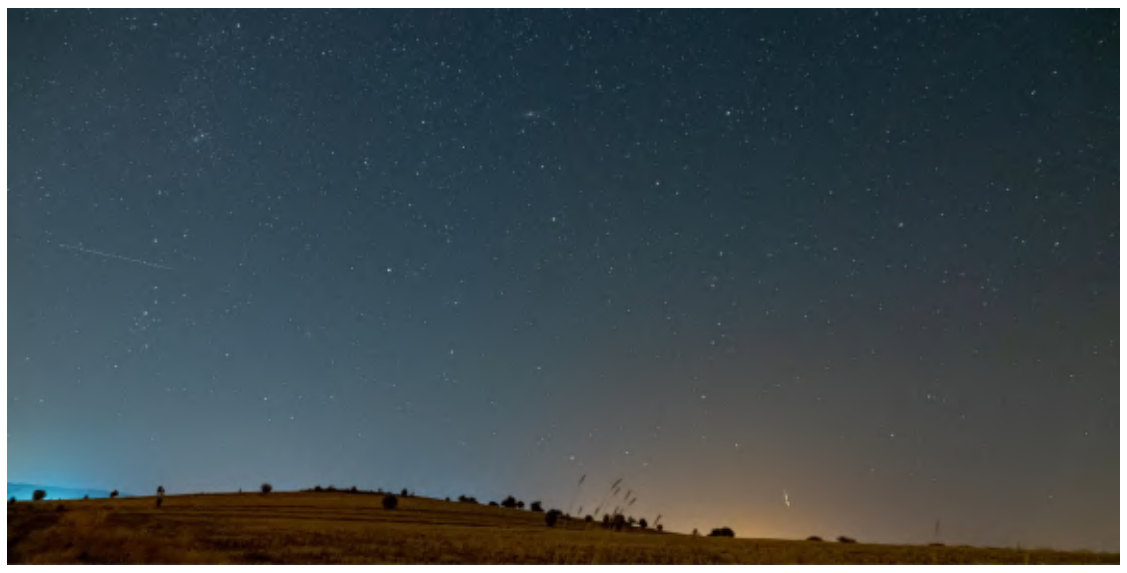

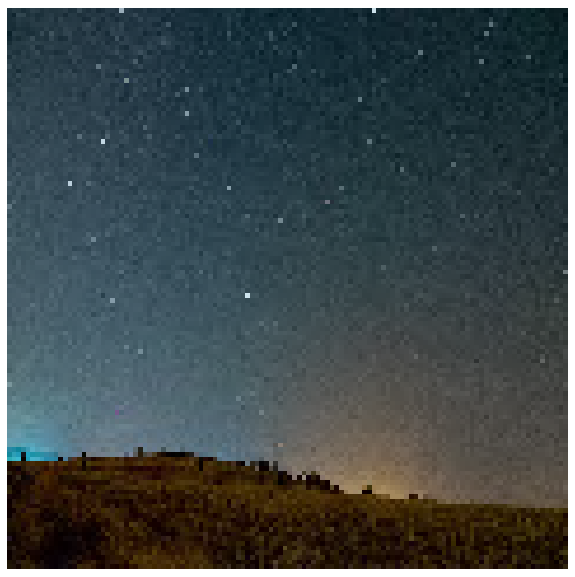

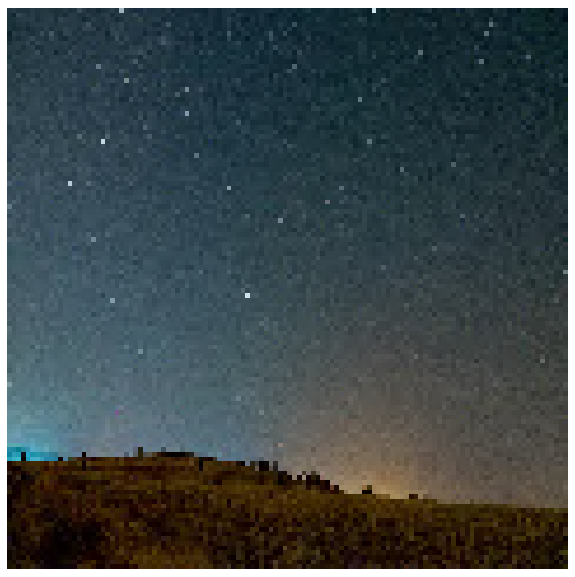

(a) (b) (c)

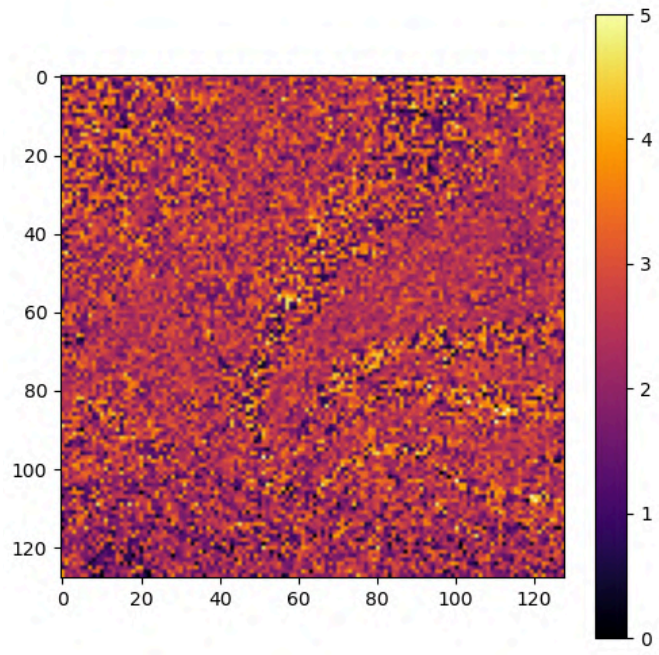

(d)

**Fig 18. (a) Original Frame(0) of Video-5 (Resolution 3840*2160, 30FPS), (b) & (c) Canonical Frame(0), Decoded Frame (0) (Resolution 120*120, 30FPS), (d) Error heatmap for Canonical and Decoded Frame (0) of Video-5.**

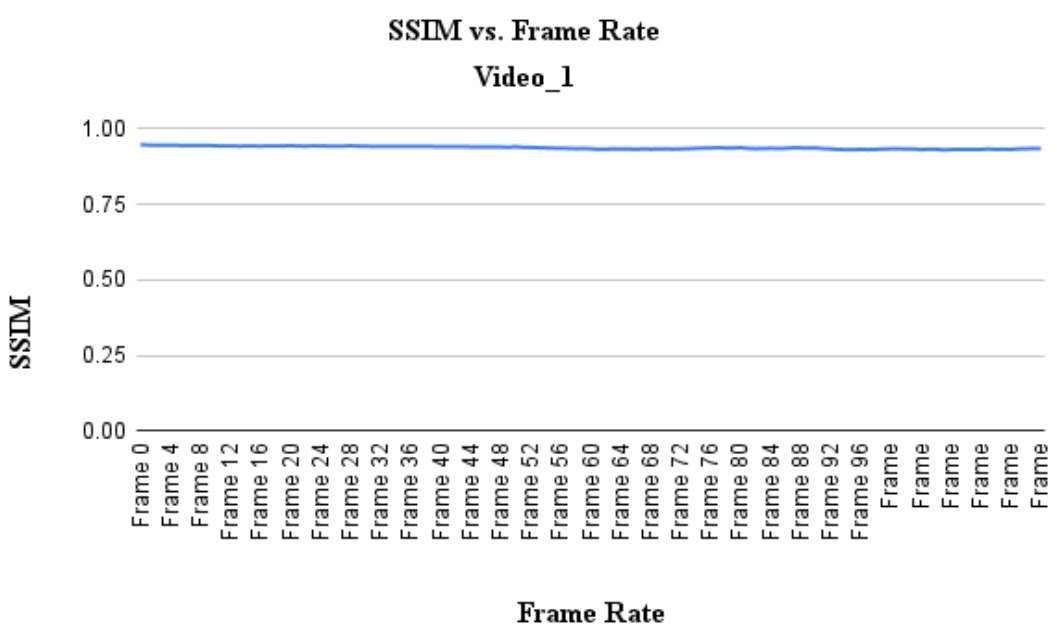

**(a)**

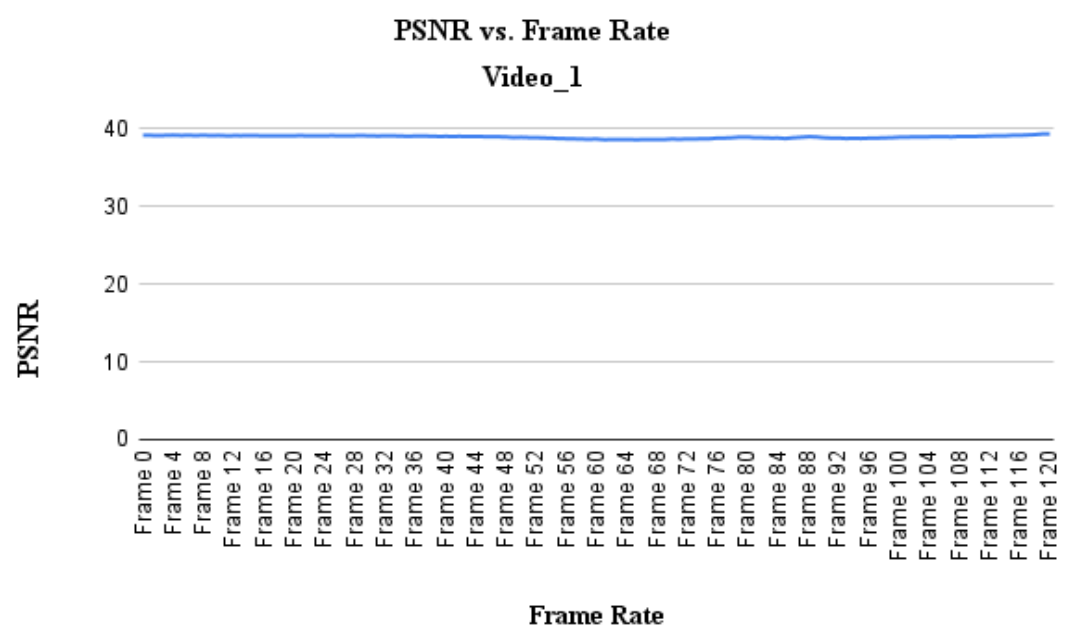

**(b)**

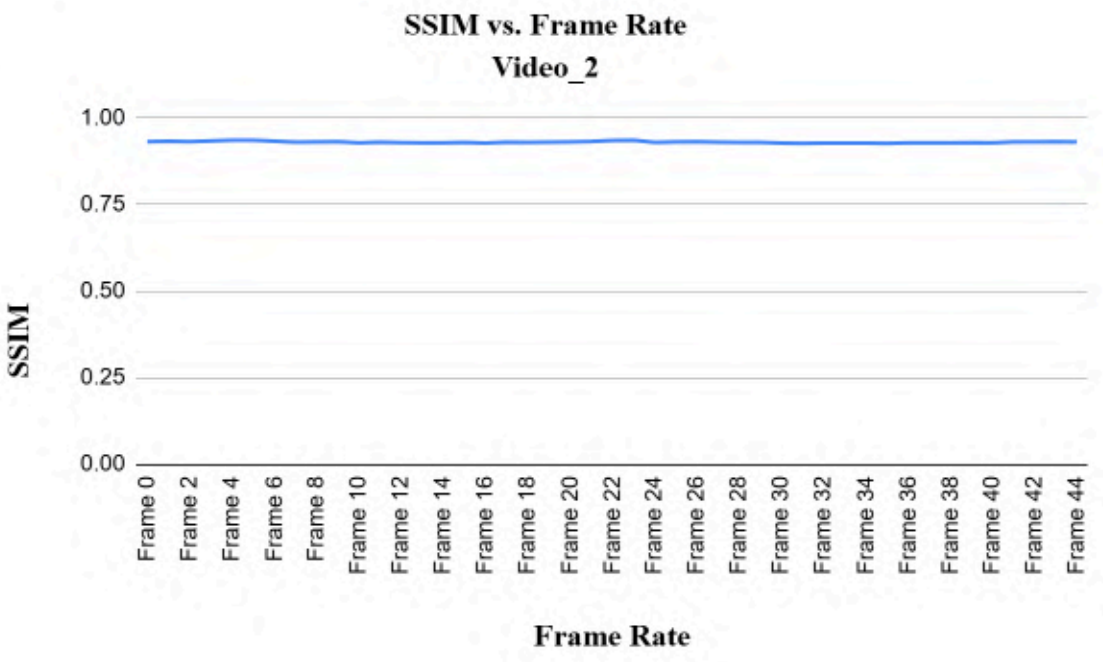

**(c)**

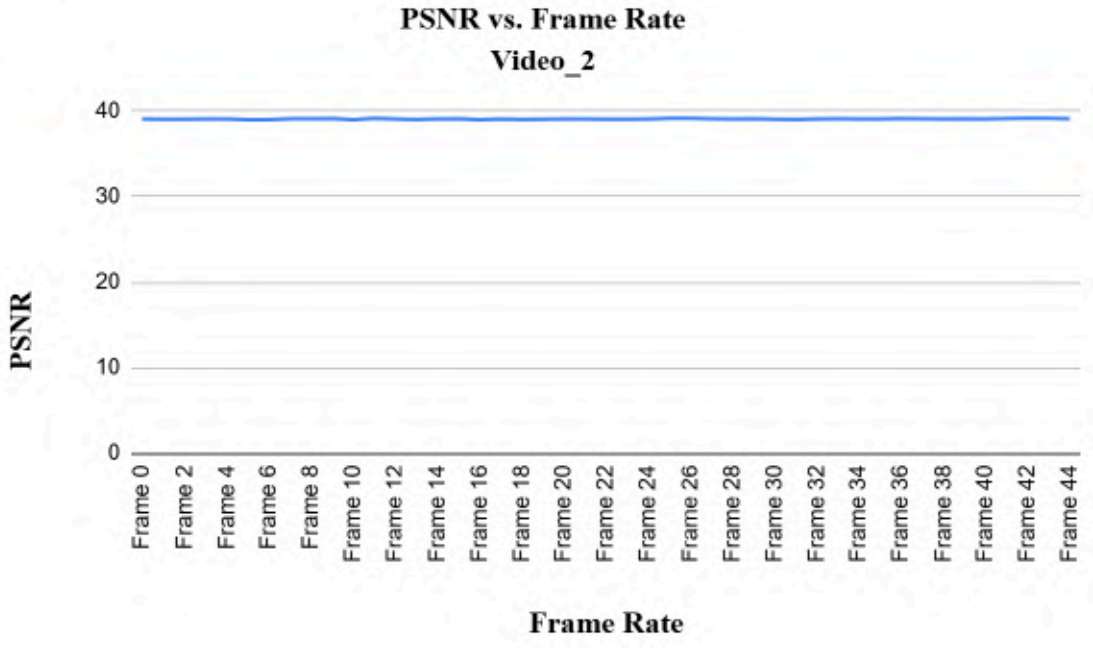

**(d)**

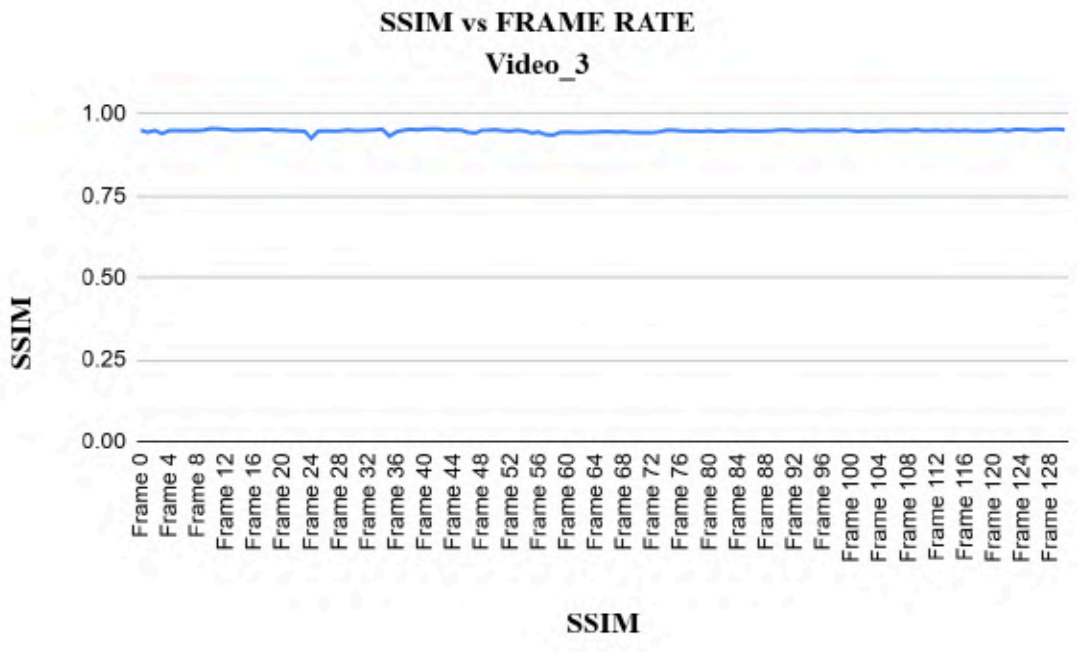

**(e)**

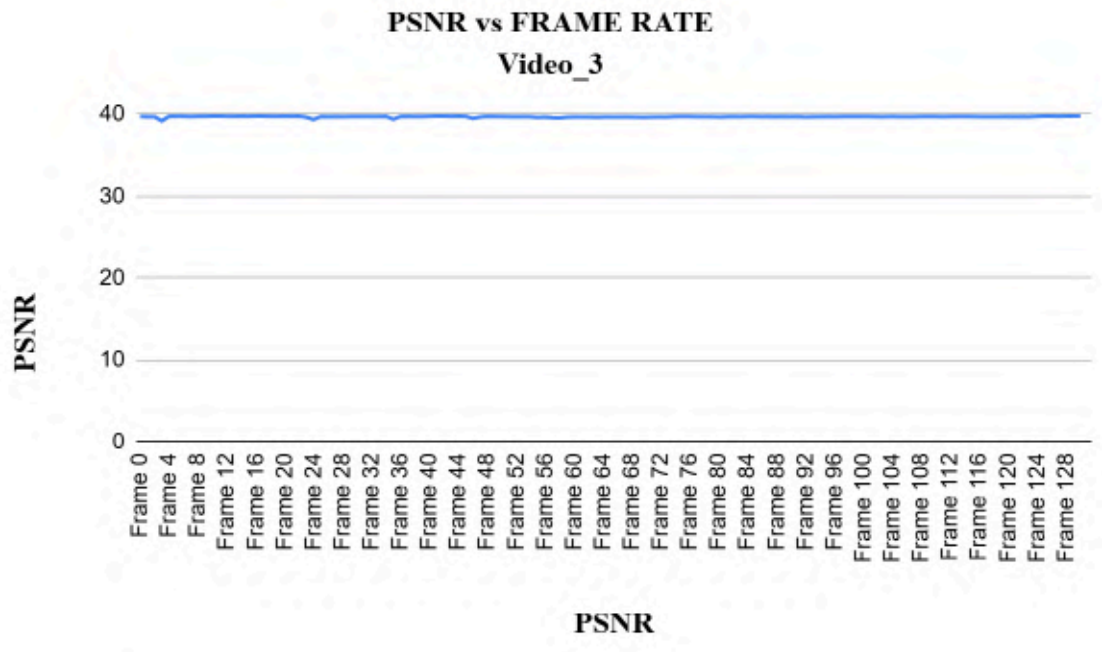

**(f)**

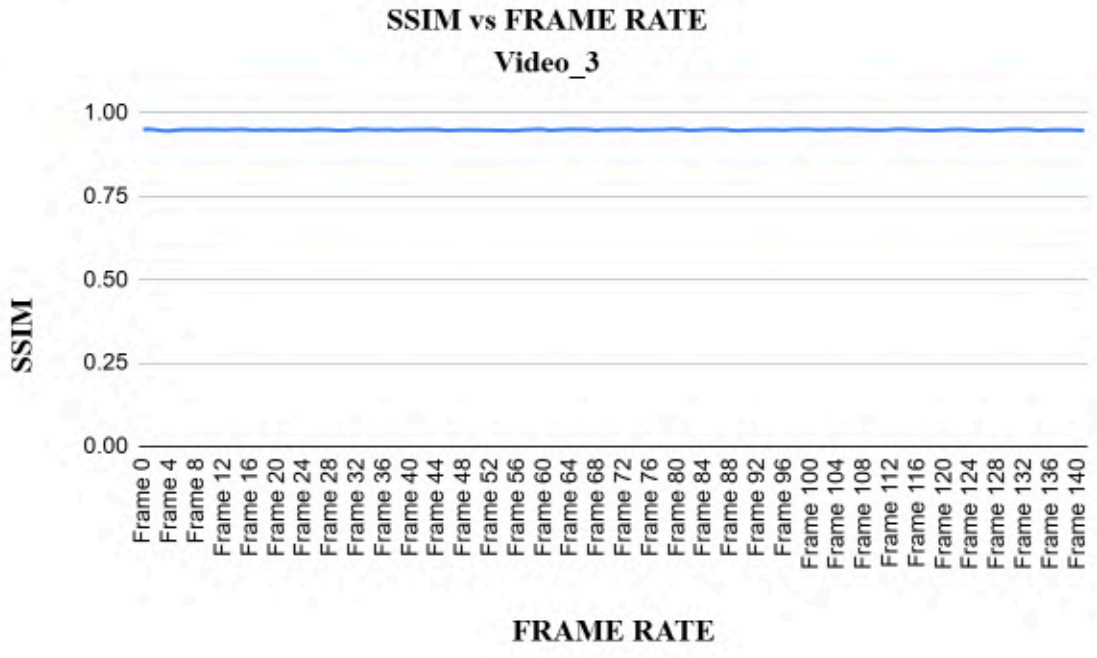

**(g)**

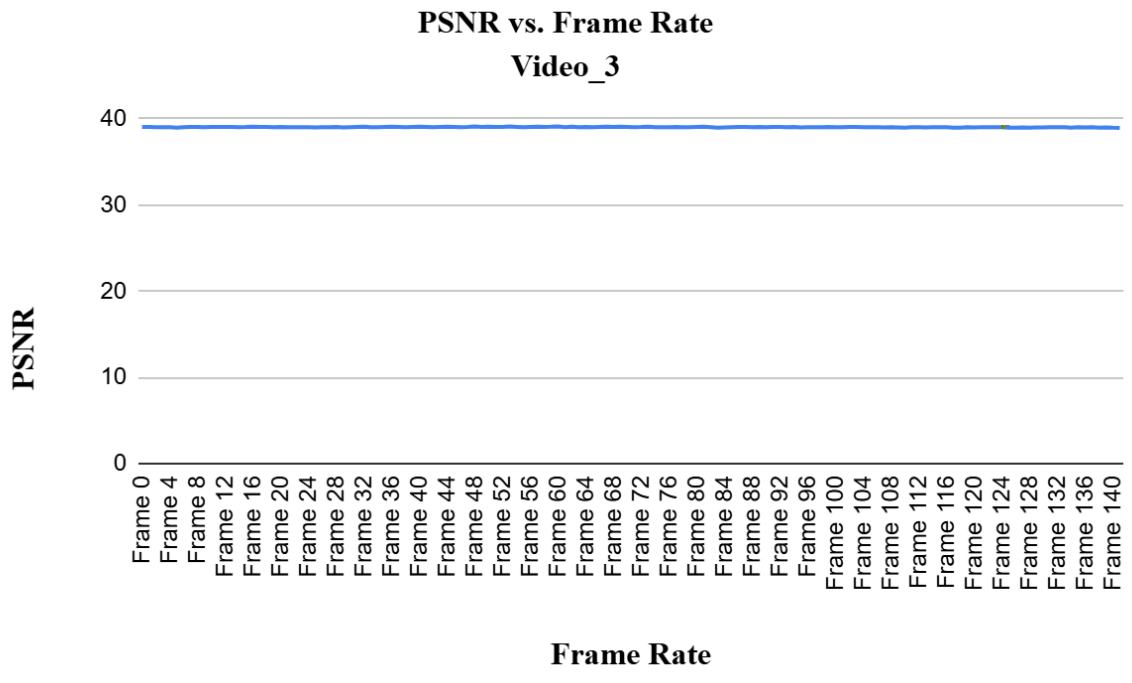

**(h)**

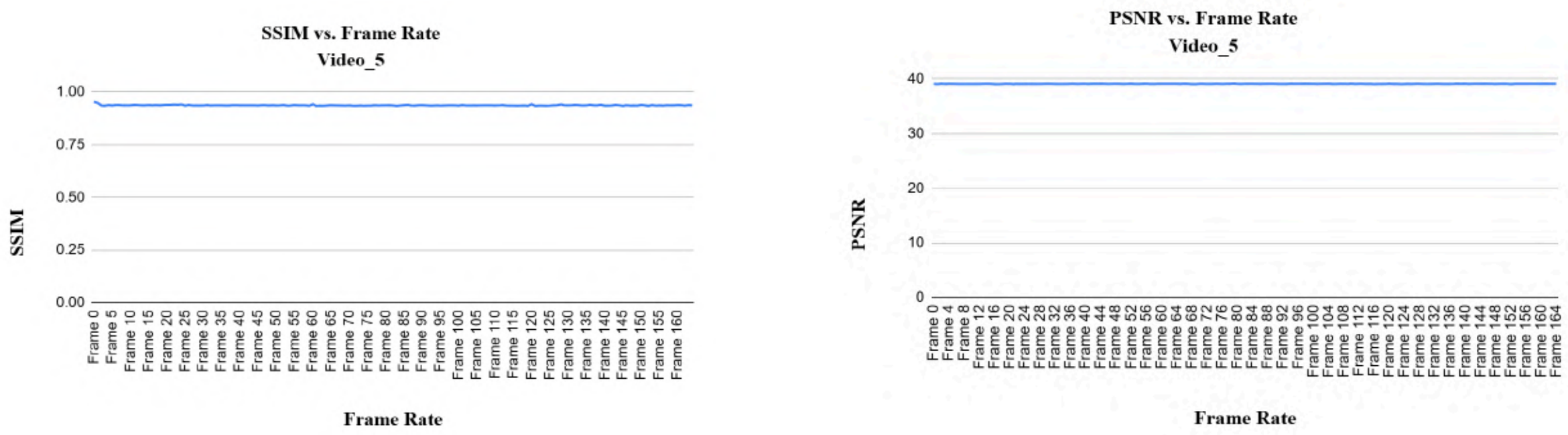

**(i)** **(j)**

**Fig 19. SSIM & PSNR vs Frame Rate plots (a-b): Video-1, (c-d): Video-2, (e-f): Video-3, (g-h): Video-4, (i-j): Video-5**

## 6. MODALITY-WISE ADVANTAGES AND LIMITATIONS:

### 6.1. TEXT-TO-TEXT:

Among the four modalities, text functions as the ideal carrier because glyph interiors provide a deterministic, high capacity, and perfectly reversible perturbation space. The encoding rule remains identical across all modalities, yet text offers a substantial margin between required and available capacity due to dense interior stroke regions. Perturbations are confined strictly within glyph interiors and never touch contours, producing encoded renderings that are visually indistinguishable from canonical text. Under deterministic rendering, this yields 0% character error rate (CER) and 0% bit error rate (BER) across all evaluated words, indicating that the channel reaches its theoretical upper bound when the input itself is discrete.

Limitations arise primarily from uneven per glyph capacity due to thin characters such as I, L, or T containing far fewer interior pixels, restricting payload density in very short cover texts. Decoding reliability also depends on exact raster determinism, if any variation in font hinting, anti-aliasing, or compression may corrupt the perturbation field. Nevertheless, within fixed rendering pipelines, text to text embedding is effectively lossless and could be further strengthened using multi pass raster fusion or adaptive perturbation scheduling.

### 6.2. IMAGE-TO-TEXT:

When extended to images, the perturbation as integer quantizer interpretation remains valid. Each pixel intensity is mapped into a bounded integer range and preserved as a perturbation count within glyph interiors. The channel behaves as a stable, modality independent quantizer where reconstructed emoji images retain structural integrity. Canonical, encoded, and decoded raster tiles remain visually aligned, confirming that interior perturbations introduce no contour distortion.

The principal limitation is the fixed 26 level quantization ceiling. Continuous tonal gradients collapse into discrete bands, and small ±1 deviations manifest as uniform artifacts across smooth regions. This limitation is architectural rather than fundamental and few other techniques such as multi glyph fusion, channel wise dithering, or vector quantization could significantly increase effective dynamic range while preserving visual stealth.

### 6.3. AUDIO-TO-TEXT:

Audio represents the most demanding modality within the proposed framework, as perceptual information is distributed across fine grained temporal dynamics and frequency dependent structure. While the raster domain approach preserves its core advantages uniform encoding logic across modalities, deterministic raster decoding, and stable frame level alignment, the reconstruction quality in the audio-to-text setting is primarily constrained by the low cardinality (0-26) scalar quantization used to form the payload.

Decoded audio signals consistently preserve global temporal alignment and coarse amplitude envelopes derived from frame level RMS values. However, fine spectral detail, including higher frequency components and transient structure, is attenuated due to the limited resolution of the scalar encoding. This behavior reflects a fundamental mismatch between continuous audio waveforms and a low cardinality mechanism, rather than instability or failure of the raster glyph perturbation mechanism itself.

Despite this limitation, the results demonstrate that RMS based scalar encoding provides a stable and well matched interface between audio signals and the raster perturbation channel, enabling reliable cross modal recovery of coarse energy structure. Further improvements in perceptual fidelity are likely through enhanced audio specific preprocessing and postprocessing. These directions represent a clear path toward higher quality audio reconstructions while retaining the unified low cardinality design philosophy of the framework.

### 6.4. VIDEO-TO-TEXT:

Video inherits the spatial advantages of image embedding while introducing temporal consistency requirements. The uniform encoding principle again proves effective, with each frame independently quantized and mapped into glyph perturbation counts using a fully deterministic raster pipeline. Because frames are encoded without inter frame dependencies, no accumulation of error occurs over time. Reconstructed videos exhibit stable quality across hundreds of frames, with no observable frame to frame drift or temporal instability, and the rendered carrier text remains visually unchanged throughout.

As with images, the dominant limitation arises from the fixed low dynamic range of the 0-26 quantization, which flattens subtle textures and smooth gradients, particularly in regions of low contrast. Hierarchical quantization strategies, chroma luma separation, motion aware preprocessing, or lightweight learned post filters could substantially improve reconstruction fidelity without altering the underlying perturbation rule or raster embedding mechanism.

## 7. DETERMINISM CONSTRAINT (CROSS-MODAL LIMITATION):

A central limitation of the perturbation cardinality channel is its strict dependence on deterministic rasterization. Decoding assumes identical glyph geometry, interior fill topology, and aliasing behavior between encoding and decoding. Any deviation such as platform dependent font hinting, subpixel anti-aliasing, PDF vector reflow, lossy compression (JPEG, HEIC), or OCR systems that reconstruct stroke outlines destroys perturbation counts and renders recovery impossible. Even minor rendering differences across operating systems, browsers, or document viewers can alter interior pixel topology.

Consequently, the method currently operates reliably only when the text carrier is preserved in a pixel exact form (e.g., PNG transport, fixed layout documents, or controlled rendering pipelines). This constraint simultaneously enables strong visual stealth since perturbations remain imperceptible and limits deployment to environments where raster integrity is maintained. Addressing robustness to reflow, lossy compression, or print scan processes constitutes an engineering challenge orthogonal to the core contribution.

## 8. FUTURE WORK:

Several extensions follow naturally from this study. Increasing the dynamic range beyond 26 levels through multi glyph fusion or hierarchical perturbation encoding could substantially reduce quantization error in image, audio, and video modalities. Preprocessing pipelines such as μ-law companding, LPC coding, vector quantized mel spectra, or motion compensated video intensities may yield significantly higher perceptual fidelity while preserving the core raster domain perturbation rule. Robustness to real world document transformations, including PDF export, OCR, and lossy compression, may be improved through glyph redundancy, error correcting codes, or multi render consensus decoding. Additionally, learned perturbation placement strategies such as grayscale dithering models or font adaptive networks could further increase capacity while maintaining visual stealth. Collectively, these directions suggest that raster domain steganography is not fundamentally capacity limited but representation limited, and offers a broad design space for future investigation.

## 9. CONCLUSION:

This work introduced a unified perturbation cardinality channel for embedding and recovering data across four fundamentally different modalities text, images, audio, and video using deterministic glyph rasterization alone. The core mechanism is intentionally simple, where each unit of information is mapped to an integer within a fixed 0-26 range and encoded as the count of minimally perturbed interior pixels within a rendered glyph. Despite this simplicity, the channel behaves as a stable, reversible, and modality independent quantizer. Textual payloads are recovered with perfect accuracy, while spatial modalities such as images and video retain high structural fidelity, consistently achieving PSNR values above 39 dB and SSIM values above 0.93. Even in the more demanding audio domain, the system preserves global temporal structure, indicating that the perturbation field remains coherent under aggressive quantization. The limitations observed in the experiments arise primarily from the constrained dynamic range imposed by the 26-level encoding. This manifests as predictable ±1 intensity errors in spatial modalities and a loss of fine spectral detail in audio. These constraints are architectural rather than fundamental. The channel can be extended through richer quantizers, hierarchical or multi glyph encoding, and modality aware pre- and post-processing, such as LPC or mel-spectral compression for audio and perceptually weighted color quantization for images and video. These extensions would increase expressive capacity without altering the underlying raster domain principle. More broadly, the results demonstrate that deterministic text rendering can function as a reliable, high capacity universal carrier when the symbol to perturbation mapping is preserved. By operating entirely in the raster domain, the method bypasses the linguistic rigidity that constrains conventional text steganography and instead exploits the geometric redundancy of glyph interiors.

Overall, this work serves as a proof of concept within a fully deterministic rendering environment. Practical deployment, robustness to non ideal pipelines, and optimized modality specific refinements are left for future study. Nevertheless, the findings establish that text treated not as a linguistic construct but as a geometric field of pixels can act as a cross modal transport medium capable of embedding structured data with strong visual stealth and predictable reconstruction behavior. This perspective opens a path toward flexible, interpretable, and deterministic data through text systems built on raster domain principles.

**CODE AVAILABILITY:**

The reference implementation of the proposed raster domain embedding framework, along with scripts for reproducing all experiments reported in this paper, is publicly available at: https://github.com/Udaykandhala/Glyph-Perturbation-Cardinality-Mechanism/.

The repository includes instructions for reproducing the audio, image, video, and text embedding experiments, as well as parameter settings corresponding to the figures shown in the main paper and appendix.

**ACKNOWLEDGEMENT:**

This research was conducted independently by the author. No external funding, institutional grants, or departmental resources were used in the development of the framework, methodology, or experimental results etc.

**CONFLICTS OF INTERESTS:**

The author declares that there are no conflicts of interests with any persons or organizations that could influence this work.

**AUTHOR CONTRIBUTIONS:**

A V Uday Kiran Kandala is the sole contributor to the conceptualization, methodology, software development, paper drafting, editing and formal analysis of the “Raster Domain Text Steganography” framework.

**APPENDIX A:**

**A.1. AUDIO TO TEXT (Additional Results):**

Appendix A.1, shows the additional experimentations in audio to text steganographic pipeline.

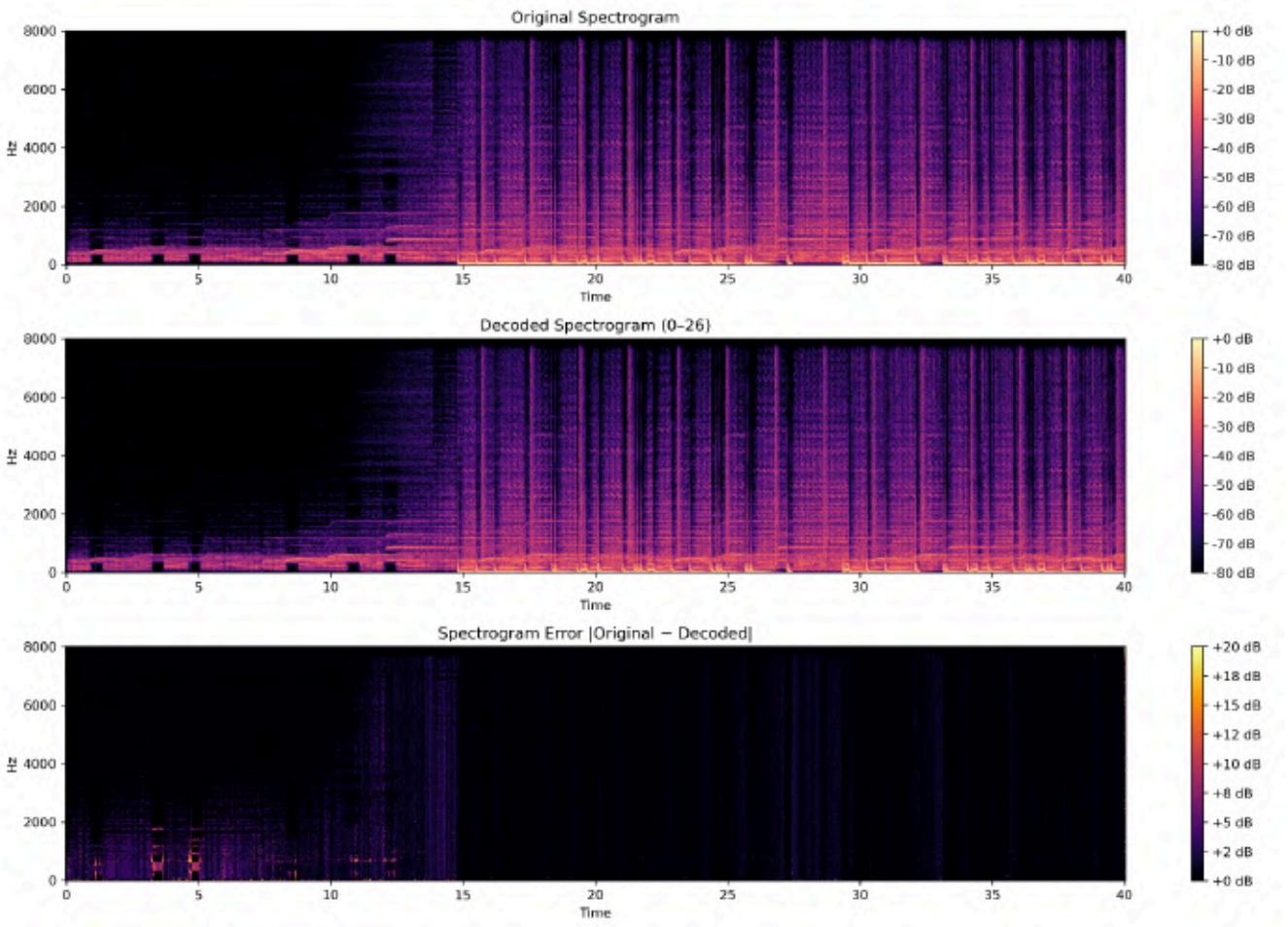

**(a)**

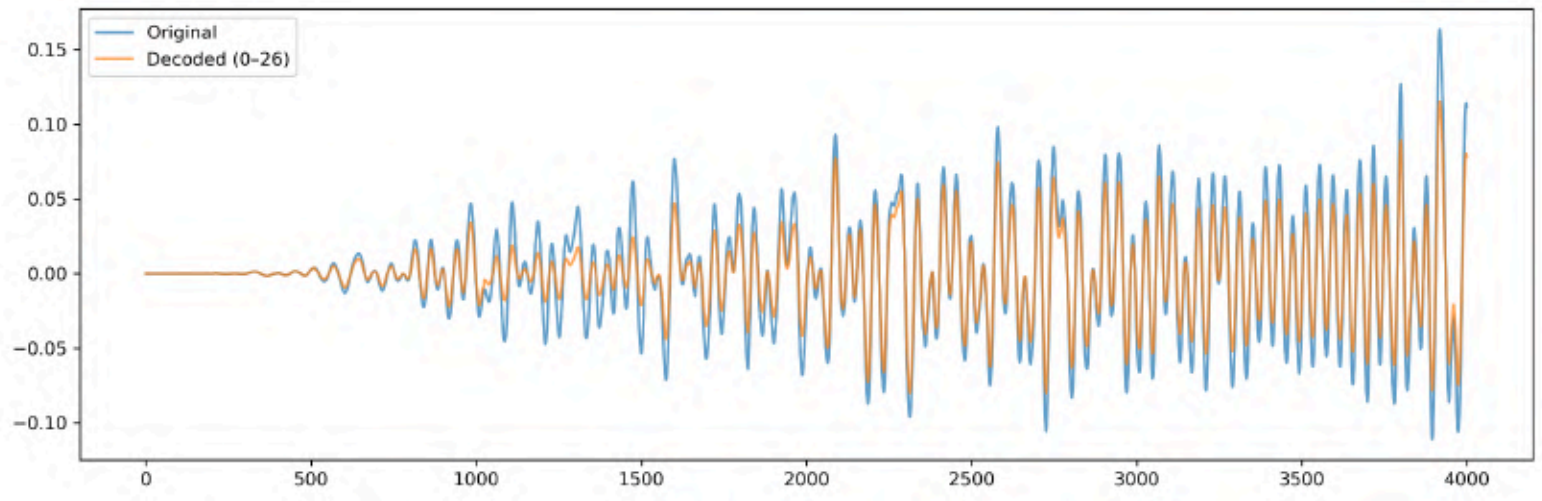

**(b)**

**Fig20: Audio-6. (a) Spectrograms and (b) Waveforms**

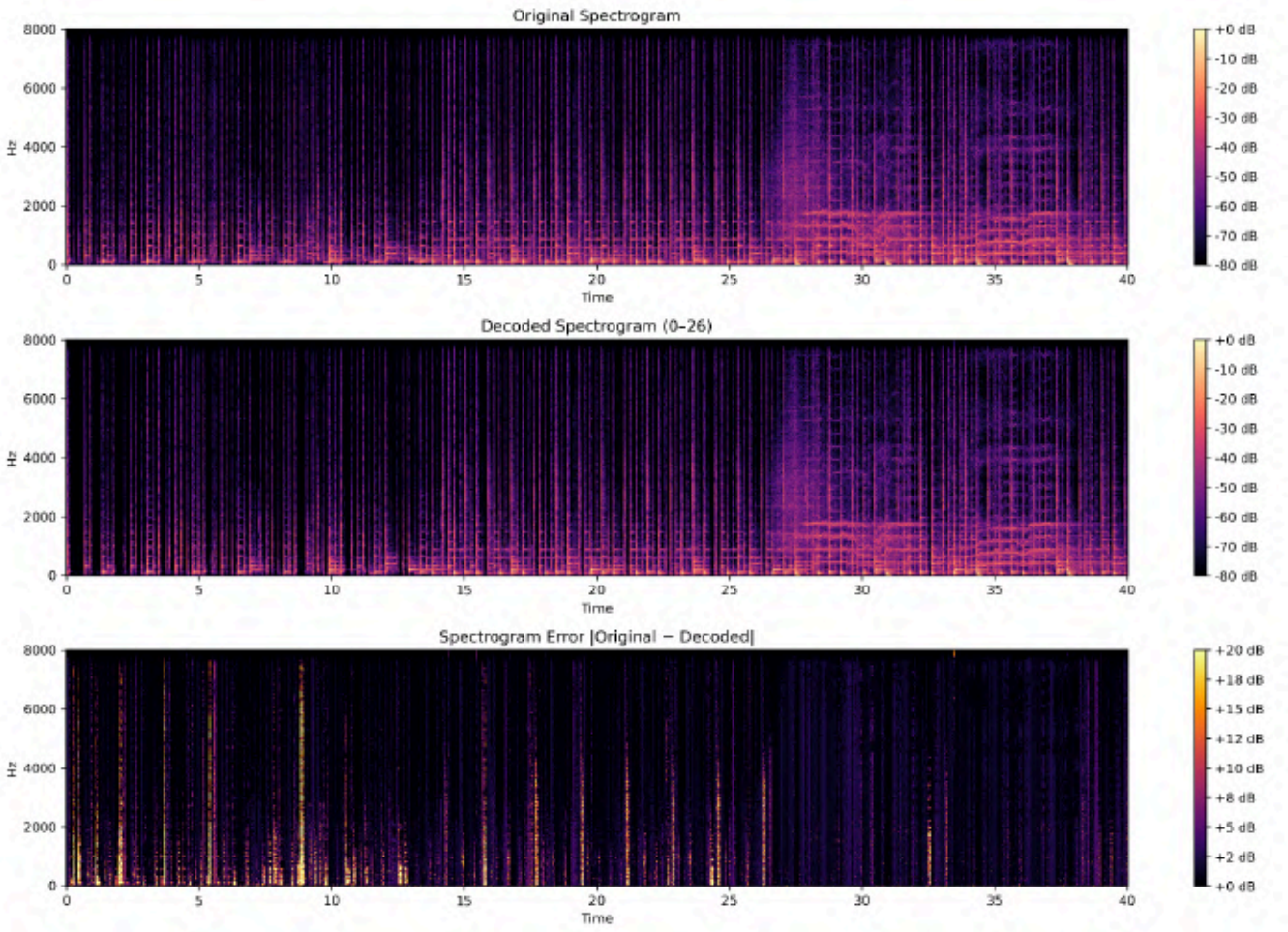

**(a)**

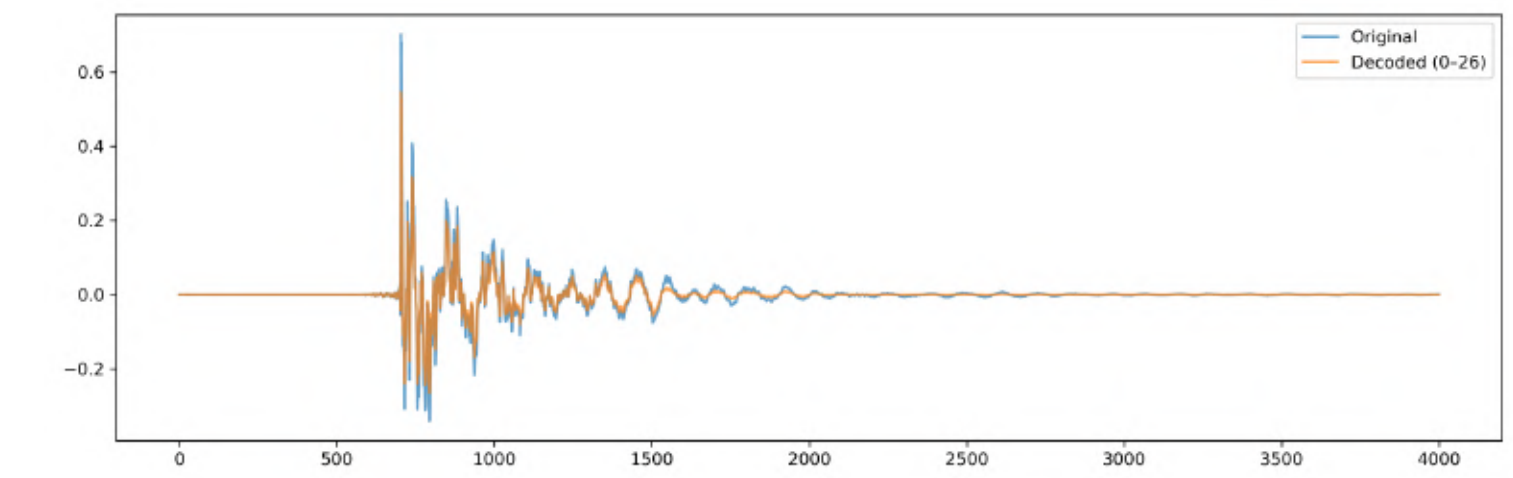

**(b)**

**Fig 21: Audio-7. (a) Spectrograms and (b) Waveforms**

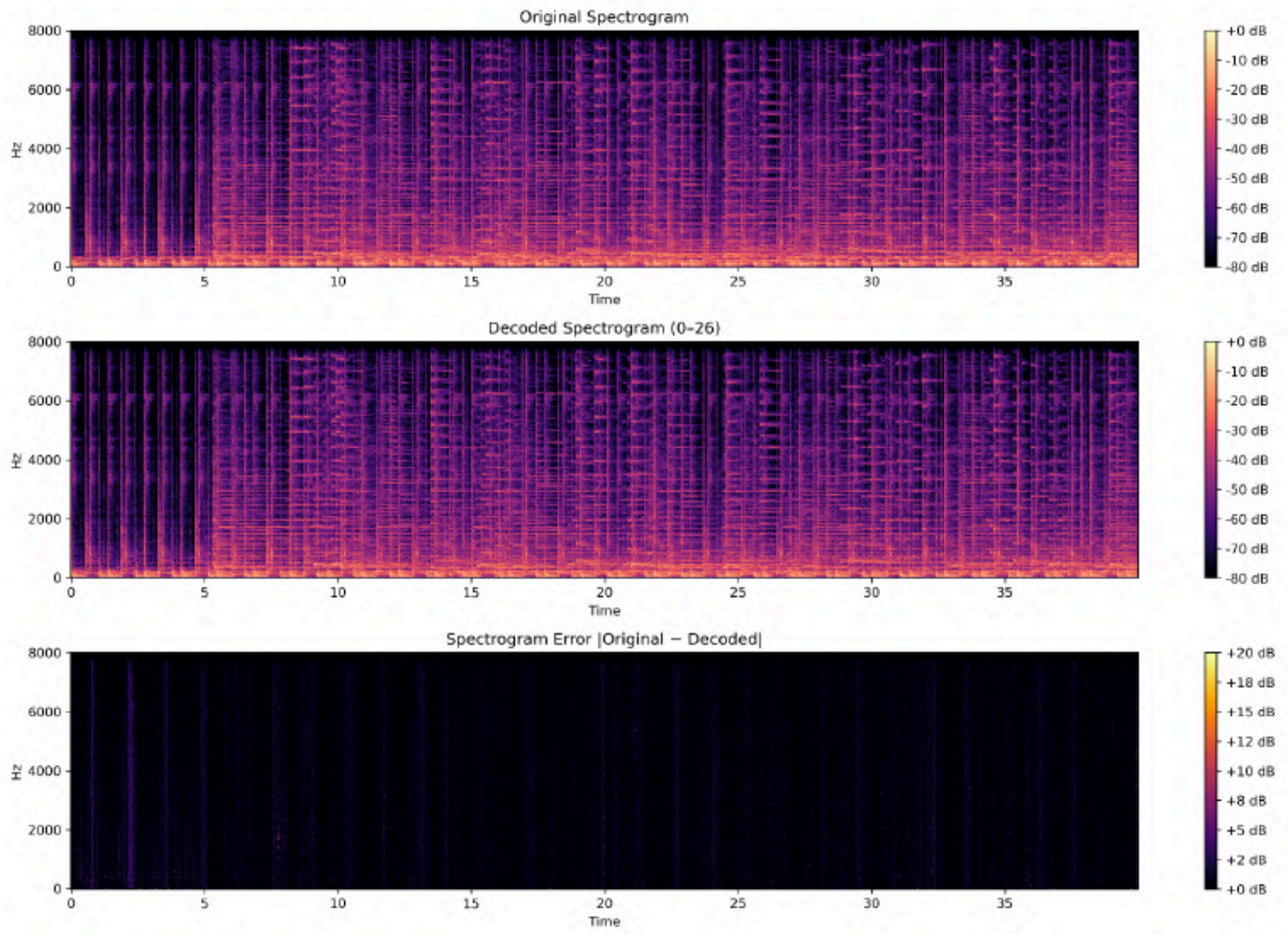

**(a)**

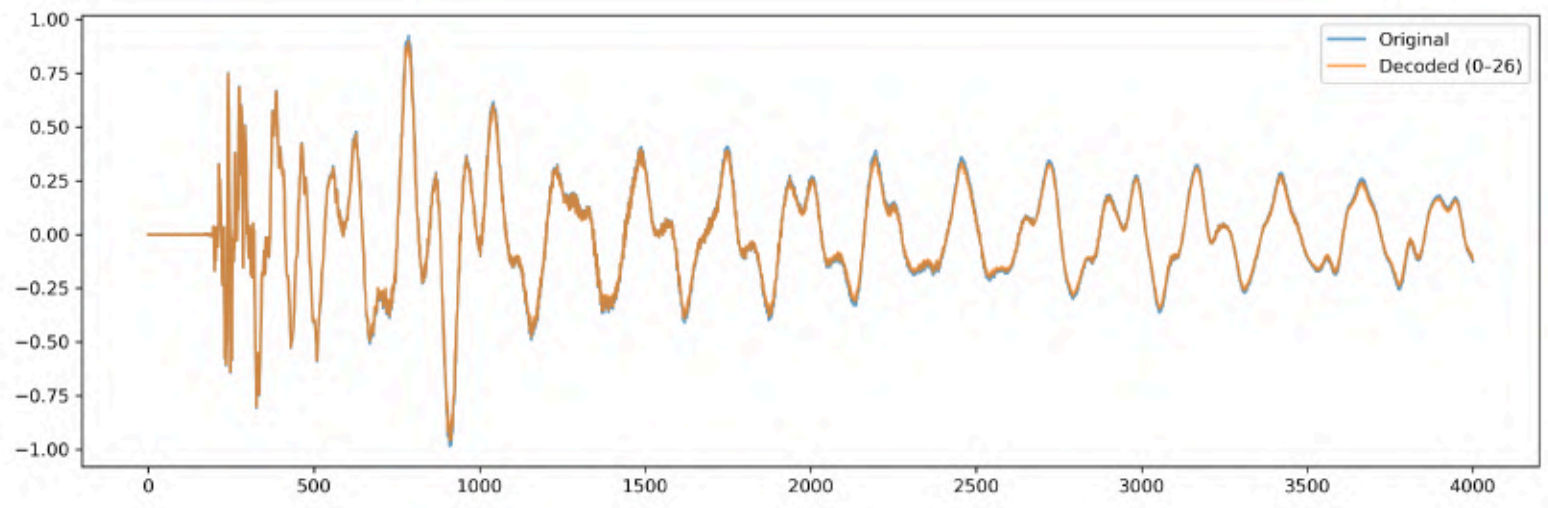

**(b)**

**Fig 22: Audio-8. (a) Spectrograms and (b) Waveforms**

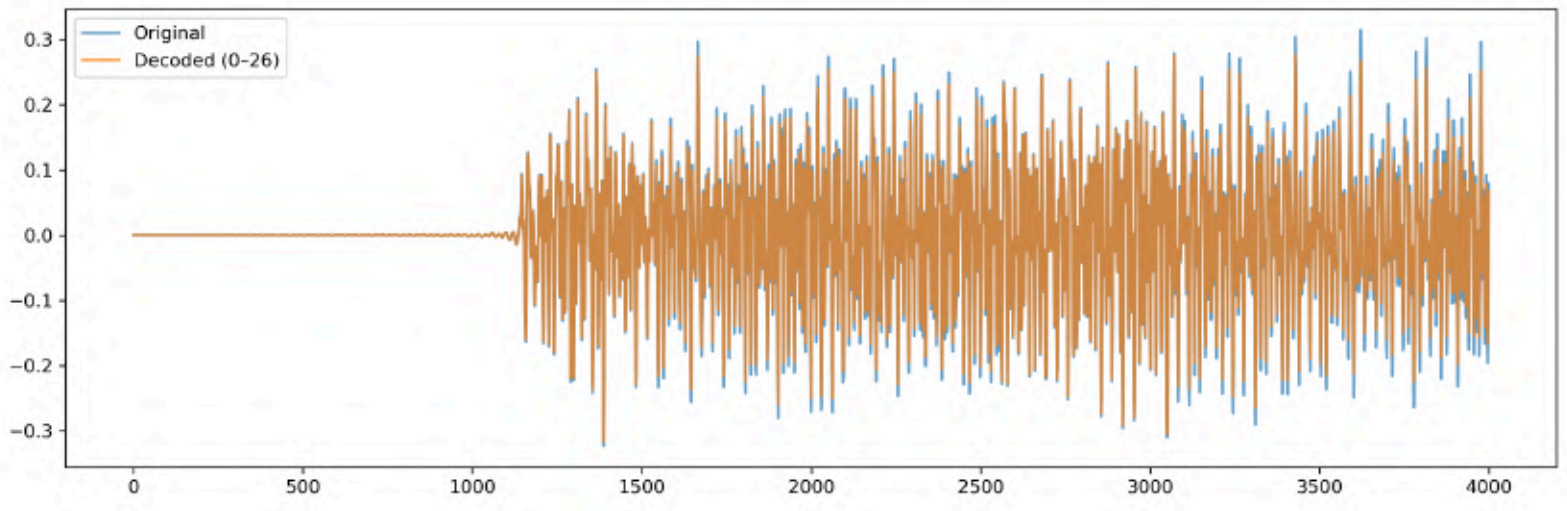

**(a)**

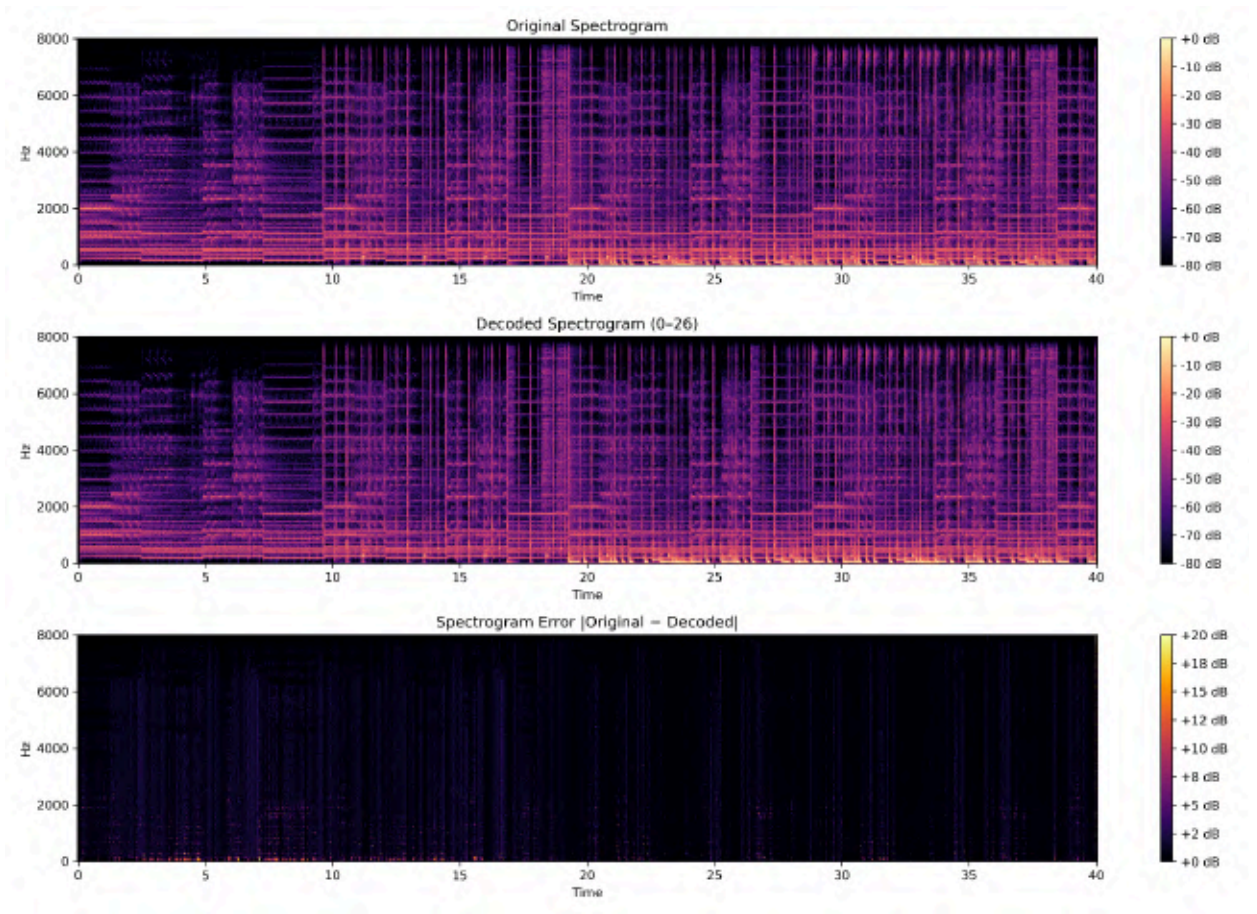

**(b)**

**Fig 23: Audio-9. (a) Spectrograms and (b) Waveforms**

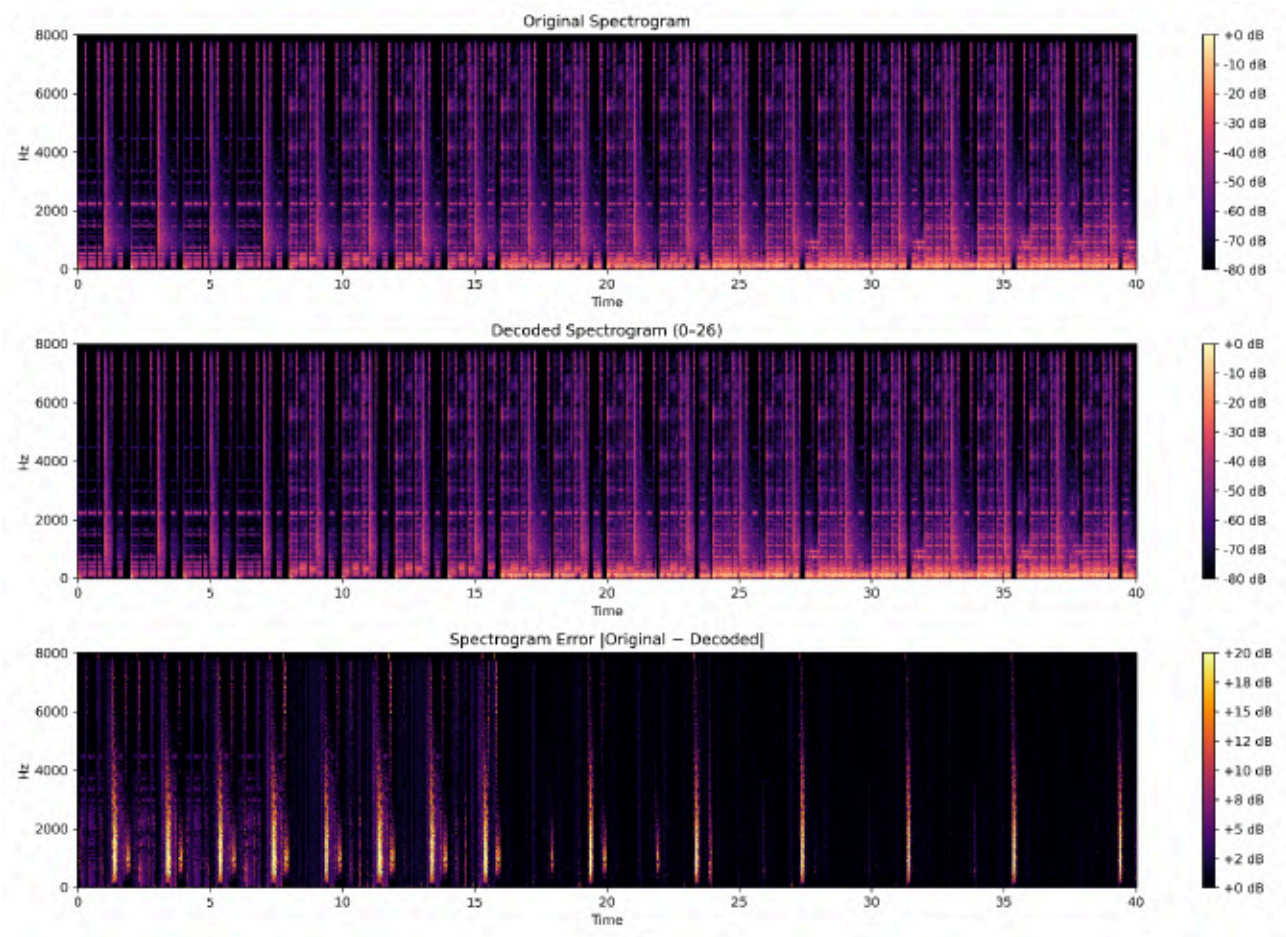

**(a)**

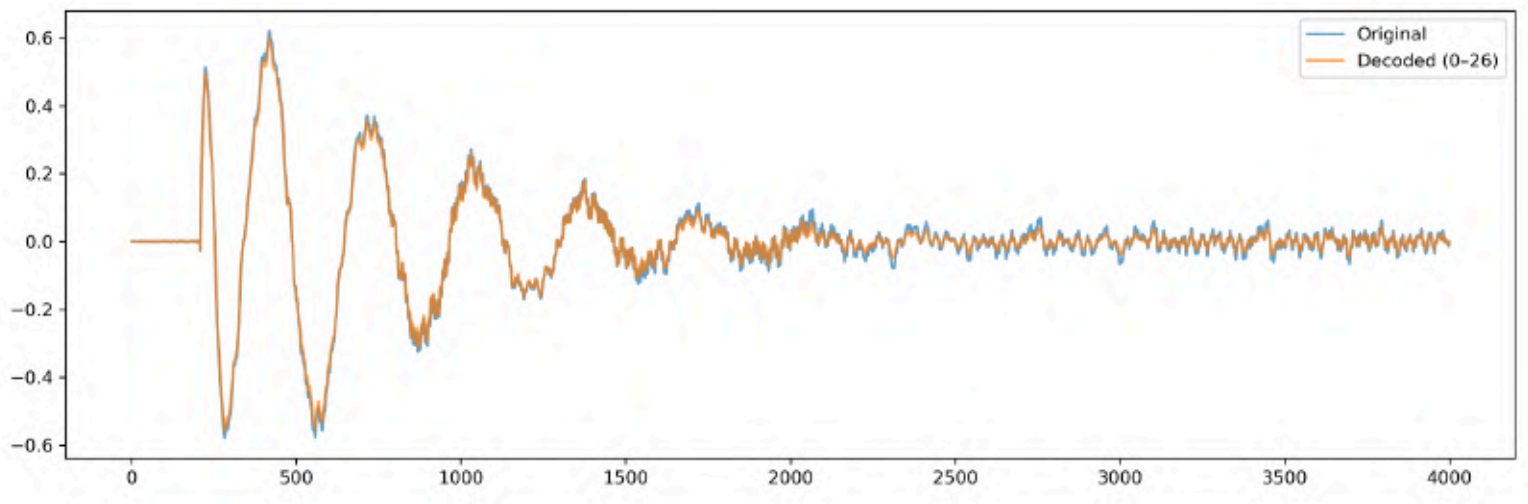

**(b)**

**Fig 24: Audio-10. (a) Spectrograms and (b) Waveforms**

**A.2. IMAGE TO TEXT (Additional Results):**

Appendix A.2, shows the additional experimentations in image to text steganographic pipeline.

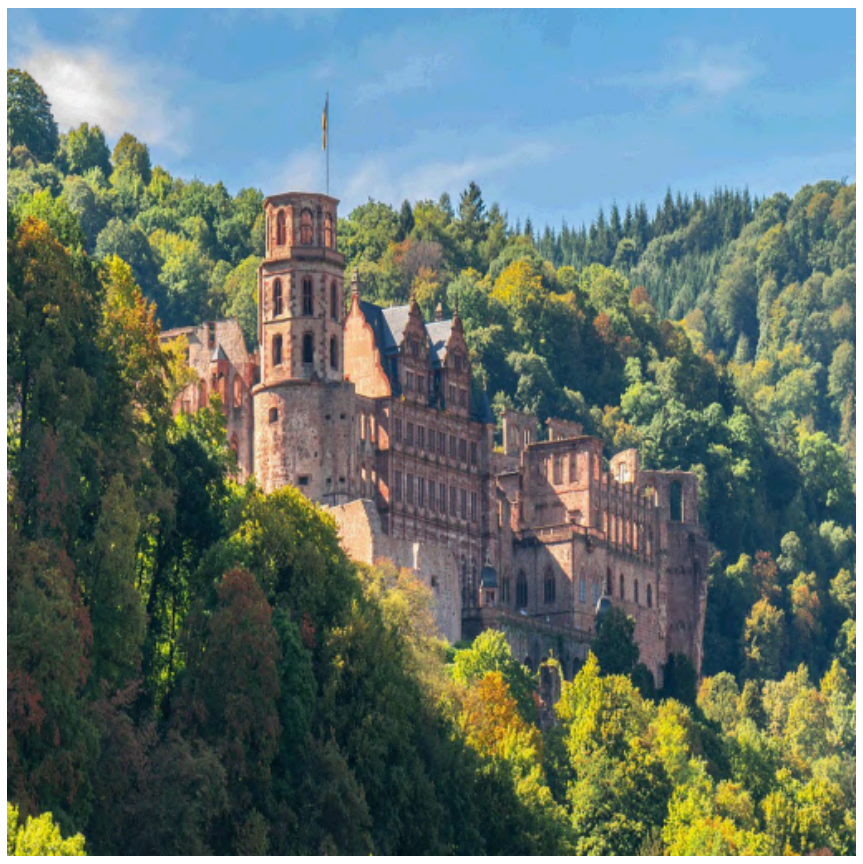

**Original Image**

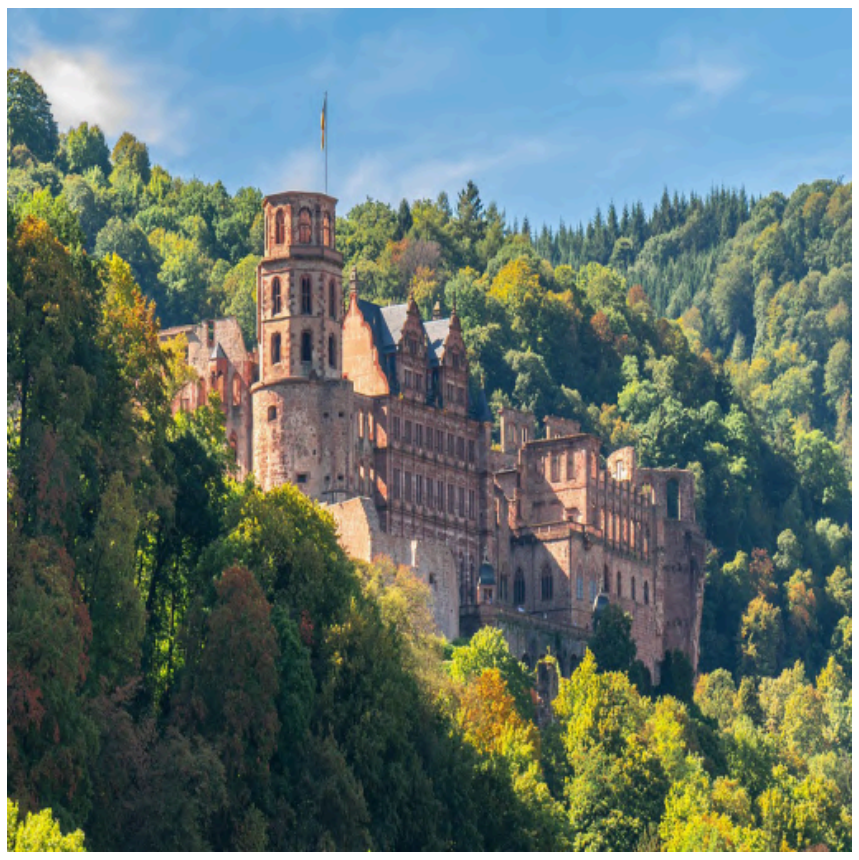

**Decoded Image**

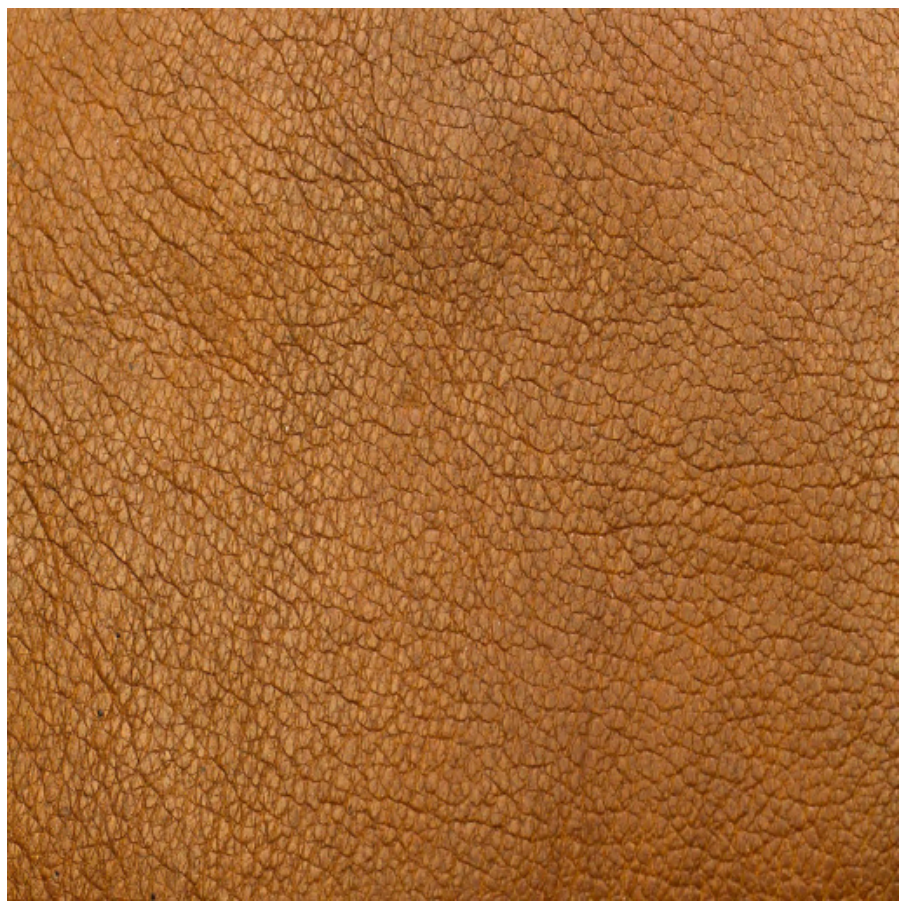

**Original Image**

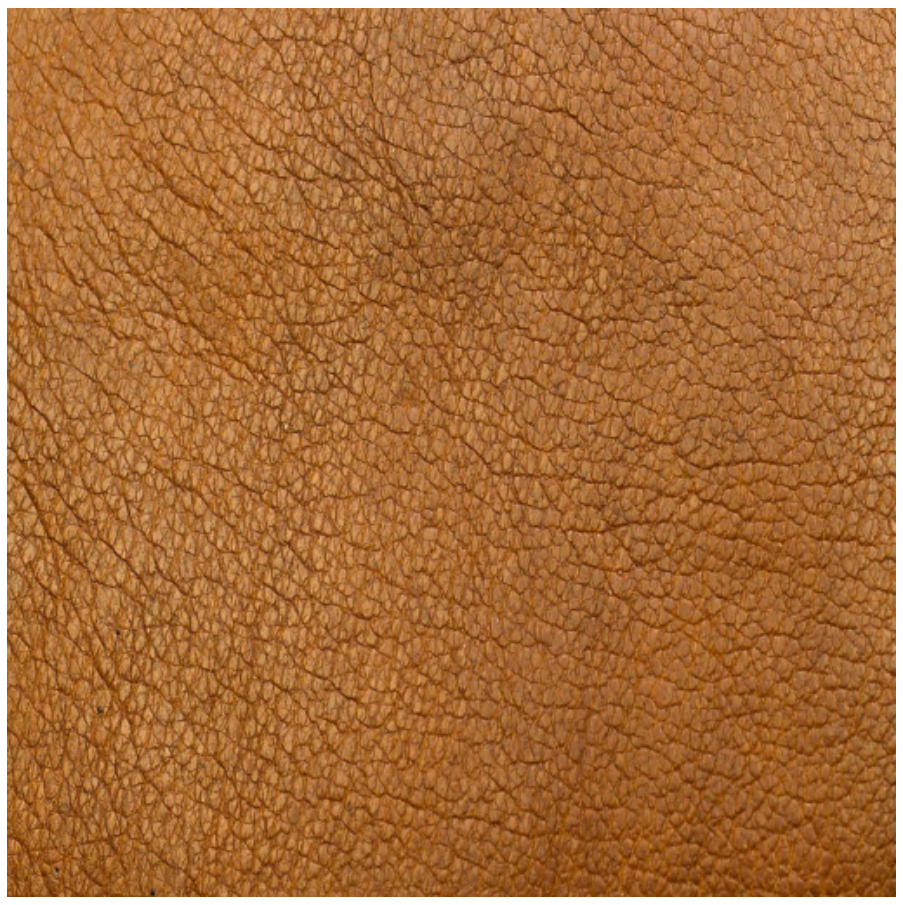

**Decoded Image**

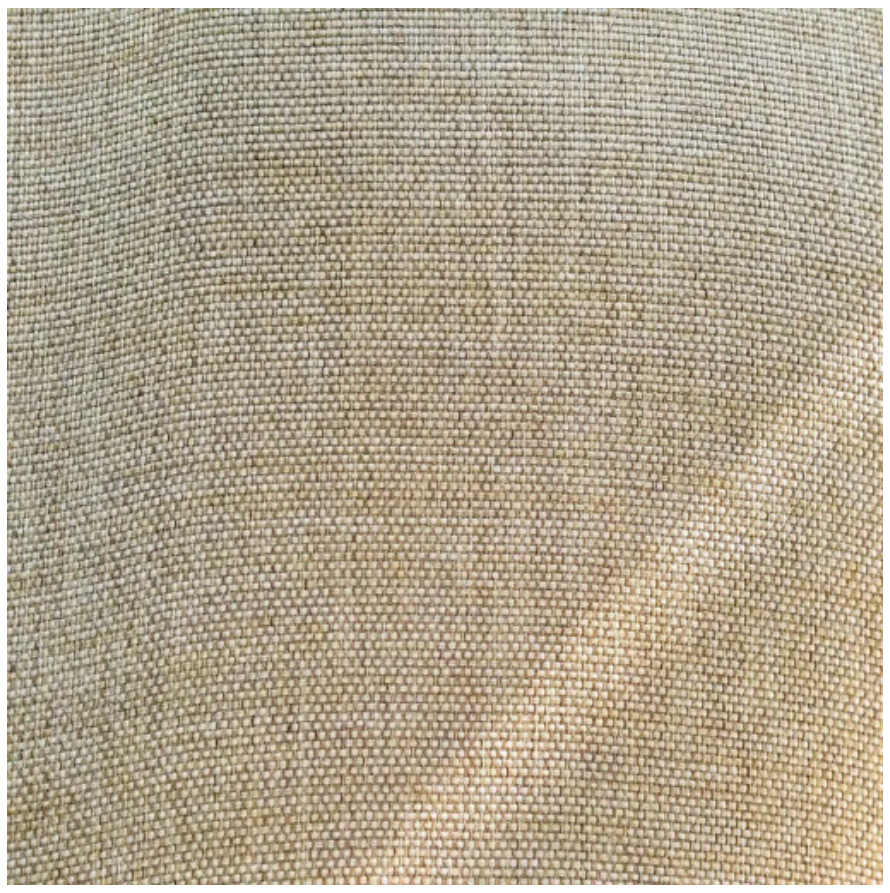

**Original Image**

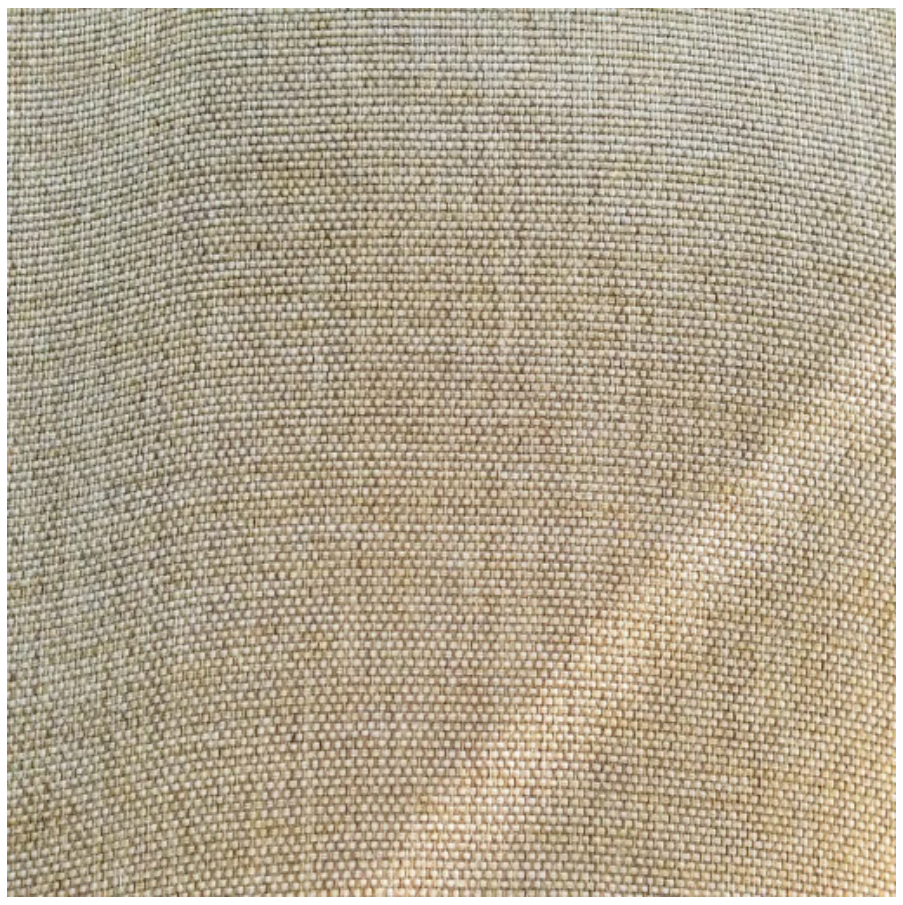

**Decoded Image**

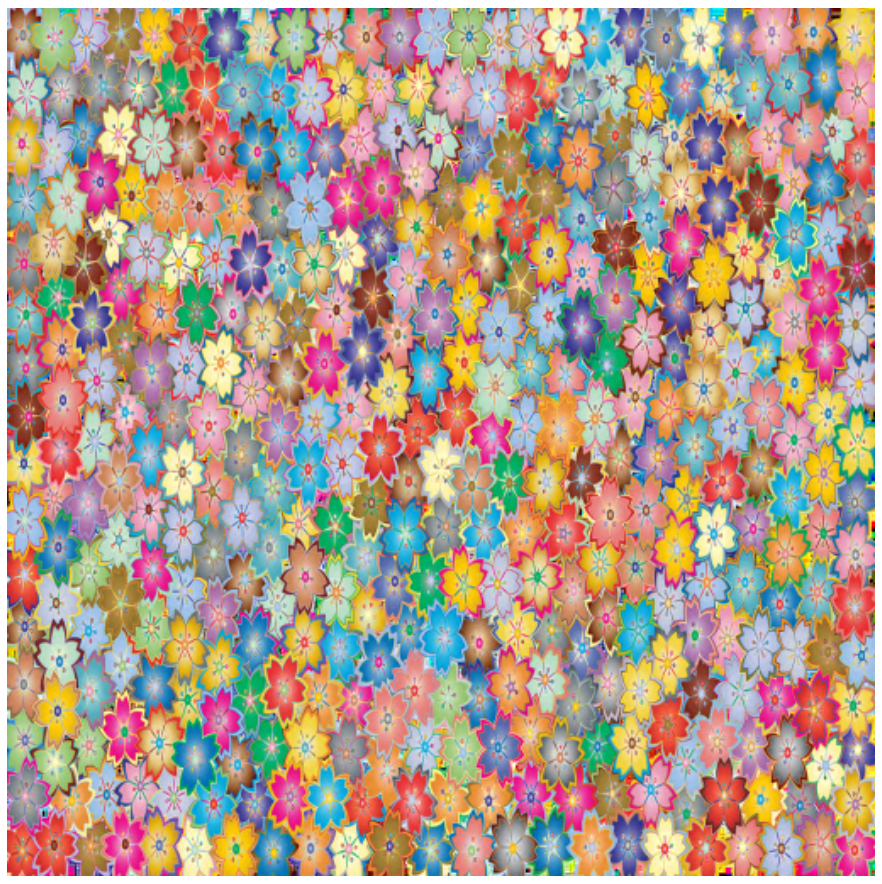

**Original Image**

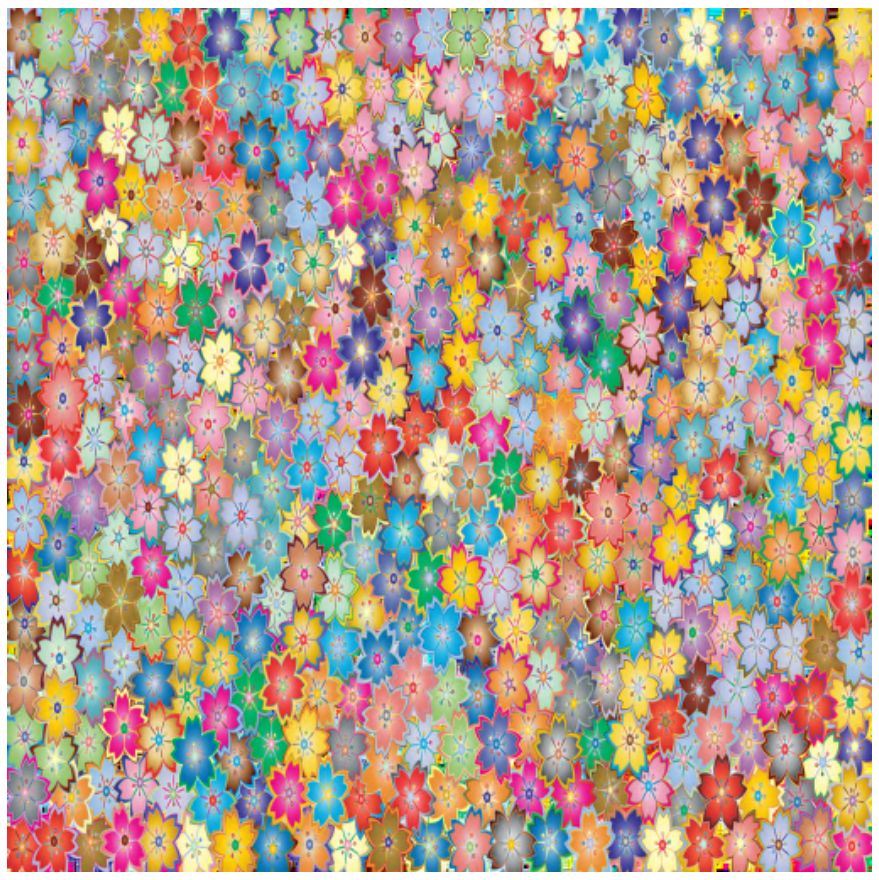

**Decoded Image**

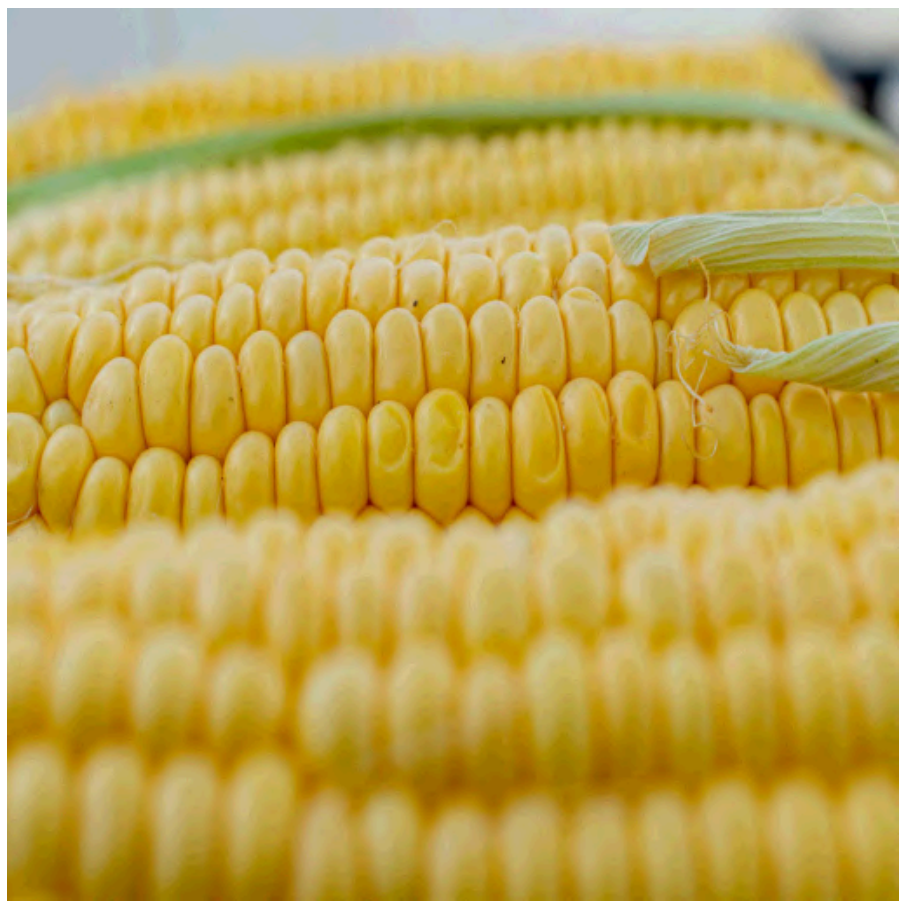

**Original Image**

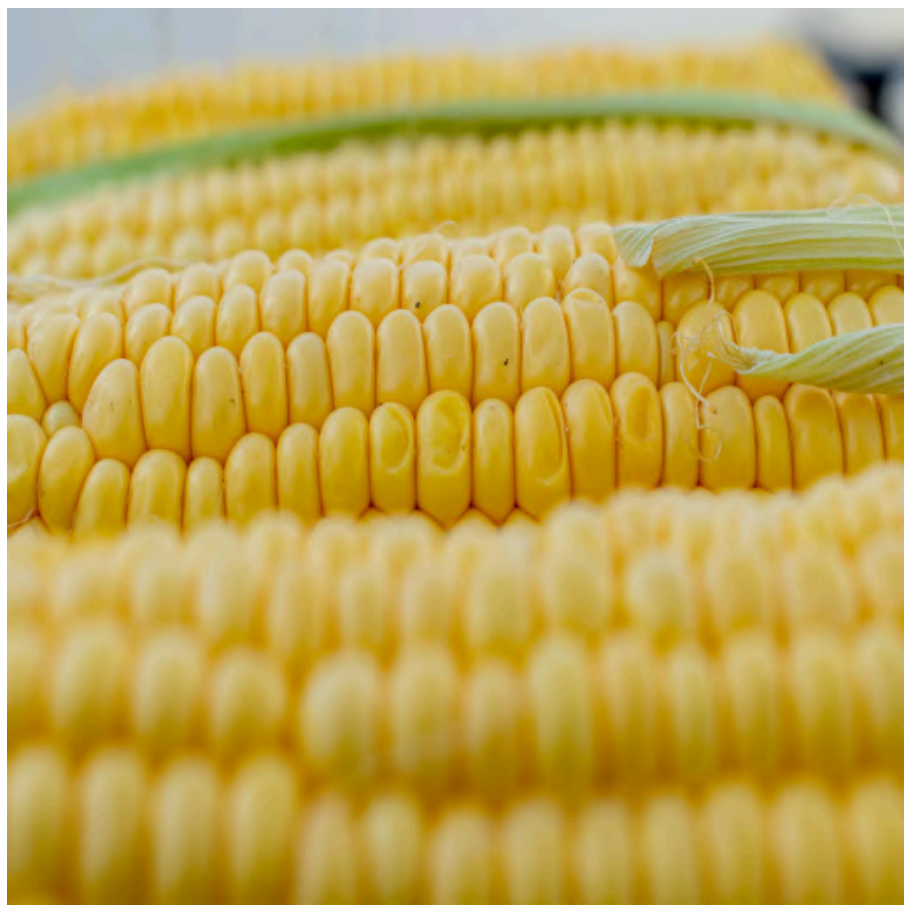

**Decoded Image**

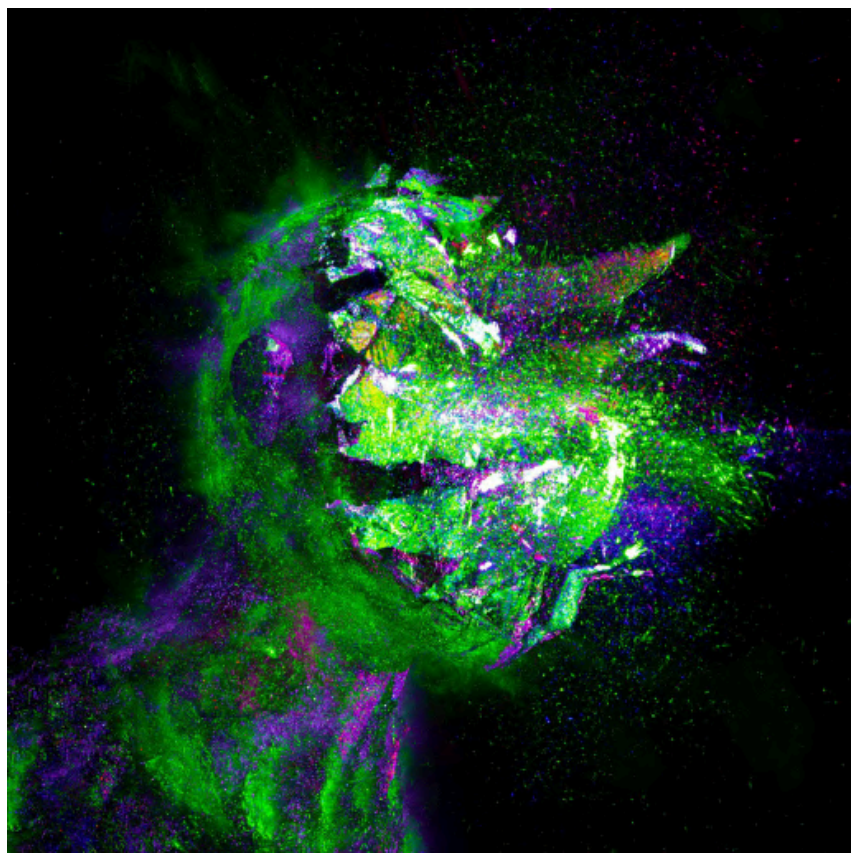

**Original Image**

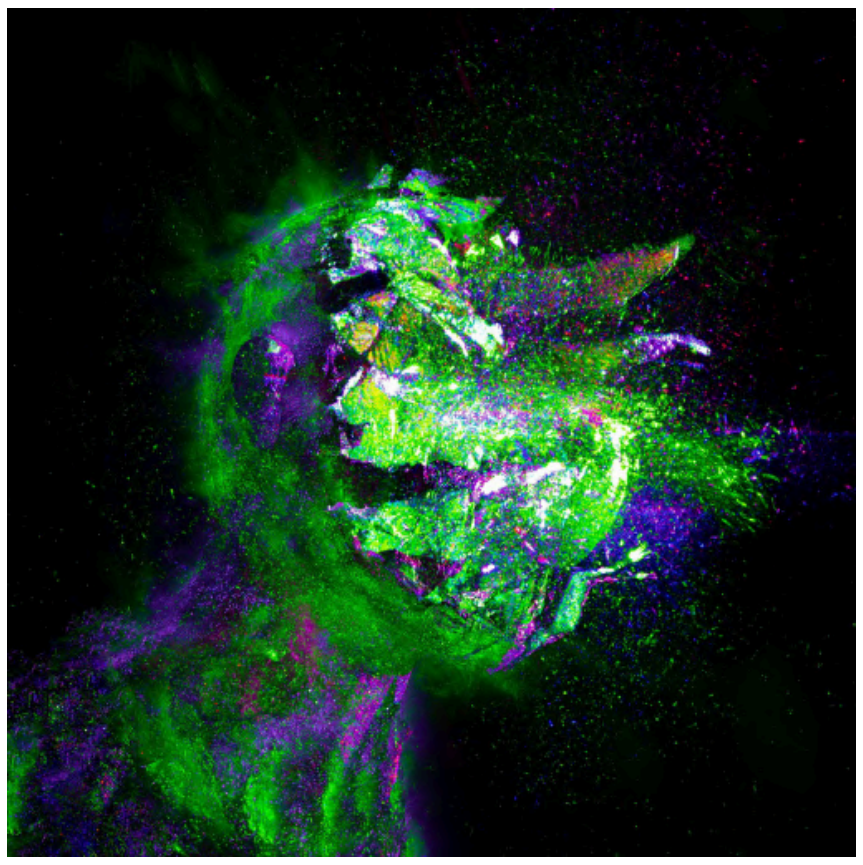

**Decoded Image**

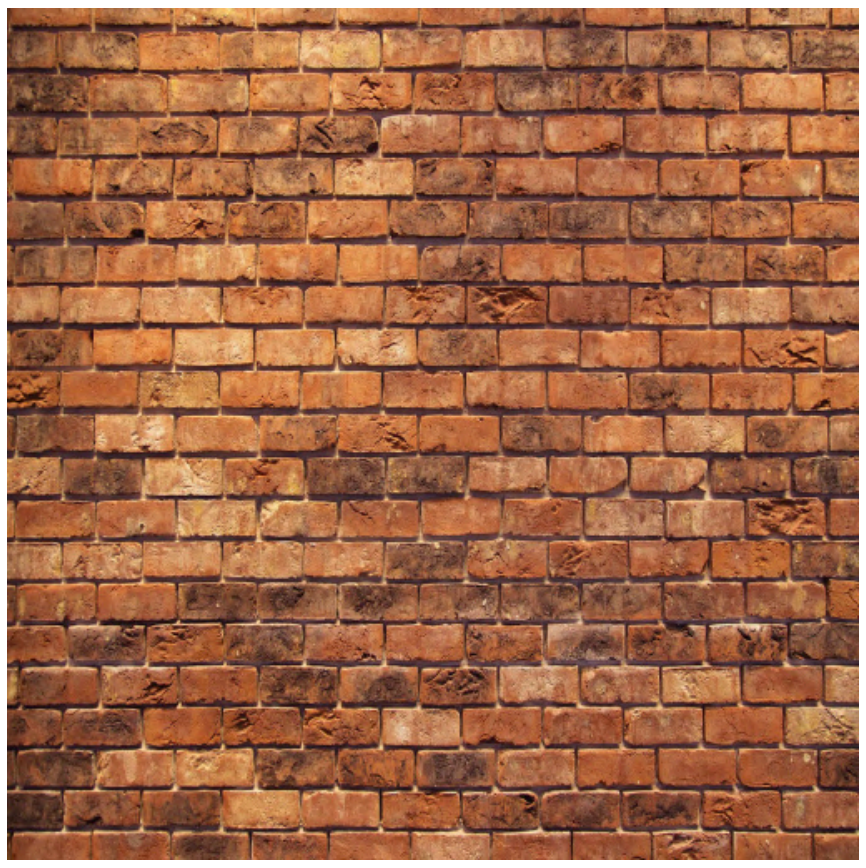

**Original Image**

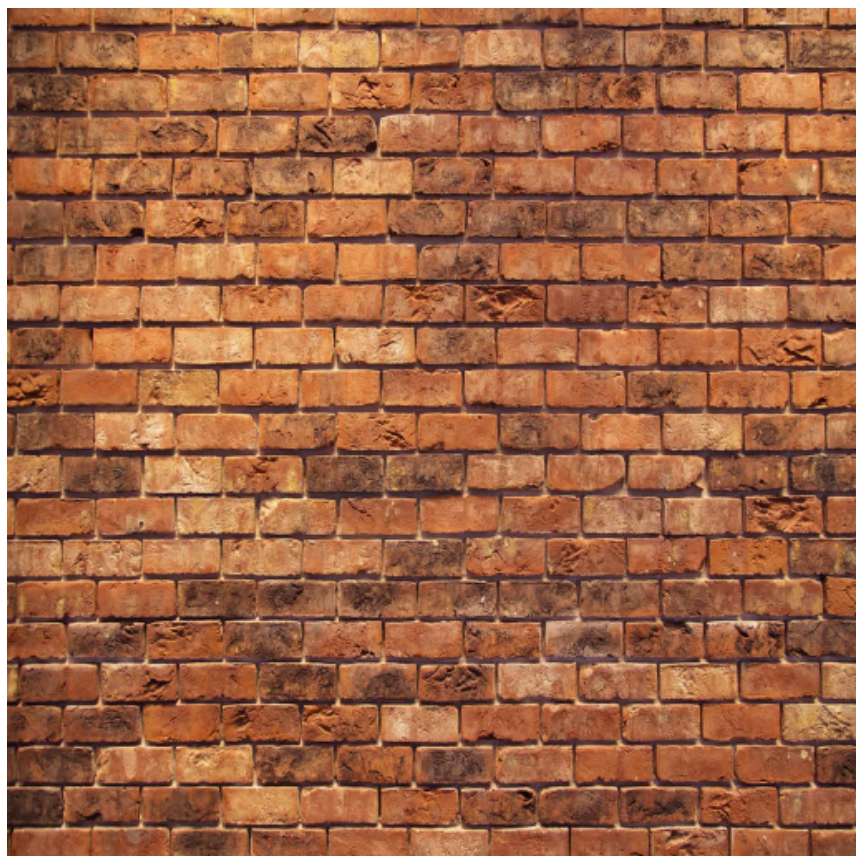

**Decoded Image**

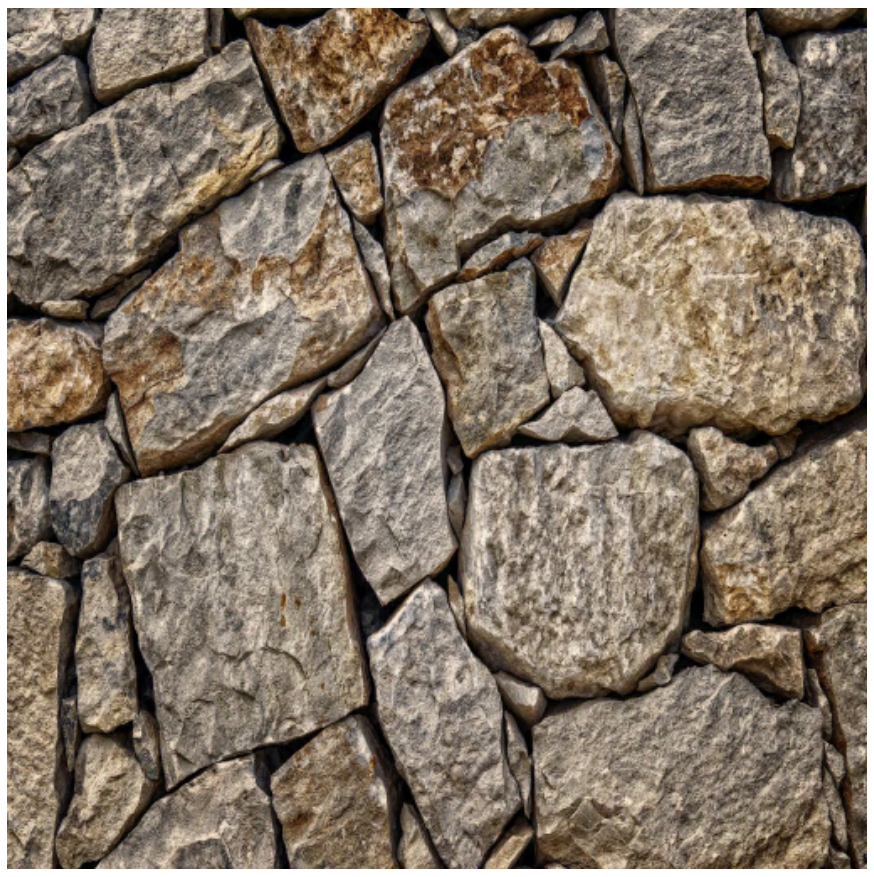

**Original Image**

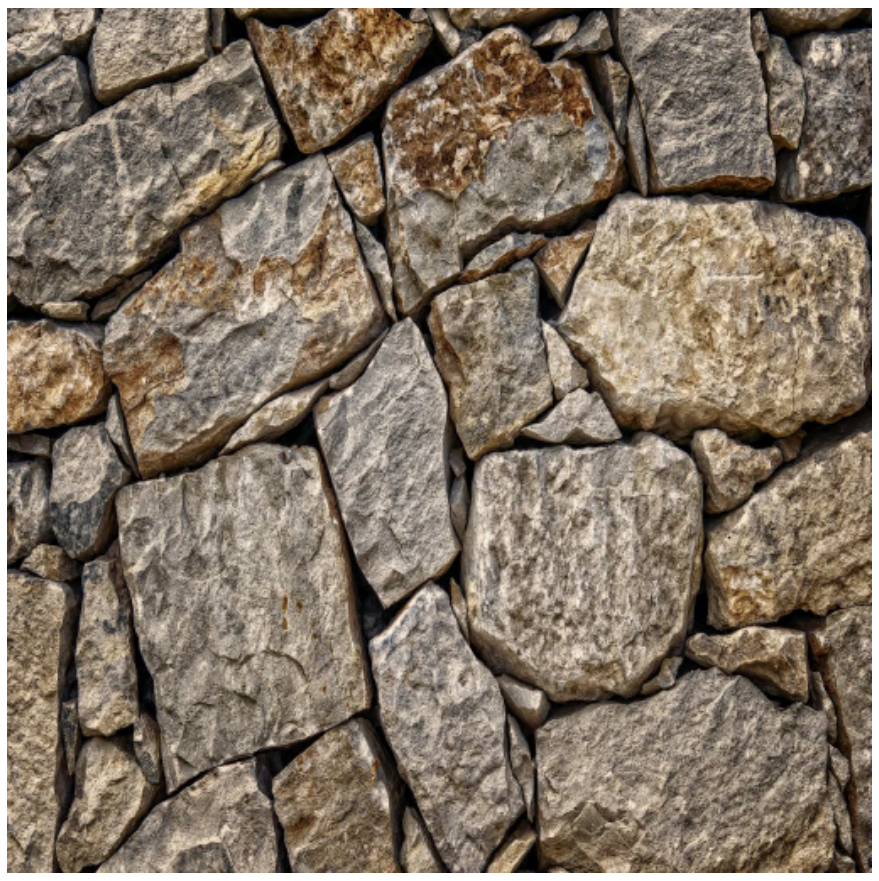

**Decoded Image**

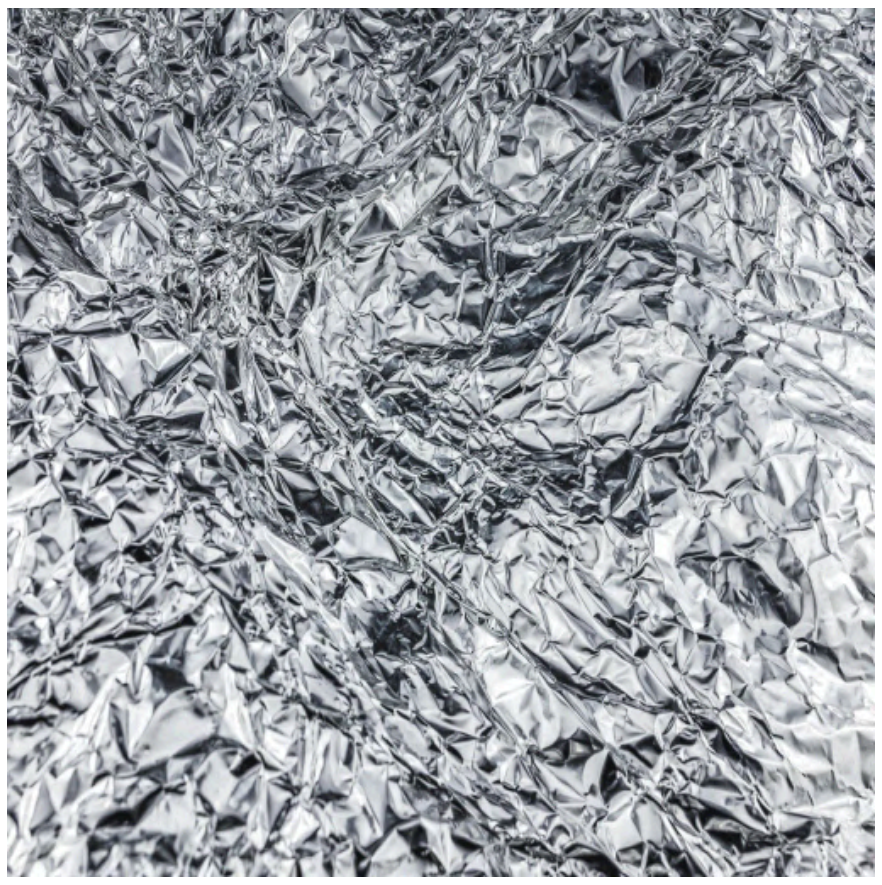

**Original Image**

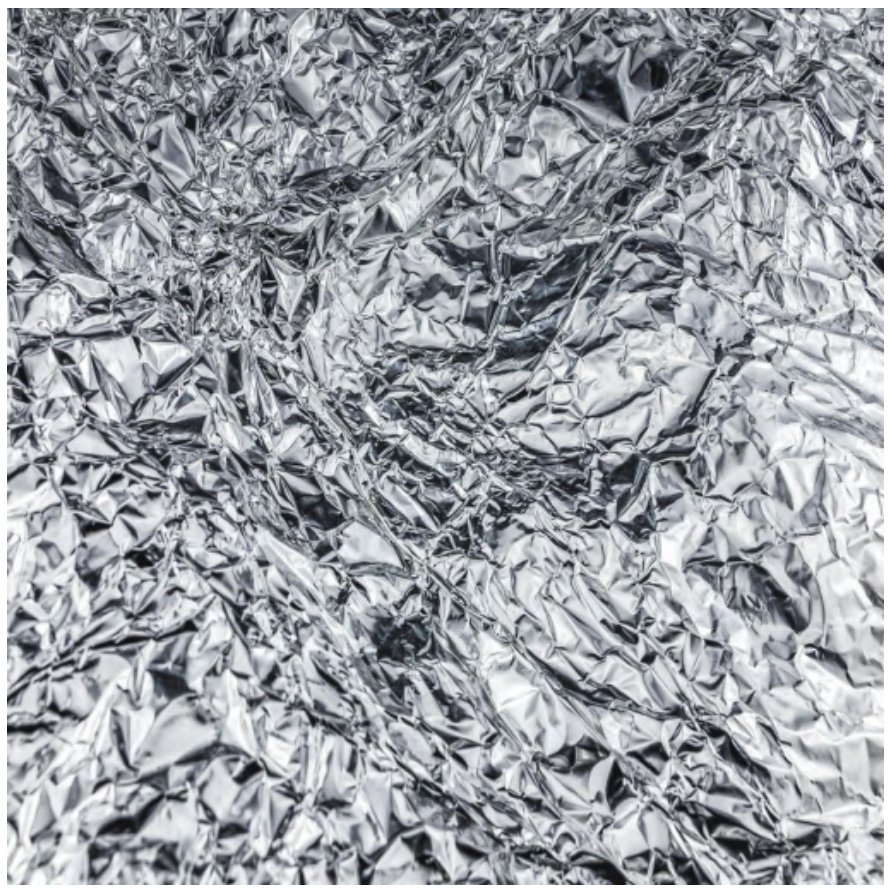

**Decoded Image**

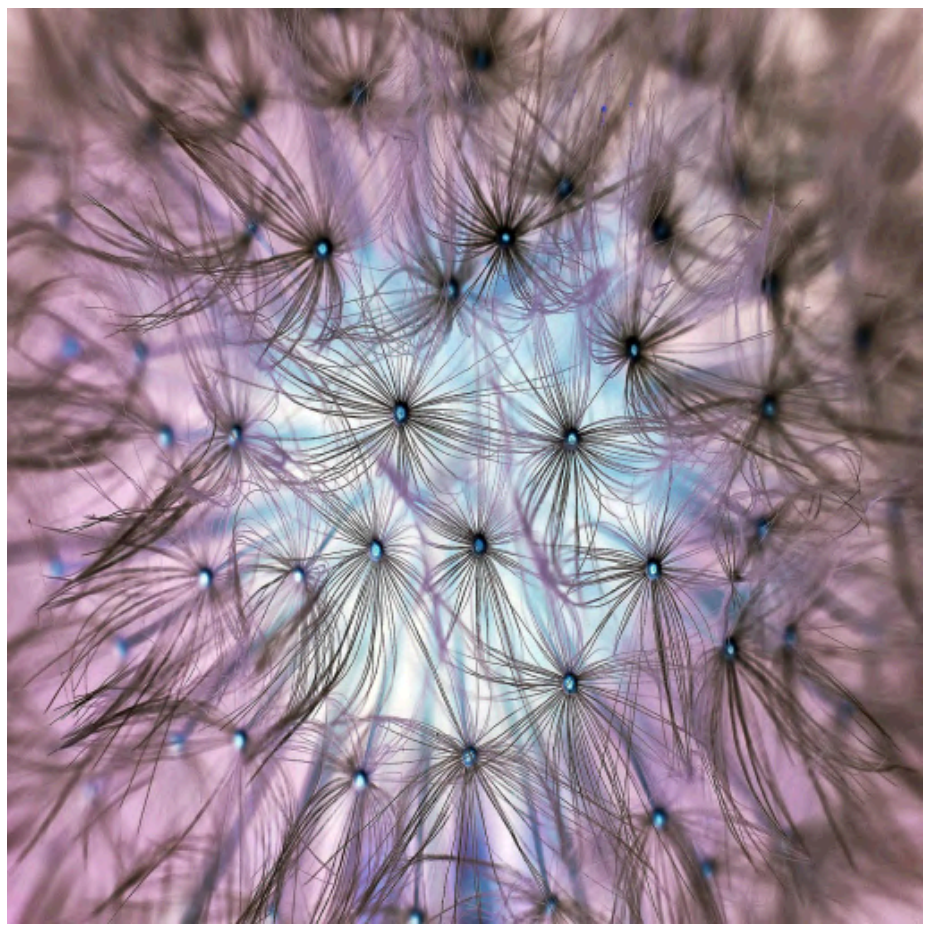

Original Image

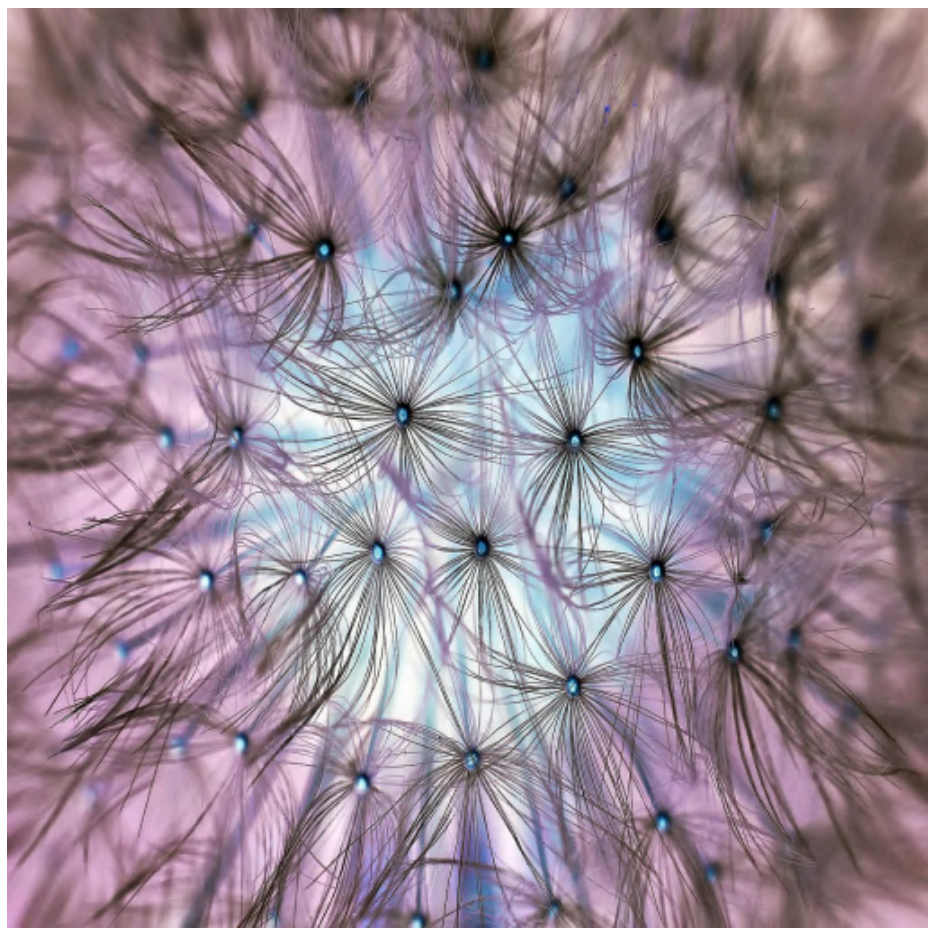

Decoded Image

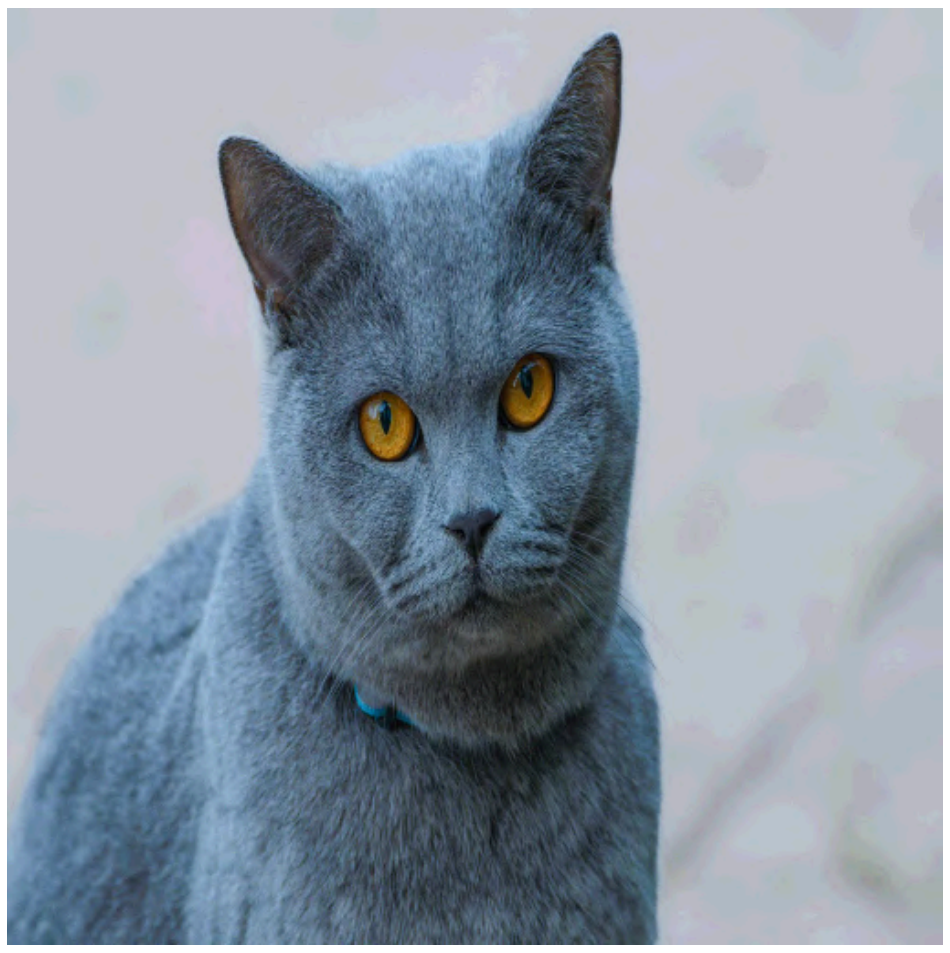

Original Image

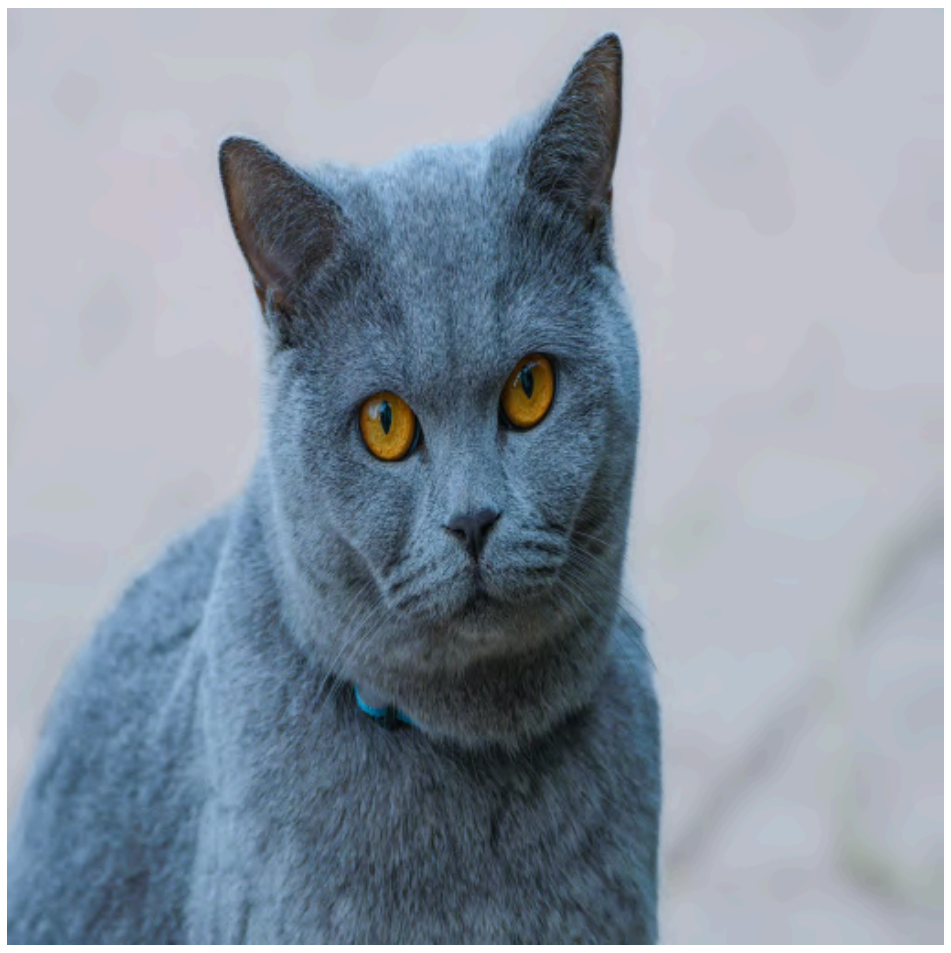

Decoded Image

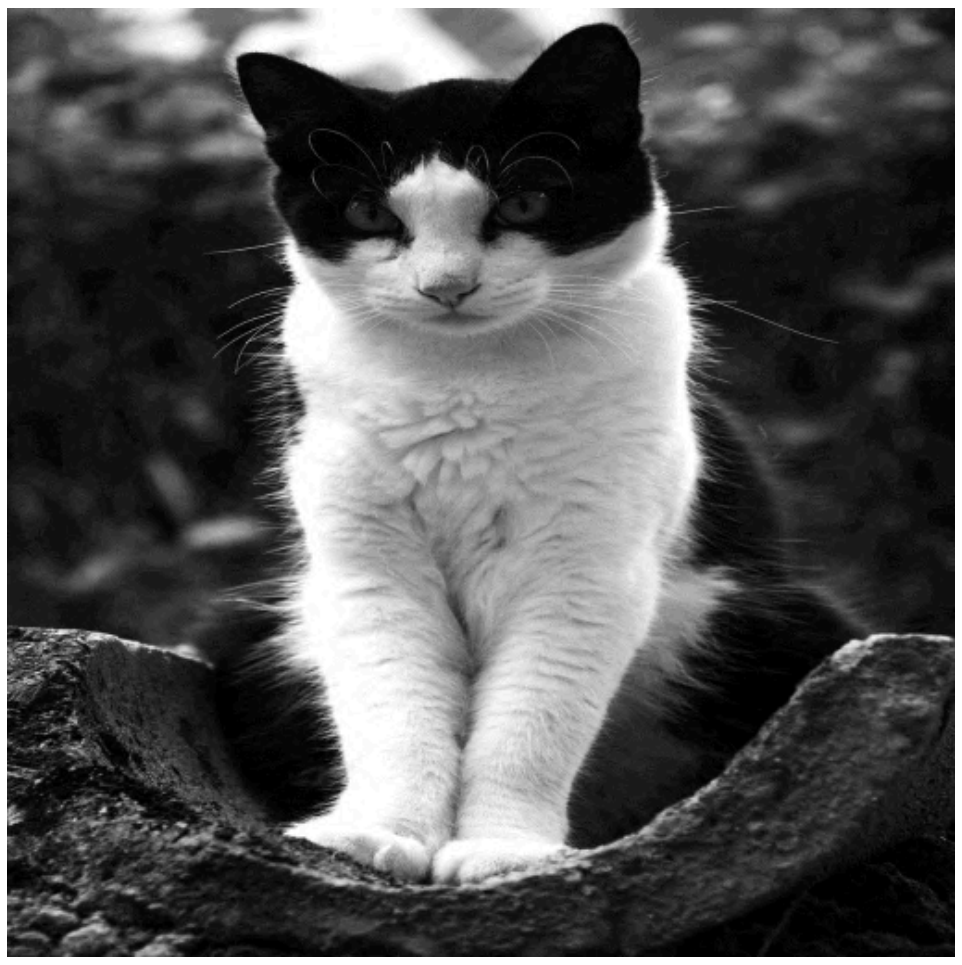

Original Image

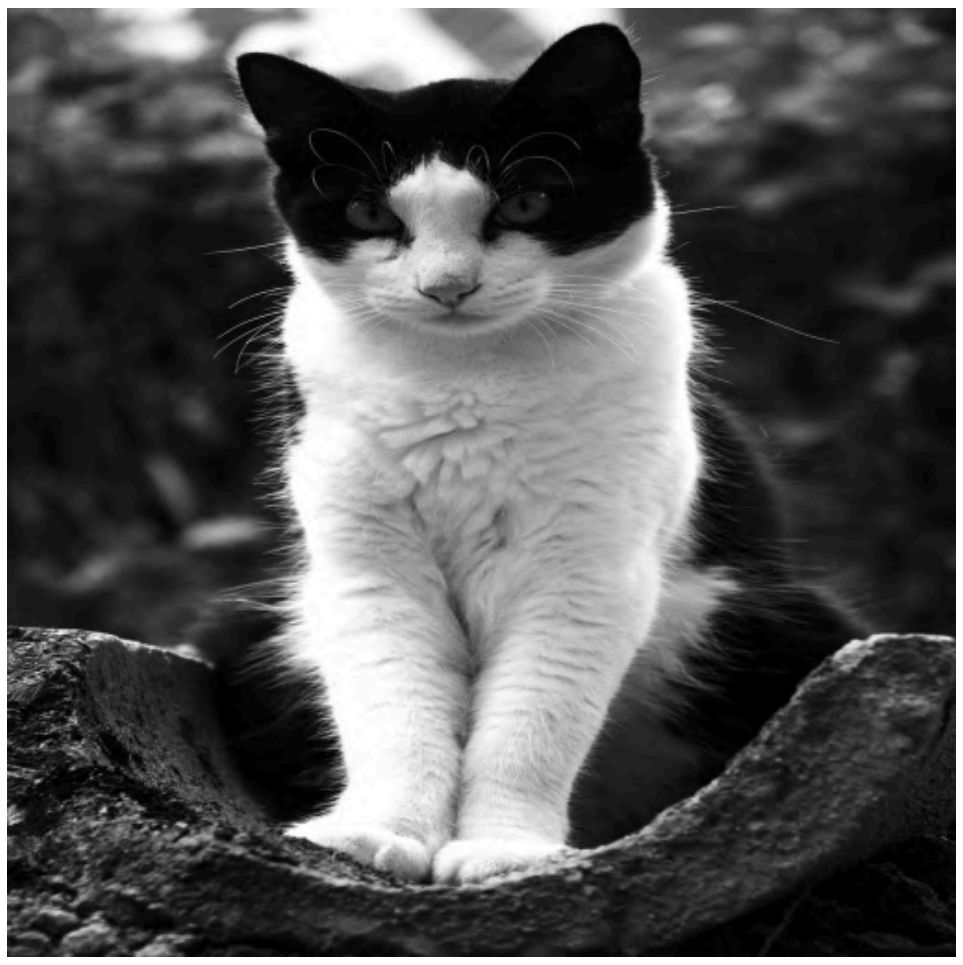

Decoded Image

**A.3. VIDEO TO TEXT (Additional Results):**

Appendix A.3, shows the additional experimentations in video to text steganographic pipeline.

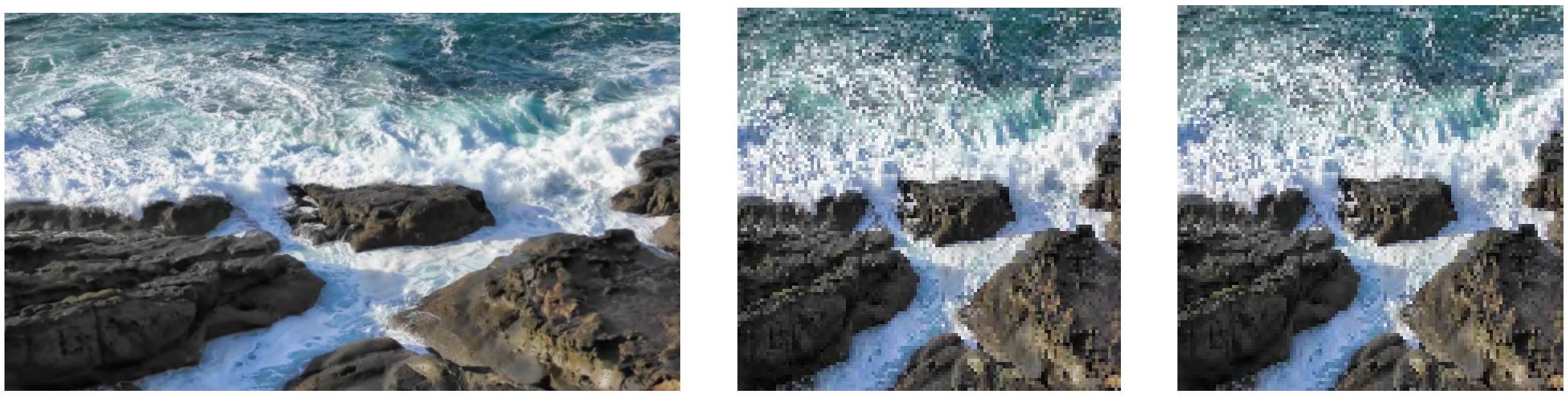

**Original Frame(0) of Video-6 (Resolution 1920×1080, 30FPS), Canonical & Decoded Frame (0) (Resolution 120×120, 30FPS)**

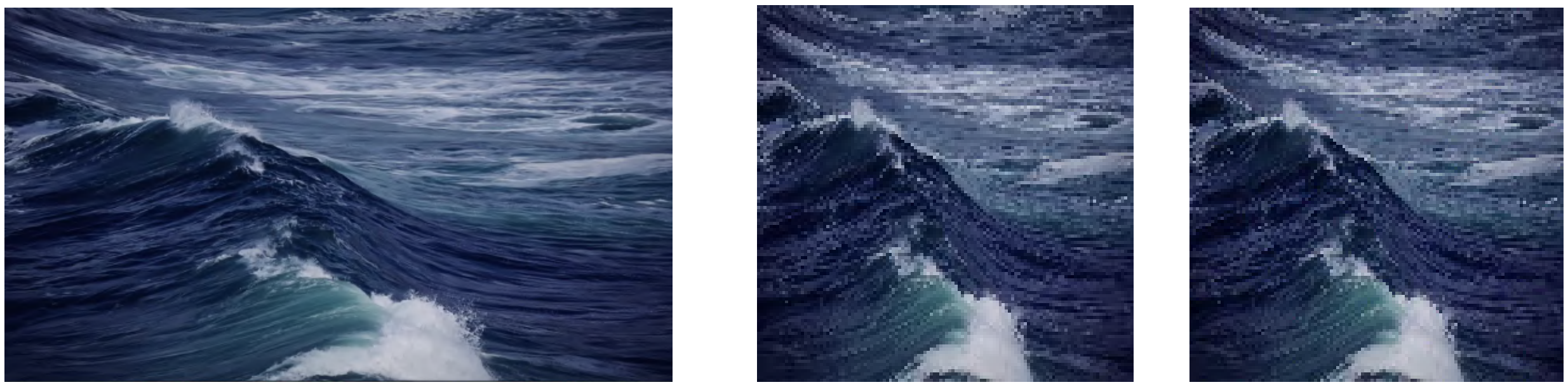

**Original Frame(0) of Video-7 (Resolution 1920×1080, 30FPS), Canonical & Decoded Frame (0) (Resolution 120×120, 30FPS)**

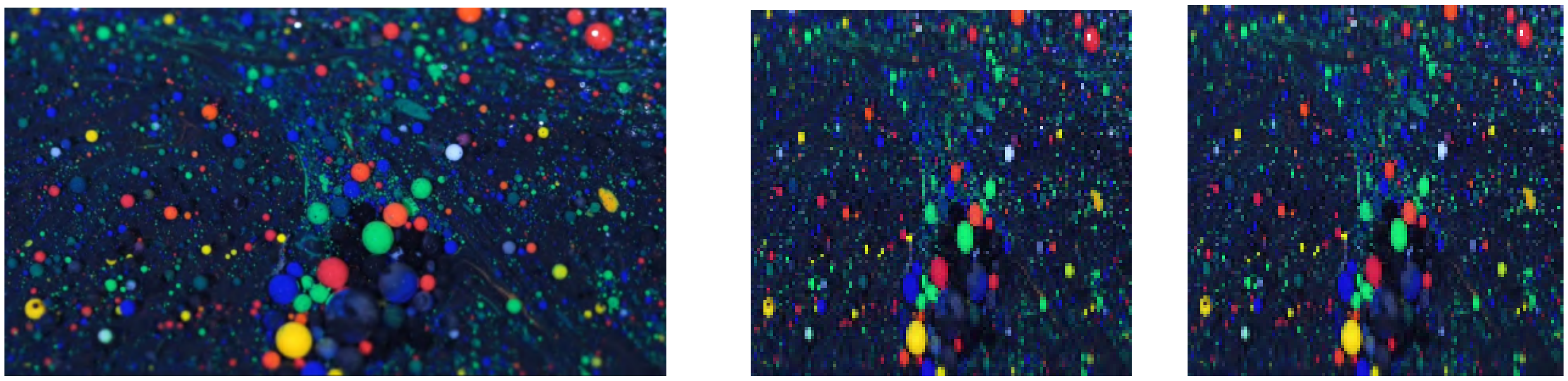

**Original Frame(0) of Video-8 (Resolution 1920×1080, 30FPS), Canonical & Decoded Frame (0) (Resolution 120×120, 30FPS)**

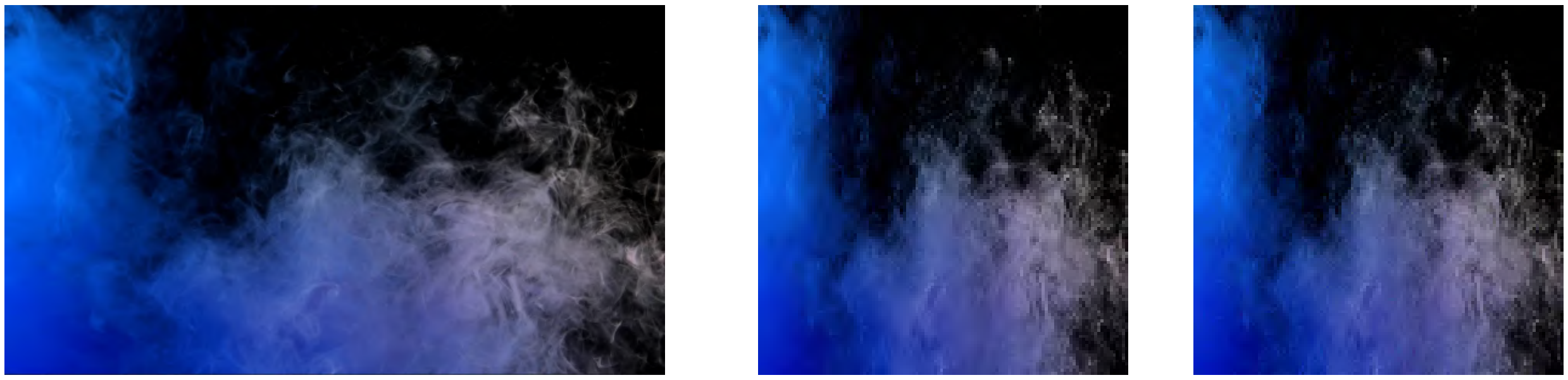

**Original Frame(0) of Video-9 (Resolution 1920×1080, 30FPS), Canonical & Decoded Frame (0) (Resolution 120×120, 30FPS)**

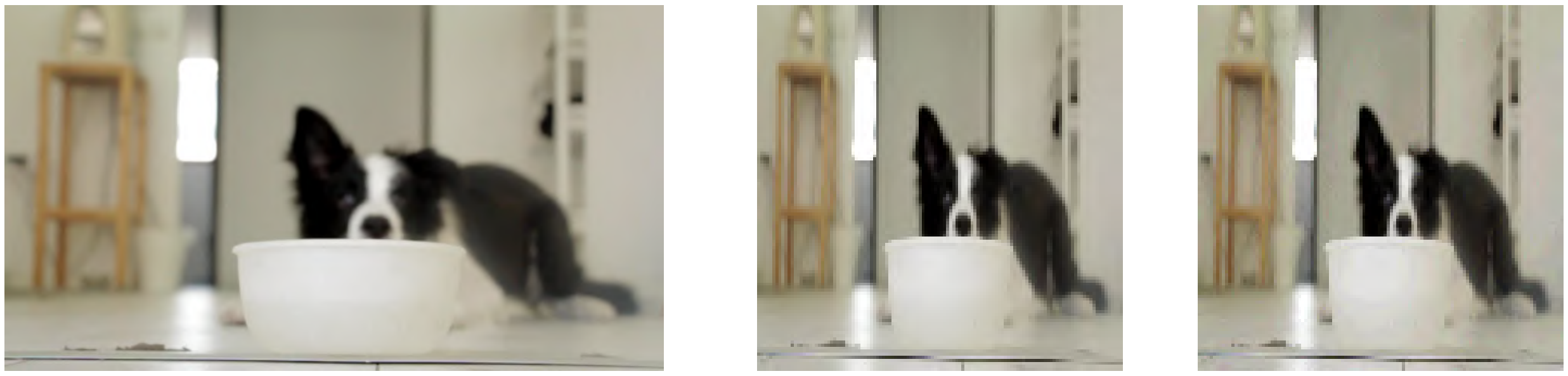

**Original Frame(0) of Video-10 (Resolution 1920×1080, 30FPS), Canonical & Decoded Frame (0) (Resolution 120×120, 30FPS)**

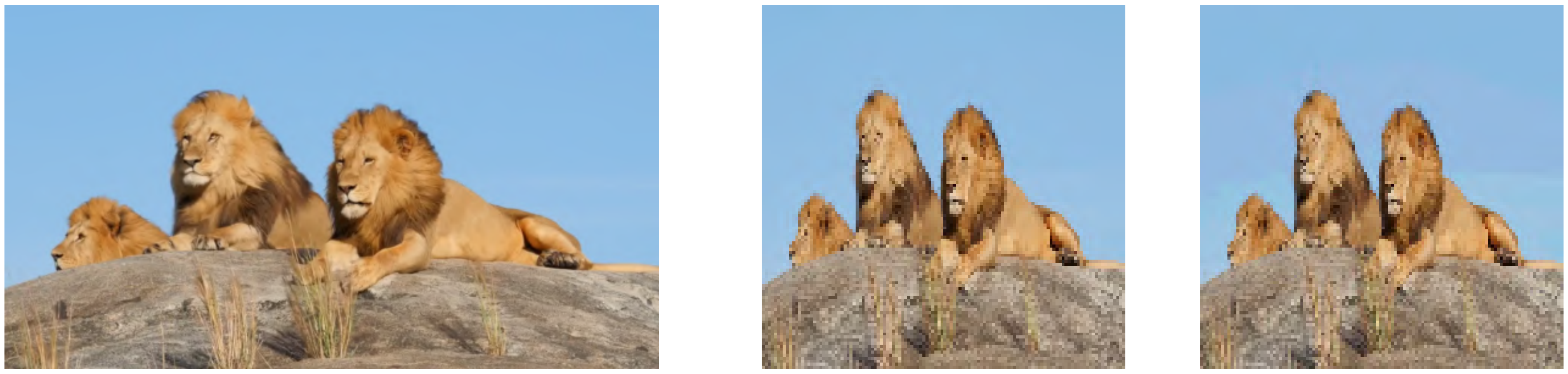

**Original Frame(0) of Video-11 (Resolution 1920×1080, 30FPS), Canonical & Decoded Frame (0) (Resolution 120×120, 30FPS)**

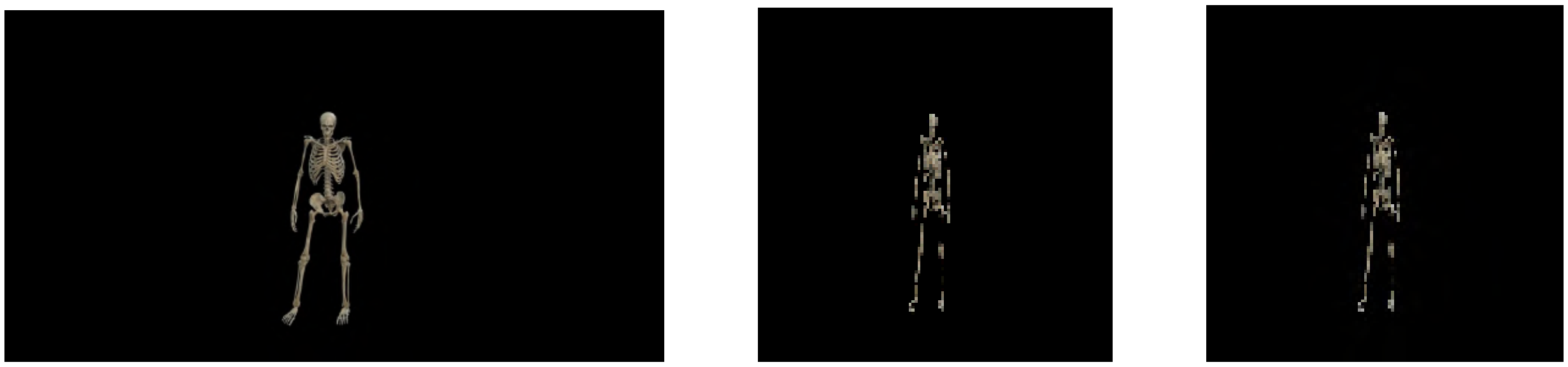

**Original Frame(0) of Video-12 (Resolution 1920×1080, 30FPS), Canonical & Decoded Frame (0) (Resolution 120×120, 30FPS)**

**A.4. TEXT TO TEXT (Additional Results):**

Appendix A.4, shows the additional experimentations in text to text steganographic pipeline.

T E X T D A T A

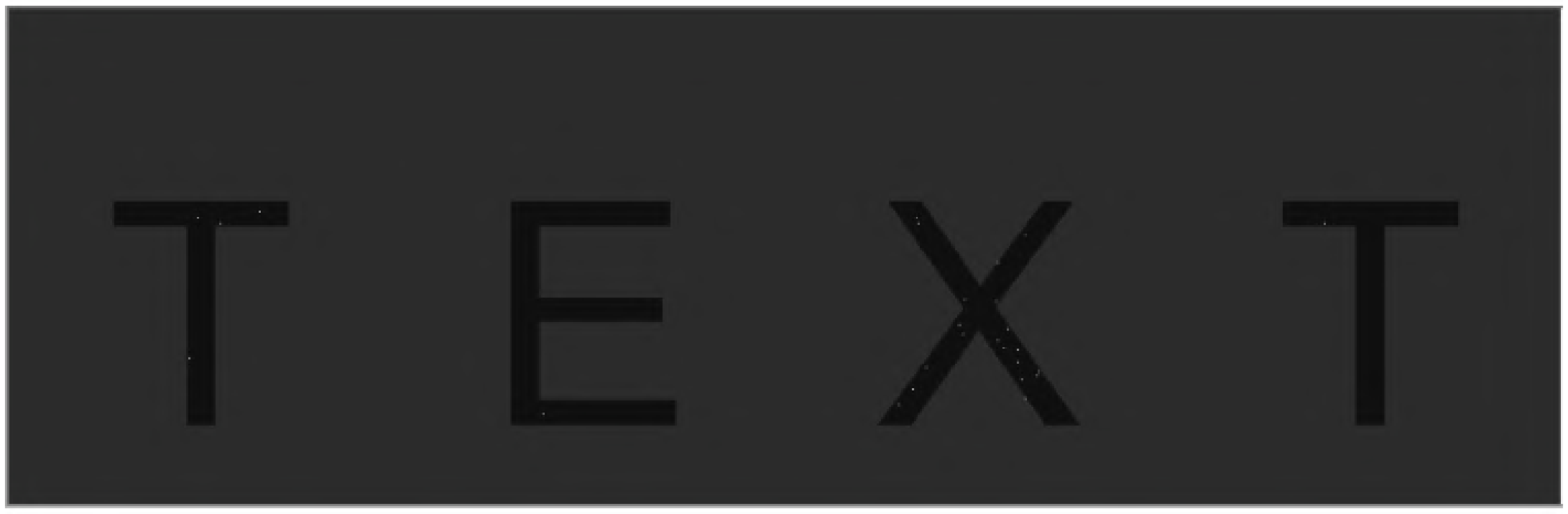

**Figure shows the perturbation of secret text (DATA) into cover text (TEXT).**

S E C R E T H A L L O W

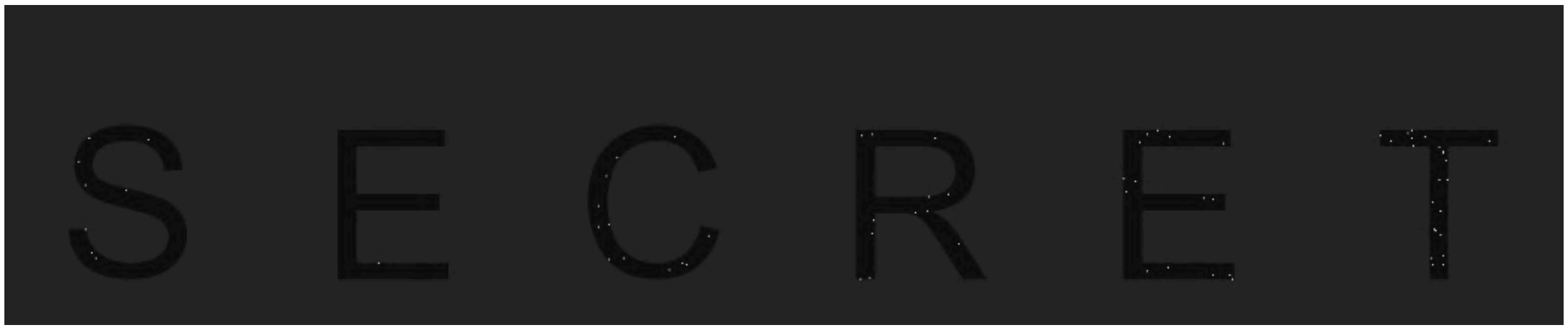

**Figure shows the perturbation of secret text (HALLOW) into cover text (SECRET).**

R E N D E R L E T T E R

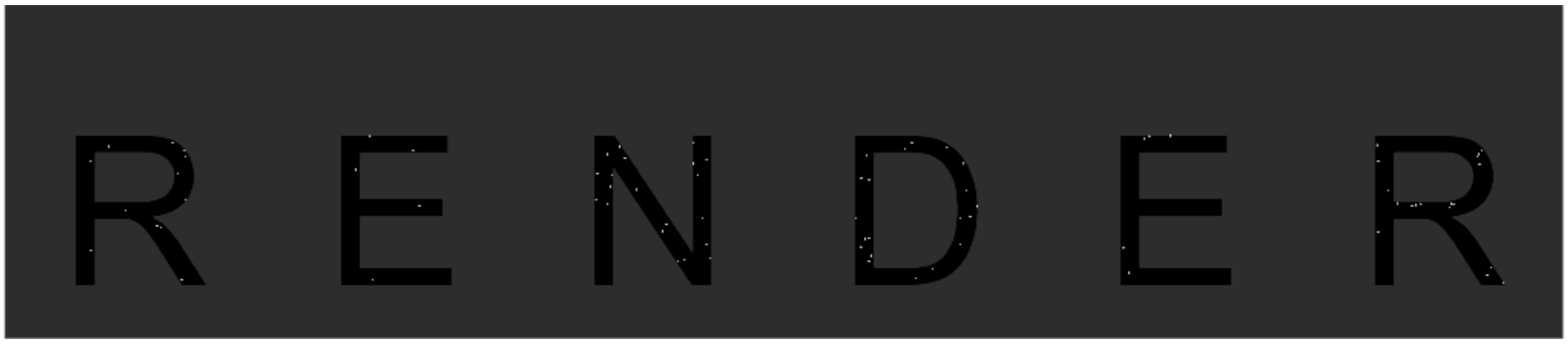

**Figure shows the perturbation of secret text (LETTER) into cover text (RENDER).**

H O R I Z O N C O N T A I N

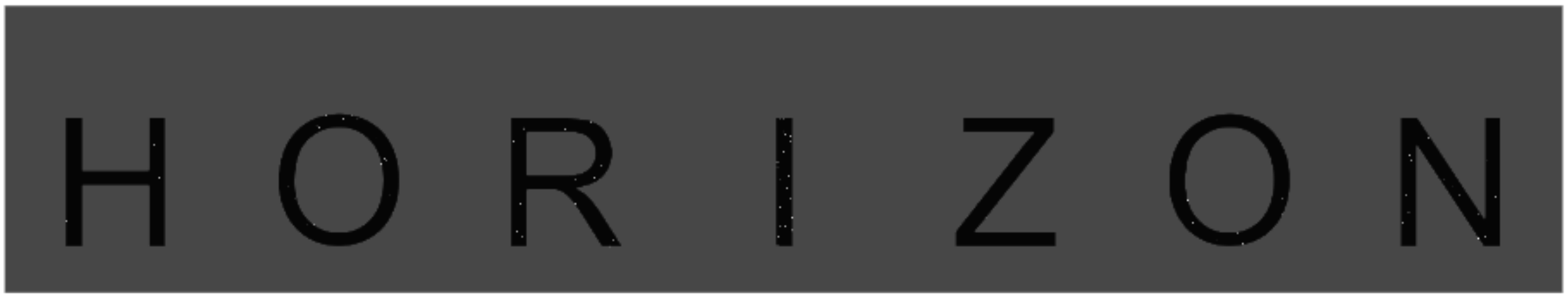

**Figure shows the perturbation of secret text (CONTAIN) into cover text (HORIZON).**

**APPENDIX B: IMAGE SOURCE (PIXABAY):**

Image Credits : Image by <a href="https://pixabay.com/users/chiemseherin-1425977/?utm_source=link-attribution&utm_medium=referral&utm_campaign=image&utm_content=3683860">🌼Christel🌼</a> from <a href="https://pixabay.com//?utm_source=link-attribution&utm_medium=referral&utm_campaign=image&utm_content=3683860">Pixabay</a>

Image by <a href="https://pixabay.com/users/pexels-2286921/?utm_source=link-attribution&utm_medium=referral&utm_campaign=image&utm_content=1838494">Pexels</a> from <a href="https://pixabay.com//?utm_source=link-attribution&utm_medium=referral&utm_campaign=image&utm_content=1838494">Pixabay</a>

Image by <a href="https://pixabay.com/users/pexels-2286921/?utm_source=link-attribution&utm_medium=referral&utm_campaign=image&utm_content=1866908">Pexels</a> from <a href="https://pixabay.com//?utm_source=link-attribution&utm_medium=referral&utm_campaign=image&utm_content=1866908">Pixabay</a>

Image by <a href="https://pixabay.com/users/gdj-1086657/?utm_source=link-attribution&utm_medium=referral&utm_campaign=image&utm_content=2069810">Gordon Johnson</a> from <a href="https://pixabay.com//?utm_source=link-attribution&utm_medium=referral&utm_campaign=image&utm_content=2069810">Pixabay</a>

Image by <a href="https://pixabay.com/users/coernl-4096898/?utm_source=link-attribution&utm_medium=referral&utm_campaign=image&utm_content=5199393">Cornell Frühauf</a> from <a href="https://pixabay.com//?utm_source=link-attribution&utm_medium=referral&utm_campaign=image&utm_content=5199393">Pixabay</a>

Image by <a href="https://pixabay.com/users/merlinlightpainting-19833603/?utm_source=link-attribution&utm_medium=referral&utm_campaign=image&utm_content=7192239">Merlin Lightpainting</a> from <a

href="https://pixabay.com//?utm_source=link-attribution&utm_medium=referral&utm_campaign=image&utm_content=7192239">Pixabay</a>

Image by <a href="https://pixabay.com/users/michael_laut-3533441/?utm_source=link-attribution&utm_medium=referral&utm_campaign=image&utm_content=1916752">Michael Laut</a> from <a href="https://pixabay.com//?utm_source=link-attribution&utm_medium=referral&utm_campaign=image&utm_content=1916752">Pixabay</a>

Image by <a href="https://pixabay.com/users/tama66-1032521/?utm_source=link-attribution&utm_medium=referral&utm_campaign=image&utm_content=3630911">Peter H</a> from <a href="https://pixabay.com//?utm_source=link-attribution&utm_medium=referral&utm_campaign=image&utm_content=3630911">Pixabay</a>

Image by <a href="https://pixabay.com/users/analogicus-8164369/?utm_source=link-attribution&utm_medium=referral&utm_campaign=image&utm_content=6961638">Tom</a> from <a href="https://pixabay.com//?utm_source=link-attribution&utm_medium=referral&utm_campaign=image&utm_content=6961638">Pixabay</a>

Image by <a href="https://pixabay.com/users/kie-ker-2367988/?utm_source=link-attribution&utm_medium=referral&utm_campaign=image&utm_content=1346727">esiuL</a> from <a href="https://pixabay.com//?utm_source=link-attribution&utm_medium=referral&utm_campaign=image&utm_content=1346727">Pixabay</a>

Image by <a href="https://pixabay.com/users/gruendercoach-13177285/?utm_source=link-attribution&utm_medium=referral&utm_campaign=image&utm_content=8540772">Siegfried Poepperl</a> from <a href="https://pixabay.com//?utm_source=link-attribution&utm_medium=referral&utm_campaign=image&utm_content=8540772">Pixabay</a>

Image by <a href="https://pixabay.com/users/pexels-2286921/?utm_source=link-attribution&utm_medium=referral&utm_campaign=image&utm_content=2178696">Pexels</a> from <a href="https://pixabay.com//?utm_source=link-attribution&utm_medium=referral&utm_campaign=image&utm_content=2178696">Pixabay</a>

**APPENDIX C: VIDEO SOURCES (PIXABAY):**

Video by <a href="https://pixabay.com/users/stefwithanf-795875/?utm_source=link-attribution&utm_medium=referral

&utm_campaign=video&utm_content=13100">Stef Bzt</a> from <a href="https://pixabay.com//?utm_source=link-attribution&utm_medium=referral&utm_campaign=video&utm_content=13100">Pixabay</a>

Video by <a href="https://pixabay.com/users/coverr-free-footage-1281706/?utm_source=link-attribution&utm_medium=referral&utm_campaign=video&utm_content=5631">Coverr-Free-Footage</a> from <a href="https://pixabay.com//?utm_source=link-attribution&utm_medium=referral&utm_campaign=video&utm_content=5631">Pixabay</a>

Video by <a href="https://pixabay.com/users/u_7ifvkc4oku-44955538/?utm_source=link-attribution&utm_medium=referral&utm_campaign=video&utm_content=221469">u_7ifvkc4oku</a> from <a href="https://pixabay.com//?utm_source=link-attribution&utm_medium=referral&utm_campaign=video&utm_content=221469">Pixabay</a>

Video by <a href="https://pixabay.com/users/patw64-16142356/?utm_source=link-attribution&utm_medium=referral&utm_campaign=video&utm_content=133143">Patrick Williams</a> from <a href="https://pixabay.com//?utm_source=link-attribution&utm_medium=referral&utm_campaign=video&utm_content=133143">Pixabay</a>

Video by <a href="https://pixabay.com/users/blendertimer-9538909/?utm_source=link-attribution&utm_medium=referral&utm_campaign=video&utm_content=283533">Daniel Roberts</a> from <a href="https://pixabay.com//?utm_source=link-attribution&utm_medium=referral&utm_campaign=video&utm_content=283533">Pixabay</a>

Video by <a href="https://pixabay.com/users/mcpix22-26289376/?utm_source=link-attribution&utm_medium=referral&utm_campaign=video&utm_content=217115">Daniel McWilliams</a> from <a href="https://pixabay.com//?utm_source=link-attribution&utm_medium=referral&utm_campaign=video&utm_content=217115">Pixabay</a>

Video by <a href="https://pixabay.com/users/engin_akyurt-3656355/?utm_source=link-attribution&utm_medium=referral&utm_campaign=video&utm_content=67358">Engin Akyurt</a> from <a href="https://pixabay.com//?utm_source=link-attribution&utm_medium=referral&utm_campaign=video&utm_content=67358">Pixabay</a>

## APPENDIX D: AUDIO SOURCES (PIXABAY):

https://pixabay.com/music/alternative-hip-hop-vlog-beat-background-349853/

**-END OF THE PAPER-**